\documentclass[%
 reprint,
 amsmath,amssymb,
 aps,
prb,
]{revtex4-2}
\usepackage{natbib}
\usepackage{graphicx}
\usepackage{dcolumn}
\usepackage{bm}
\usepackage{color}
\usepackage{tabularx}
\usepackage{hyperref}

\newcommand{\rmd}{{\rm d}}
\newcommand{\rmi}{{\rm i}}

\usepackage{braket}

\begin{document}


\title{Electrically driven insulator-to-metal transition in a correlated insulator: \\Electronic mechanism and thermal description}
\author{Manuel I. D\'iaz}
\email{manuel.diaz@phys.ens.fr}
\affiliation{Laboratoire de Physique, \'Ecole Normale Sup\'erieure, CNRS, Universit\'e PSL, Sorbonne Universit\'e, Universit\'e de Paris, 75005 Paris, France}
\author{Jong E. Han}
\email{jonghan@buffalo.edu}
\affiliation{Department of Physics, State University of New York at Buffalo, Buffalo, New York 14260, USA}
\author{Camille Aron}
\email{aron@ens.fr}
\affiliation{Laboratoire de Physique, \'Ecole Normale Sup\'erieure, CNRS, Universit\'e PSL, Sorbonne Universit\'e, Universit\'e de Paris, 75005 Paris, France}
 \affiliation{Institute of Physics, \'Ecole Polytechnique F\'ed\'erale de Lausanne (EPFL), CH-1015 Lausanne, Switzerland}

\date{\today}

\begin{abstract}
Motivated by the resistive switchings in transition-metal oxides (TMOs) induced by a voltage bias, we study the far-from-equilibrium dynamics of an electric-field-driven strongly-correlated model featuring a first-order insulator-to-metal transition at equilibrium, namely the dimer-Hubbard model. We use a non-equilibrium implementation of the dynamical cluster approximation to access the steady-state spectral and transport properties. We show that the electric field can drive both  metal-to-insulator and insulator-to-metal transitions. 
While they proceed by quite distinct mechanisms, specifically simple heating of the metal versus non-equilibrium effects in the correlated charge gap, we show that both of these non-equilibrium transitions can be unified in a single framework once the excitations are accounted for in terms of an effective temperature.
This conceptual advance brings together the two sides of the long-lasting debate over the origins of the electrically-driven resistive switching in TMOs.
\end{abstract}

\maketitle


\section{Introduction}

Insulator-to-metal transitions (IMTs) in transition-metal compounds are among the most abrupt first-order phase transitions in nature: temperature variations of a few Kelvins can cause dramatic resistivity drops of three to six orders of magnitude, on timescales of only tens of nanoseconds. For over half a century, theories and experiments have examined the different mechanisms behind these transitions, in particular the role of strong correlation among the electrons~\cite{cox2010,maekawa2004,khomskii2014}. 
These resistive switchings can also be triggered electrically, at ambient temperature and pressure, with relatively small electric fields, $E_{\rm IMT} \sim10^2$ – $10^3$~kV/cm~\cite{sawa2008,sung2015}. This makes them excellent candidates for modern electronic switches with reduced response time and power consumption suitable, in particular, to neuromorphic computing~\cite{cope1968,lin2018,marcelo2018, marcelo2022}.
They proceed with the creation of metallic filaments; these heterogeneous dynamics are a manifestation of a bistable phase where both the insulator and the metal can coexist~\cite{Ridley1963,kogan1969,duchene1971}. This bistability also accounts for the hysteretic behavior by which the metal-to-insulator transition (MIT) occurs at lower threshold fields $E_{\rm MIT} \ll E_{\rm IMT}$.

The colossal energy mismatch between the electron-Volt (eV) electronic scales and the meV scales corresponding to the measured threshold fields  has fuelled a long-standing debate on whether electrically-driven resistive switchings  follow the same mechanisms as in equilibrium, with a temperature increase in the filament driven by Joule heating effects, or whether intrinsically non-equilibrium electronic effects are at play~\cite{stefanovich2000,debate2009,janod2015}.
The case of vanadium oxides has been particularly scrutinized as their thermally-driven IMT is often presented as archetypal of the physics of strongly-correlated electrons. 
Experimental evidence seems to indicate that the resistive switching in pristine Vanadium dioxide (VO$_2$) follows the Joule-heating scenario, whereas the one in Vanadium sesquioxide (V$_2$O$_3$) is thought to result from non-thermal effects~\cite{higgins1977,zimmers2011,zimmers2013,jouleh2016,schuller2017,overheating2020,marcelo2020,schuller2021}. However, the controversy is still fierce and elements of both scenarios could participate towards the transition.

From a theoretical perspective, it is a great challenge to address the electrically-driven resistive switchings in these correlated insulators. On the one hand, we often start with a partial understanding of the \emph{equilibrium} IMT, and on the other hand, it requires solving far-from-equilibrium dynamics of complex open quantum many-body systems.

Conceptual progress has been made in the context of elementary correlated band insulators treated by means of static mean-field techniques, allowing access to analytical solutions~\cite{sugimoto2008,CamilleJongMF}.
The ubiquity of non-equilibrium first-order switchings was unveiled by showing that even when the equilibrium system features a continuous phase transition, such as in V$_3$O$_5$, a finite electric field can open a bistable regime between the metal and the insulator, and turn the resistive switching into a non-equilibrium  first-order transition.
Moreover, it was proposed to unify the two competing scenarios of resistive switching in a single framework. The pivotal concept that has been put forward is the notion of effective temperature, $T_{\rm eff}$, which quantifies the number of electronic excitations irrespective of their thermal or purely non-equilibrium origin~\cite{millis2008}.
In that view, IMT and MIT may proceed by distinct mechanisms, at much different threshold fields, but one should recover the equilibrium phase diagram once the non-equilibrium phase diagram is parameterized in terms of $T_{\rm eff}$.

The last decade has also seen methodological progress with the development of non-equilibrium formulations of Dynamical Mean-Field Theory (DMFT) allowing to treat both the finite electronic interaction in driven-dissipative correlated lattices and their distance to equilibrium in a non-perturbative fashion~\cite{review1996,Freericks2006,okamoto2008,nessdmft2014}. In particular, non-equilibrium steady-state (NESS) implementations offer to bypass time-dependent transient regimes that are otherwise numerically intensive to resolve~\cite{freericks2008,Floquet2008,camille2012,Arrigoni2013,camille2015}. 

Several scenarios of non-equilibrium resistive switching have been studied in the context of the Hubbard model, such as the dielectric breakdown of the Mott insulator by an intense electric field~\cite{Werner2010,camille2012b}, the role of a finite-sized sample~\cite{Mazza2015}, the role of filament formation and heterogeneities~\cite{jongfilament2017}, the role of the nature of the dissipative environment~\cite{millis2018,CamilleJongAvalanche,arrigoni2022}, the role of Hund's coupling~\cite{Hunds2020}, photo-induced resistive switching~\cite{eckstein2021}, etc.
Notably, all the DMFT studies were performed with models featuring a thermally-driven MIT at equilibrium rather than the more common IMT, excluding \textit{e.g.} the case of VO$_2$.

In this manuscript, we harness those techniques to study the electric-field-driven resistive switching of a correlated insulator that features a thermally-driven first-order IMT. Our particular model is inspired by the physics of VO$_2$, with strong intra-orbital correlation and dimerization. We devise a non-equilibrium version of a two-site cluster-DMFT approach to solve for its NESS.
We study the non-equilibrium spectral and transport properties as the electric field is varied. We identify the mechanisms behind both MITs and IMTs, discussing in particular how the self-heating effects are strongly suppressed in the insulator, giving way to truly non-equilibrium electronic mechanisms.
Importantly, we establish that the effective temperature is a fruitful concept that serves as a Rosetta stone between the equilibrium phase diagram and its non-equilibrium counterpart.

The paper is organized as follows. The driven-dissipative model is introduced in Sect. II, and its equilibrium phase diagram is discussed along Fig.~\ref{fig:coex}.
Section III is devoted to describing the non-equilibrium steady-state DMFT methodology we use to solve the Schwinger-Keldysh Green's functions. The results are presented in Sect. IV. In particular, the non-equilibrium phase diagram parameterized in terms of the effective temperature $T_{\rm eff}$ is displayed in Fig.~\ref{fig:pd05}. We discuss the outlooks and conclude in Section V.

\section{Model}

\begin{table}
\centering
\begin{tabular*}{0.95\linewidth}{@{\extracolsep{\fill}}|c|c|c|c|c|c|}
\hline\hline
$\vphantom{\Bigl(}$ $t$ &  $t^\perp$ & $U$ & $\Gamma$ & $T_{\rm IMT}$(VO$_2$) & $E_{\rm IMT}$(VO$_2$)\\
\hline
$\vphantom{\Bigl(}$ 0.25 & 0.3 & 2.5 & $10^{-3}$ & $2.9 \times 10^{-2}$ & $10^{-3}$ -- $10^{-2}$ \\
\hline \hline
\end{tabular*}
\caption{\label{tab:table1}
Typical values (in units of eV) of the parameters of the Dimer-Hubbard model (DHM) in Eq.~(\ref{eq:DHM}), which are pertinent to a description of vanadium dioxide (VO$_2$)~\cite{najera2}.
$E_{\rm IMT}$ values were extracted from Refs.~\cite{zimmers2011,schuller2017} and converted to eV using the potential drop between neighboring sites, $|q|Ea$, and the interatomic distance $a=4.5$~\AA~\cite{dist}.
}
\end{table}

We consider a correlated insulator driven out of equilibrium by a DC electric field and coupled to a heat sink. The total Hamiltonian of the electronic many-body system, its non-equilibrium drive, and its dissipative environment reads
\begin{align}
    H = H_{\mathrm{DHM}} + H_E + H_{\Gamma}\,. \label{eq:DHM}
\end{align}
The electronic system is inspired by the physics of VO$_2$. It is given by the dimer-Hubbard model (DHM) which is a variation on the standard Hubbard model incorporating two ingredients that are now firmly established to be key to the equilibrium IMT in VO$_2$: strong electronic interaction and the dimerization of the vanadium atoms~\cite{biermann2005,shao2018}. The DHM was first proposed in the context of the thermally-driven IMT in VO$_2$ in Ref.~\cite{doniach} and was recently studied in Refs.~\cite{najera1,najera2}.
We should note that a quantitative description of the thermally-driven IMT in VO$_2$ requires more sophisticated modeling~\cite{biermann2005,biermann2008,eyert2011,weber2012,brito2016} whose out-of-equilibrium treatment is simply out of reach for current state-of-the-art non-equilibrium methodologies. The DHM Hamiltonian is given by (we set $\hbar = 1$)
\begin{align}
    H_{\mathrm{DHM}} = &-t \!\! \sum_{\langle ij \rangle a\sigma} \! (c_{ia\sigma}^{\dagger}c_{ja\sigma}+\mathrm{H. c.})
    -\frac{U}{2} \! \sum_{ia\sigma}c^\dagger_{ia\sigma}c_{ia\sigma} 
    \label{dhm} \\
    &
    \!\!+ U \!\sum_{ia} c^\dagger_{ia\uparrow}c_{ia\uparrow}c^\dagger_{ia\downarrow}c_{ia\downarrow}
    -t^\perp \! \sum_{i\sigma}(c^\dagger_{i1\sigma}c_{i2\sigma}+\mathrm{H. c.}). \nonumber
\end{align}
The $c_{ia\sigma}^\dagger$ ($c_{ia\sigma}$) are the creation (annihilation) operators of electrons at the site $i$ of a two-dimensional square lattice, in the orbital $a = 1,2$, with spin $\sigma=\;\uparrow,\downarrow$. In two dimensions, the DHM is also referred to as the bilayer-Hubbard model~\cite{bhm1,bhm2,bhm3,bhm4}.
$t>0$ sets the hopping amplitude between nearest neighboring sites.
$U>0$ sets the strength of the local electronic interaction originating from the intra-orbital Coulombic repulsion. 
$t^\perp > 0$ couples the two orbitals within each dimer site.
At $t^\perp=0$, one recovers two uncoupled copies of the standard single-orbital Hubbard model (SOHM).
A finite $t^\perp$ tends to dimerize the overall system by favoring the formation of local spin singlets at each site.
This model is symmetric under orbital and spin permutations, $1 \leftrightarrow 2$ and $\uparrow\;\leftrightarrow\;\downarrow$ respectively. Furthermore, the presence of the quadratic term in $U/2$ ensures the particle-hole symmetry of the DHM.
Below, we assume that these symmetries are not spontaneously broken. 
\medskip

The electrons are driven out of equilibrium by a constant and uniform external electric field $\boldsymbol{E}$ which we choose to be aligned along the $x$-axis of the square lattice: $\boldsymbol{E} = E \boldsymbol{u}_x$.
After a transient regime, this is expected to generate a steady electric current $\boldsymbol{J} = J \boldsymbol{u}_x$.
We work in the Coulomb gauge where the electric field enters the problem as a ramp potential:
\begin{align}
    H_E = -|q|E\sum_{ia\sigma}x_ic^\dagger_{ia\sigma}c_{ia\sigma}\,,
    \label{he}
\end{align}
where $x_i$ is the spatial coordinate along the $x$-axis of the site $i$ and $-|q|$ is the charge of the electron. We set the interatomic distance $a=1$ and $|q|=1$.

\begin{figure}
\includegraphics[width=\columnwidth]{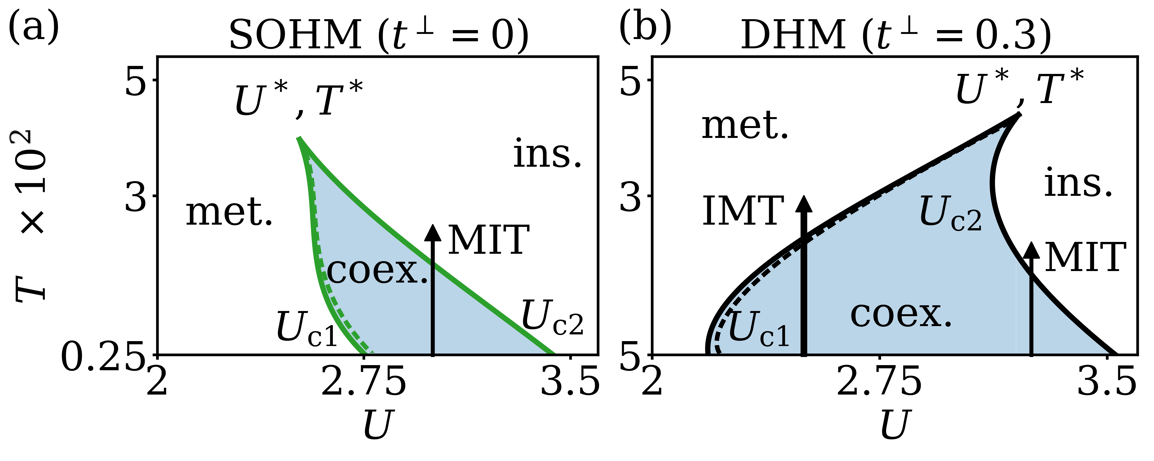}
\caption{\label{fig:coex}
Equilibrium phase diagram in the $U-T$ plane of (a) the paramagnetic single-orbital Hubbard model (SOHM, $t^\perp = 0$) ; and (b) the Dimer-Hubbard model (DHM, $t^\perp = 0.3$) in Eq.~(\ref{eq:DHM}). The solid first-order transition lines delimit the extent of the coexistence region between the metal and the insulator. {The dashed lines are the spinodal curves.} The SOHM paramagnet features a coexistence region whose left-leaning tilt supports temperature-driven metal-to-insulator transitions (MIT). The opposite right-leaning tilt of the DHM brings temperature-driven insulator-to-metal transitions (IMT).
($\Gamma=7.5\times10^{-4}$).}
\end{figure}

\medskip

To allow for non-trivial steady states, it is crucial to include an energy dissipation channel. Otherwise, the work performed on the electronic system per unit of time and volume, $\mathcal{W} = \boldsymbol{J} \cdot \boldsymbol{E} > 0$, would lead to a trivial infinite-temperature steady state.
In practice, we couple the orbitals to independent reservoirs of non-interacting electrons~\cite{buttiker1985, oka2009, camille2012,amaricci2012,jong2013a, jong2013,werner2018,arrigoni2022}
\begin{align}
    H_{\Gamma} = \gamma \sum_{ia\sigma\,l}  c_{ia\sigma}^\dagger b_{ia\sigma l} + \sum_{ia\sigma\, l}  (\epsilon_l - |q|E x_i) b_{ia\sigma l}^\dagger b_{ia\sigma l} \,,
\end{align}
where $\gamma$ sets the hopping amplitude to the reservoirs and $\epsilon_l$ are the many energy levels of the reservoirs. 
We impose that the fermionic reservoirs act as a good thermal bath with no back-action from the system: independently of the state of the system, they remain in equilibrium at the temperature $T$ and chemical potential $\mu - |q|E x_i$.
In practice, we take reservoirs with a flat density of states, $\sum_l \delta(\omega - \epsilon_l) = \rho_{\rm b}$, providing dissipation channels at all energies.
The electronic reservoirs are controlling the electronic filling of the DHM: we set $\mu = 0$ to work at half-filling, \textit{i.e.} with one electron per orbital on average.
Owing to their non-interacting nature, the reservoir degrees of freedom can be explicitly integrated out. Eventually, this simple form of dissipation enters the problem via two energy scales: the bath temperature $T$ (we set $k_{\rm B} =1$) and $\Gamma = \pi\gamma^2\rho_{\rm b}$ which sets the rate at which electrons are exchanged with the environment.
We stress that this choice of a fermionic environment should be seen as a simple heuristic way to single-handedly account for the different channels of energy dissipation present in actual physical systems.

\medskip

We work in units of $4t$, which corresponds to the half-bandwidth of the system at $t^\perp = U = 0$. Incidentally, this roughly corresponds to working in units of eV.
In Table~\ref{tab:table1} we collect the typical values of the model parameters that are experimentally pertinent and that we use throughout this manuscript. The dissipative rate $\Gamma$ was estimated using experimental data, see App.~\ref{app:gamma} for more details.

\begin{figure}
\includegraphics[width=0.8\columnwidth]{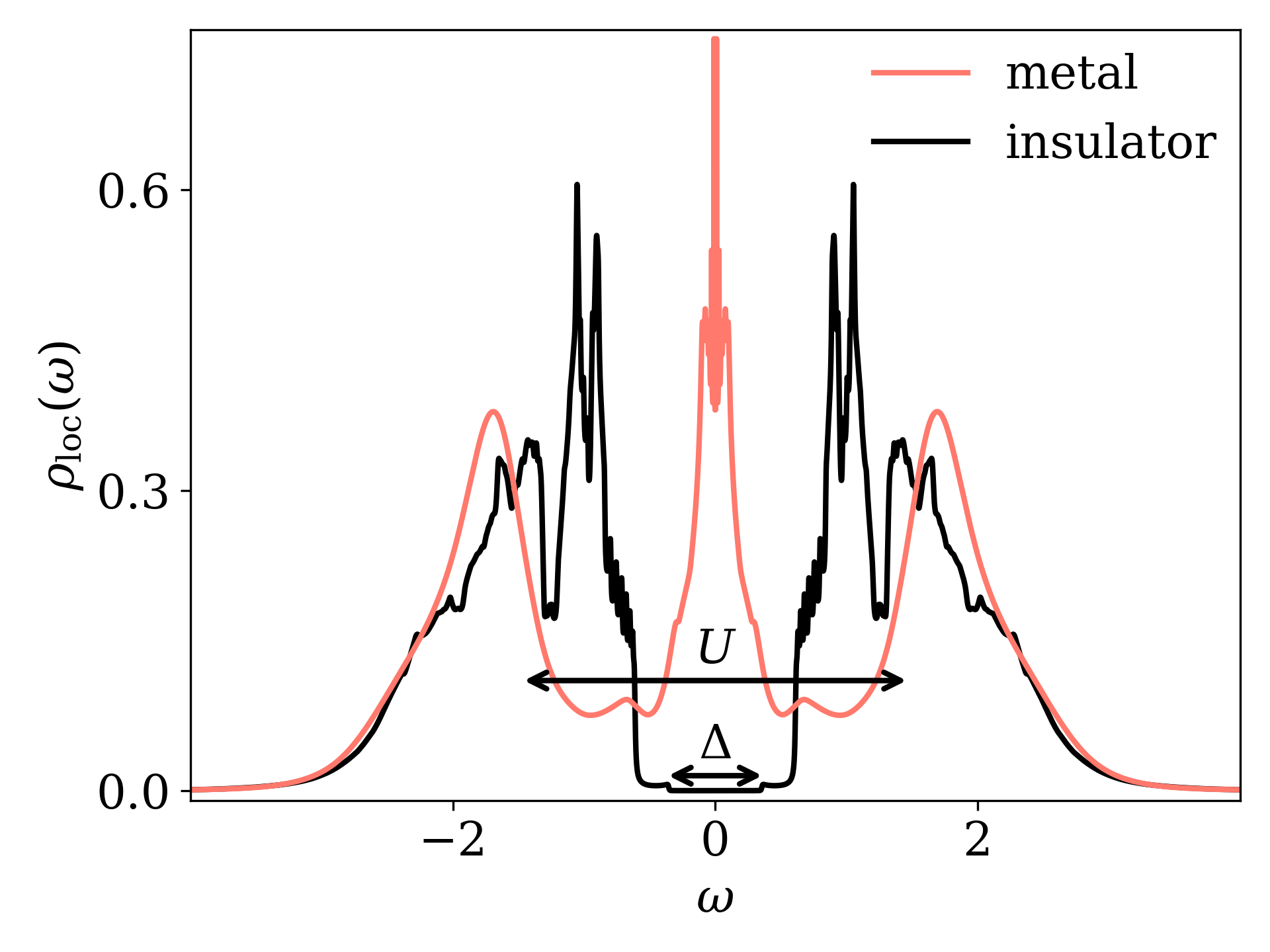}
\caption{\label{fig:coex_spectrum}
Metal/insulator coexistence: equilibrium local density of states (DOS) of the DHM in Eq.~(\ref{eq:DHM}) for a set of parameters corresponding to the coexistence region. In red: the metallic solution. In black: the insulating solution, characterized by a Mott pseudogap controlled by the electronic interaction $U$ which features a cleaner gap $\Delta$ controlled by $t^\perp$, see the close up in Fig.~\ref{fig:gap}~(a). The parameters are $U=3$, $t^\perp=0.3$, $T=2.5\times10^{-3}$, and $\Gamma=7.5\times10^{-4}$.
}
\end{figure}

\medskip

The equilibrium phase diagram (at $E=0$ and $\Gamma=0$) of the DHM has been mapped out in Refs.~\cite{rkkyMott1999,Millis1997,bhm3,najera1,najera2} by means of dynamical mean-field theory methods. 
At $U=0$, the non-interacting Hamiltonian can be diagonalized exactly. For small values of $t^\perp$, the non-interacting system is a metal. For large $t^\perp > 4t$, the system experiences a Peierls transition from a metal to an insulator with the opening of band gap $\Delta^\perp = 2t^\perp-8t$. In this work, following the typical parameter values collected in Table~\ref{tab:table1}, we stay below this transition.

In Figure~\ref{fig:coex}, we report the $U-T$ equilibrium phase diagrams of the SOHM ($t^\perp = 0$) and of the DHM ($0 < t^\perp < 4t$) computed by means of single-site DMFT utilizing a simple impurity solver (namely Iterated Perturbation Theory), where short-range correlations are neglected (see details below). Directly relevant to resistive switching is the metal-to-insulator transition (MIT) driven by a finite electronic repulsion $U$. The corresponding Mott transition proceeds as a first-order phase transition at low temperatures, with the opening of a correlated gap in the density of states.
$U_{c1}$ marks the insulator-to-metal transition (IMT) below which the system is a stable metal, while $U_{c2} > U_{c1}$ marks the MIT transition above which the system is a stable insulator.
In the intermediate region, $U_{c1} < U < U_{c2}$, both the metal and the insulator may coexist.
This is illustrated in Fig.~\ref{fig:coex_spectrum} where we display the local density of states of the coexisting metal and insulator. This coexistence region subsists until a critical temperature $T^*$, where $U_{c1}=U_{c2}=U^*$. For temperatures $T>T^*$, the electronic interaction drives a smooth crossover between a (bad) metal and a (dirty) insulator.

Importantly, within single-site DMFT, the shape and the nature of the coexistence region depends crucially on whether $t^\perp = 0$ or $t^\perp$ is finite. The resulting qualitative picture has been confirmed using a more sophisticated single-site solver~\cite{najera1}.
At $t^\perp = 0$, the overall triangular shape of the coexistence region in the $U$-$T$ plane allows for metal-to-insulator transitions: a metal prepared in the metastable region can be heated up to experience an MIT.
At finite $t^\perp$, the overall tilt of the coexistence region is reversed: an insulator prepared in the metastable region can be heated up so as to cross the IMT line. This presence of a thermally-driven first-order IMT corresponds to the equilibrium physics of VO$_2$.

The difference in the overall tilt of the coexistence region between the (half-filled) paramagnetic SOHM at $t^\perp = 0$ and the DHM at $t^\perp > 0$ is related to the different nature of their respective insulating ground state~\footnote{This difference in the nature of the insulating state is analogous to the difference between the insulating states of the SOHM computed by means of single-site DMFT versus cluster-DMFT. In the latter, short-range correlations are kept such as to allow for a lower entropy insulating ground state.}~\cite{kotliar2008}.
In the SOHM, neglecting short-range correlations, the paramagnetic insulator consists of a collection of independent spins at every site. This is responsible for a macroscopic zero-temperature residual entropy (of value $\log{2}$ per site). 
Comparing the free energies as the temperature $T$ is increased, this favors the insulating solution against the metallic solution for which the Fermi-liquid entropy only grows linearly in $T$.
In the DHM, the insulating ground state corresponds to a collection of pairs of spins locked in spin singlets at every site. This corresponds to a zero entropy state, as for the metal. However, as the temperature $T$ is increased, the entropy of the Mott insulator increases slower than that of the metal because excitations require tunneling through the charge gap. Hence, the free energy of the metal is favored and the DHM undergoes a thermally-driven IMT. 

\begin{figure}
\includegraphics[width=0.47\columnwidth]{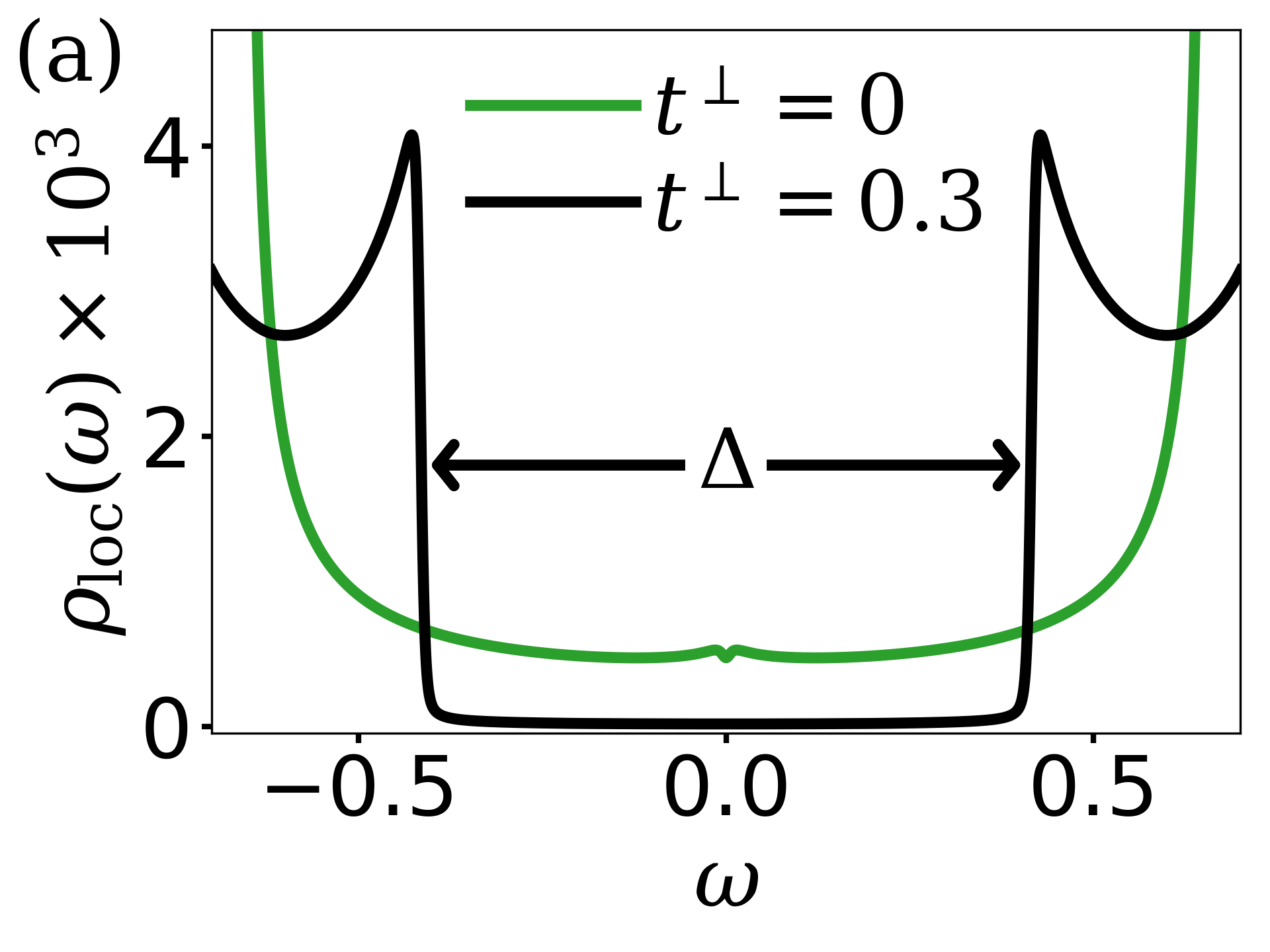}
\includegraphics[width=0.47\columnwidth]{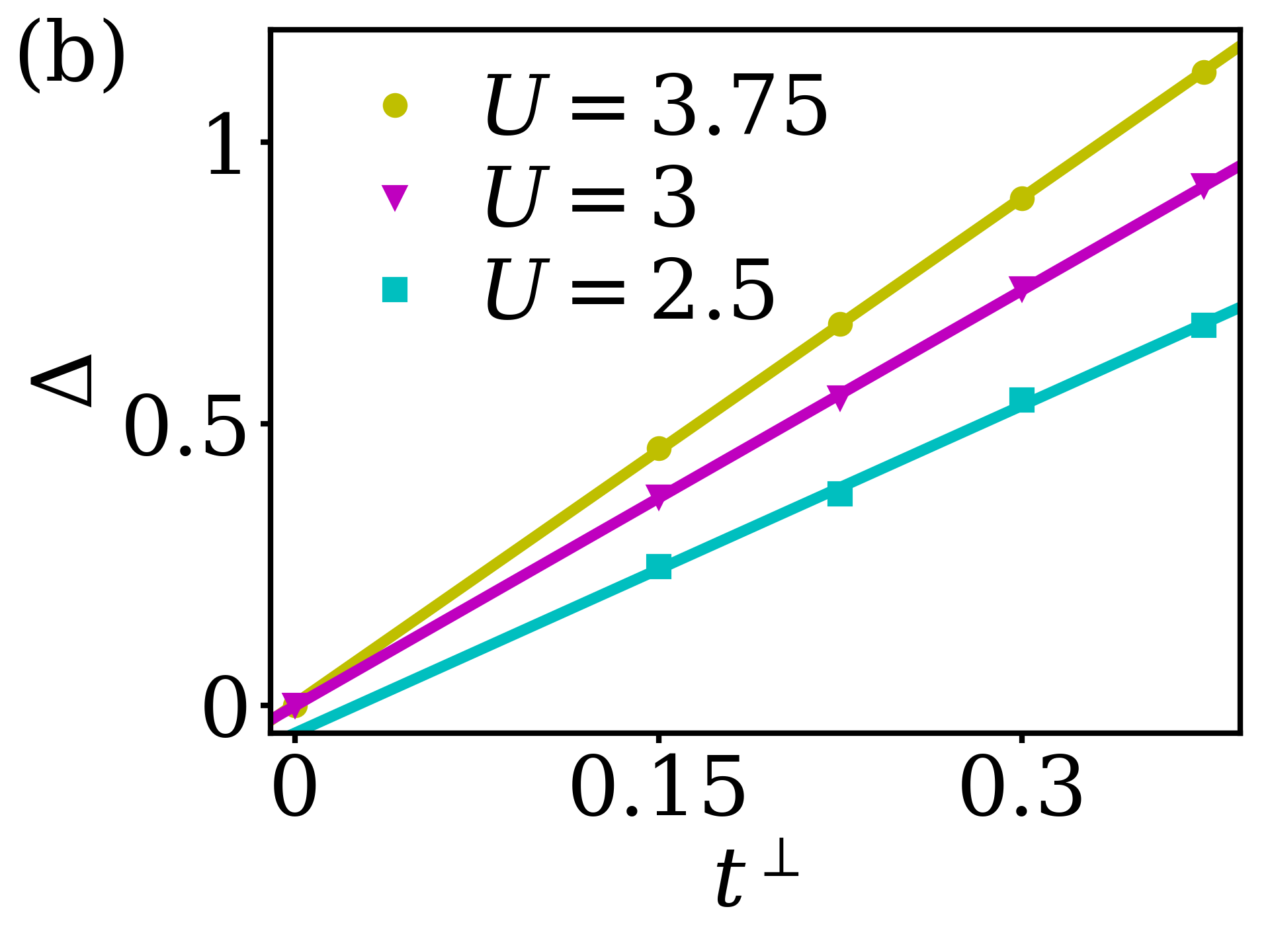}
\caption{\label{fig:gap}
(a) Close-up of the gapped local density of states (DOS) of the SOHM ($t^\perp = 0$) and of the DHM ($t^\perp = 0.3$) at equilibrium for the same $U=3.5$.
The level of the local DOS at the Fermi energy of the DHM is 25 times smaller than for the SOHM.
(b) Width of the gap $\Delta$ as a function of $t^\perp$. The solid lines are linear fits. The parameters are $T=2.5\times10^{-3}$ and $\Gamma = 7.5\times10^{-4}$.}
\end{figure}

This difference in the nature of the insulating state between the paramagnetic SOHM and the DHM has also important consequences on their spectral features. While both insulators exhibit a residual in-gap density of states due to the dissipation $\Gamma >0$, the one of the DHM is smaller by at least one order of magnitude. This difference can be understood as follows.
The in-gap states of the SOHM are predominantly caused by the dissipative leaking of the Hubbard lobes inside the Mott gap: $\rho(\omega  = 0) \sim \Gamma / U^2$~\cite{camille2012b}.
In the DHM case, the local spin singlets are bound tighter as $t^\perp$ increases, and this results in the opening of an additional gap of width $\Delta$ within the correlated gap. 
This gap is illustrated in Fig.~\ref{fig:gap}~(a), and we report the linear dependence of $\Delta$ with respect to $t^\perp$ in Fig.~\ref{fig:gap}~(b). The residual in-gap density of states is now controlled by the local coupling of the singlets to the dissipative reservoirs, $\rho(\omega  = 0) \sim  {t^\perp}^2\Gamma/U^4$, which makes the gap much cleaner than in the SOHM case. See App.~\ref{app:rho0} for more details.

In turn, the extremely low density of states of the DHM insulator makes it impervious to perturbations around the Fermi energy, such as the micro-currents driven by an external electric field. Joule heating is therefore expected to be much less effective than in the case of the SOHM insulator.

\section{Methods}
\label{sec:methods}

We solve the electric-field-driven many-body problem directly in the non-equilibrium steady state (NESS), bypassing the transient dynamics. In practice, the non-equilibrium Green's functions are obtained by solving a Schwinger-Keldysh formulation of the Schwinger-Dyson's equations, assuming that the dynamics reach a well-defined NESS with time and space translation symmetries. Such a functional approach allows us to tackle non-equilibrium regimes far from the linear-response theory and to properly treat the quantum fluctuations of the system and its dissipative environment.

We account for the finite electronic interaction $U$ by means of non-equilibrium DMFT. This mean-field approximation simplifies the task of solving the original extended lattice model by mapping it to a self-consistently determined local impurity model. In our case, the impurity model consists of a dimer site (with two orbitals) coupled to an out-of-equilibrium fermionic bath. This constitutes a non-equilibrium implementation of cluster-DMFT~\cite{cdmft2004,cdmft2005,NeqDCA2014} with a cluster of size two.

\medskip

Given the permutation symmetry between the orbitals 1 and 2 of the DHM Hamiltonian in Eq.~(\ref{dhm}), it is simpler to work in the bonding~(B) - antibonding~(A) basis which diagonalizes the non-interacting problem. In terms of the original orbital degrees of freedom, this amounts to working with the creation operators $c^\dagger_{i{\rm B}\sigma}=(c_{i1\sigma }^\dagger+c_{i2\sigma}^\dagger)/\sqrt{2}$ and $c^\dagger_{i{\rm A}\sigma}=(c_{i1\sigma}^\dagger-c_{i2\sigma}^\dagger)/\sqrt{2}$.
The real-time retarded and Keldysh Green's functions are defined as
\begin{align}
\begin{array}{rl}
    G^{\rm R}_{ij\, m}(t,t') &= -\frac{\rmi}{2}\langle [c_{im\sigma}(t),c^\dagger_{jm\sigma}(t')]\rangle \Theta(t-t') \,,\\
    \vphantom{\biggl(} G^{\rm K}_{ij \, m}(t,t') &= -\frac{\rmi}{2}\langle \{c_{im\sigma}(t),c^\dagger_{jm\sigma}(t')\}\rangle \,,
    \end{array}
    \label{eq:keldysh}
\end{align}
respectively, where $m = {\rm B}, {\rm A}$ is the band index and $\Theta(t)$ is the Heaviside step function. Assuming paramagnetic solutions, and to simplify notations, we dropped the spin index.

\medskip

The DHM Hamiltonian is invariant under translations in the directions perpendicular to the electric field $\boldsymbol{E} = E \boldsymbol{u}_x$. This ensures that the spatial extent of the problem in the $y$-direction can be simply accounted for by the quantum number $k_y$. However, our choice to work with the Coulomb gauge implies that the translational invariance is formally broken along $\boldsymbol{u}_x$. This prompts us to work with Green's functions that are evaluated at equal coordinates along the electric field direction: $G^{\rm R/K}(\omega,k_y ;x_i = x_j)$.
Because we are targeting spatially homogeneous steady-states, all sites are assumed \emph{physically} equivalent and, for convenience, we pick $x_i = x_j = 0$. These Green's functions obey the following Dyson's equations on the lattice~\cite{okamoto2008, camille2015}
\begin{align}
\begin{array}{l}
     G^{\rm R}_{m}(\omega,k_y)^{-1}\! = \omega \!-\!\epsilon^\perp_m(k_y)\!  -\! \Sigma^{\rm R}_{m}(\omega) \!-\!  t^{ 2}F_m^{\rm R} (\omega,k_y),\\
     \vphantom{\biggl(} G^{\rm K}_{m}(\omega,k_y) = \lvert G^{\rm R}_{m}(\omega) \rvert^2 \!  \left[ \Sigma^{\rm K}_{m}(\omega)\! +\! t^{ 2}F^{\rm K}_{m}(\omega,k_y) \right]. 
     \end{array}
     \label{eq:Dyson_lattice}
\end{align}
$\epsilon^\perp_{\rm A}(k_y) = -2t \cos(k_y) + t^\perp$ and $\epsilon^\perp_{\rm B}(k_y) = -2t \cos(k_y) - t^\perp$ are the band dispersion relations in the sublattice corresponding to the directions perpendicular to the electric field.
We introduced the quantities $F^{\rm R/K}_m(\omega,k_y)=F^{\rm R/K}_{m+}(\omega+E,k_y)+F^{\rm R/K}_{m-}(\omega-E,k_y)$ which stem from the hybridization of a given site to the semi-infinite chains of its neighbors along $\pm \boldsymbol{u}_x$.
They obey the self-consistent equations
\begin{align} 
        & F^{\rm R}_{m\pm}(\omega,k_y)^{-1} \!=\! \omega \!-\! \epsilon^\perp_m(k_y)\! -\! \Sigma^{\rm R}_{m}(\omega) \!-\! t^{2}F^{\rm R}_{m\pm}(\omega\! \pm\!  E,k_y),\nonumber \\
        & F^{\rm K}_{m\pm}(\omega,k_y) \!=\! \lvert F^{\rm R}_{m\pm}(\omega,k_y) \rvert^2 \big[ \Sigma^{\rm K}_{m}(\omega) \!+\!t^{ 2}F^{\rm K}_{m\pm}(\omega\! \pm\!  E,k_y) \big]. \label{eq:F}
\end{align}

The self-energies have contributions from the dissipative environment and from the Hubbard $U$ interaction: $\Sigma = \Sigma_{\Gamma} + \Sigma_U$ where
\begin{align}
\begin{array}{rl}
\vphantom{\Bigl(} \Sigma^{\rm R}_{\Gamma}(\omega) &= -\rmi\Gamma \,, \\
\Sigma^{\rm K}_{\Gamma}(\omega) &=-2\rmi\Gamma\tanh{\left(\frac{\omega}{2T}\right)} \,,
\end{array}
\end{align}
and $\Sigma_U$ will be discussed below.
The expression of $\Sigma^{\rm K}_{\Gamma}$ is dictated by the fluctuation-dissipation theorem (FDT), $\Sigma^{\rm K}_{\Gamma}(\omega) = 2 \rmi \tanh{\left(\frac{\omega}{2T}\right)} \mathrm{Im}\, \Sigma^{\rm R}_{\Gamma}(\omega)$, which is applicable since the dissipative environment is assumed to remain in equilibrium.
In Eqs.~(\ref{eq:Dyson_lattice}) and~(\ref{eq:F}) above, we already implemented the DMFT approximation which consists in assuming that $\Sigma_U$ is local, namely that it does not depend on the momentum $k_y$ but only on the frequency $\omega$. 

\medskip

In the spirit of the DMFT, the local Green's functions $G_m^{\rm R/K}(\omega) = \sum_{k_y} G_m^{\rm R/K}(\omega,k_y) $ are identified to those of a quantum impurity problem consisting of a single dimer site coupled to an \textit{ad hoc} non-interacting non-equilibrium environment. The non-interacting Green's functions $\mathcal{G}^{\rm R/K}(\omega)$ of the impurity problem are determined by the following Schwinger-Dyson's equations,
\begin{align}
\begin{array}{rl}
    &\mathcal{G}^{\rm R}_{m}(\omega)^{-1} \! =  G^{\rm R}_{m}(\omega)^{-1} + \Sigma^{\rm R}_{U m}(\omega)\;, \\
\vphantom{\biggl(}    &\mathcal{G}^{\rm K}_{m}(\omega) = \lvert \mathcal{G}^{\rm R}_{m}(\omega)\rvert^2 \left[ \frac{G^{\rm K}_{m}(\omega)}{\lvert G^{\rm R}_{m}(\omega)\rvert^2} - \Sigma^{\rm K}_{U m}(\omega)\right].
    \end{array}
    \label{eq:Dyson_imp}
\end{align}
Note that the Green's functions and the self-energies are diagonal in both the spin and band index due to the original spin and orbital permutation symmetries of the driven-dissipative DHM.

\medskip

The local self-energy $\Sigma_U$ is computed by means of iterated perturbation theory (IPT)~\cite{GeorgesKotliar1992,review1996} which treats the electronic interaction $U$ to second order. IPT has already been used in the context of the DHM in thermal equilibrium~\cite{rkkyMott1999,najera2}.
For a compact expression in the non-equilibrium steady state, it is convenient to go back to the original Kadanoff-Baym-Keldysh contour ($\alpha, \beta =+,-$), to real time, and to the orbital basis ($a,b=1,2$). At half-filling, it reads
\begin{align}
\label{IPTc}
\Sigma^{\alpha\beta}_{U \, ab}(\tau)=
 - \alpha\beta \, U^2 \, \mathcal{G}^{\alpha\beta}_{ab}(\tau)^2\mathcal{G}^{\beta\alpha}_{ba}(-\tau) \, .
\end{align}
The corresponding expression in the Keldysh basis is given in App.~\ref{app:IPT}.

The IPT expressions are valid in the weakly-interacting regime $U/t\rightarrow 0$, and they have also been shown to be exact in the dimer limit $t/U \rightarrow 0$ at half-filling in zero-temperature equilibrium.
For intermediate, finite, values of $U/t$, the IPT provides a crude approximation to the exact impurity self-energy which has already proven to be extremely effective at capturing the qualitative aspects of the Mott equilibrium phase transition of the paramagnetic SOHM~\cite{kotliar1993, review1996}. It has also been successfully used in similar non-equilibrium contexts~\cite{camille2012, camille2012b, camille2015}. Notably, the electric-field induced dimensional crossover at strong fields also justifies the use IPT out of equilibrium since it correctly reproduces the limit $E/U \to \infty$~\cite{camille2012}.
At finite temperatures, the IPT produces spurious in-gap states in the insulating DOS of the DHM. However, we argue in App.~\ref{app:IPT} that they have a very limited impact on the equilibrium and non-equilibrium phase transitions discussed in this manuscript.

\medskip

Our NESS DMFT algorithm proceeds as follows.
We start from an educated guess for the self-energy kernels. In the metallic side, we start from the non-interacting limit $\Sigma^{\rm R}_{U}(\omega) = 0$, and in the insulating side from the non-dissipative dimer limit at zero-temperature, $\Sigma^{\rm R}_{U \, {\rm  B/A}}(\omega) = \frac{U^2}{4}\frac{1}{\omega\mp3t^\perp+\rmi 0^+}$, see App.~\ref{app:IPT}.
($i$) The $F^{\rm R/K}(\omega,k_y)$ are determined by solving self-consistently the set of equations~(\ref{eq:F}).
($ii$) The lattice Greens functions $G^{\rm R/K}(\omega, k_y)$ are computed via Dyson's Eqs.~(\ref{eq:Dyson_lattice}).
($iii$) The impurity non-interacting Green's functions $\mathcal{G}^{\rm R/K}(\omega)$ are determined via Dyson's Eqs.~(\ref{eq:Dyson_imp}).
($iv$) The self-energies $\Sigma^{\rm R/K}_U(\omega)$ are updated using the IPT expression in
Eq.~(\ref{IPTc}).
Steps ($i$) to ($iv$) are repeated until convergence is achieved. When varying the electric field, we use the previously converged solution as the educated guess for the self-energy kernels.

\section{Results}
Let us now present the fate of the driven-dissipative DHM when the electric field is varied by studying both its non-equilibrium spectral and transport properties.

\subsection{Non-equilibrium spectral features}

\begin{figure}
\includegraphics[width=1\columnwidth]{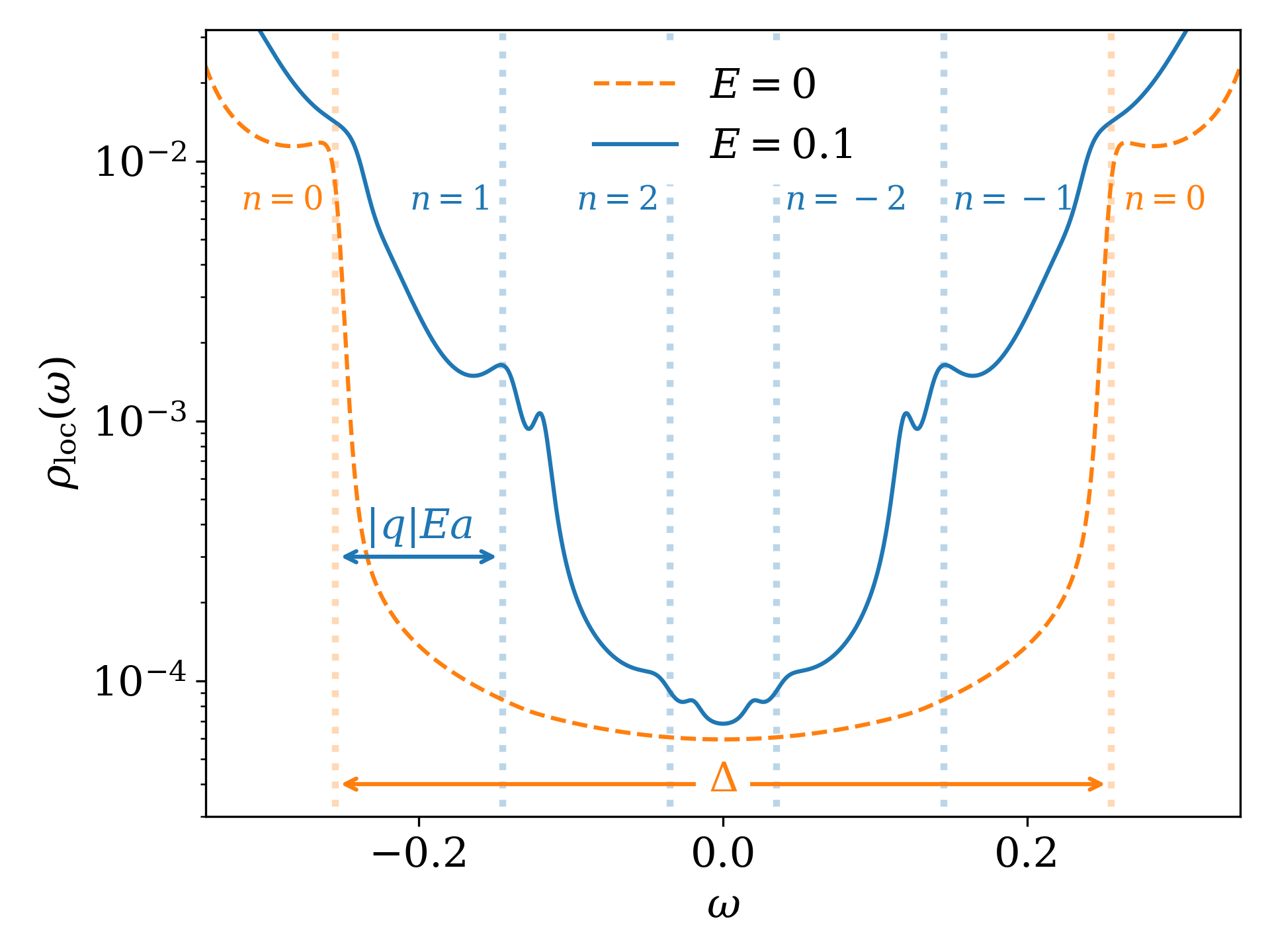}
\caption{\label{fig:islands}
Close-up of the gapped local density of states in the non-equilibrium DHM insulator (note the semi-log scale). 
At equilibrium ($E=0$, dashed line), $t^\perp$ opens a clean gap of width $\Delta \approx 0.5$ around the Fermi energy.
In the presence of the electric field ($E=0.1$, solid line), the edges of the gap are replicated into Bloch-Zener side-bands located at $\pm \Delta/2 + n |q|Ea$, $n\in\mathbb{Z}$.
The parameters are $t^\perp=0.3$, $U=2.5$, $\Gamma=7.5\times10^{-4}$, $T=2.5\times10^{-3}$.}
\end{figure}

We first discuss the local density of states of an orbital,
$\rho_{\rm loc}(\omega) =- \frac{1}{\pi} \sum_{k_y} \frac12 \sum_m \mathrm{Im}\, G_m^{\rm R}(\omega,k_y)$, when the electric field $E$ is turned on.
The electric field has little impact on the density of states of the metal.
However, it induces distinctive features in the density of states of the insulator: in-gap states are created within the correlated gap, sometimes referred to as Bloch-Zener or Wannier-Stark side-bands.
This is illustrated in Fig.~\ref{fig:islands}.
The origin of these in-gap structures is easily understood in the context of an electric-field-driven non-interacting one-dimensional band insulator: distinct regions of finite density of states appear beyond the edges of the bands. They are distant in energy by multiples of the potential drop $|q|Ea$ and are exponentially suppressed in the gap on a scale $t^{1/3}(|q|Ea)^{2/3}$~\cite{wilkins1988}. Notably, their spectral content is inherited from the spectral content of the band insulator at equilibrium but in lower dimensions: the dimensions perpendicular to the electric field.
A similar scenario was already identified and described in the context of a correlated insulator (SOHM) in Refs.~\cite{camille2012b, werner2018, arrigoni2022}. 
In the case of the DHM with a finite $t^\perp$, the edges of the gap located at $\pm \Delta/2$ are responsible for Bloch-Zener side-bands located at $\pm \Delta/2 + n |q|Ea$ ($n\in\mathbb{Z}$), see Fig.~\ref{fig:islands}. Features corresponding to values of $|n| > 2$ are strongly suppressed.

\subsection{Energy distribution function}

\begin{figure}
\includegraphics[width=0.45\columnwidth]{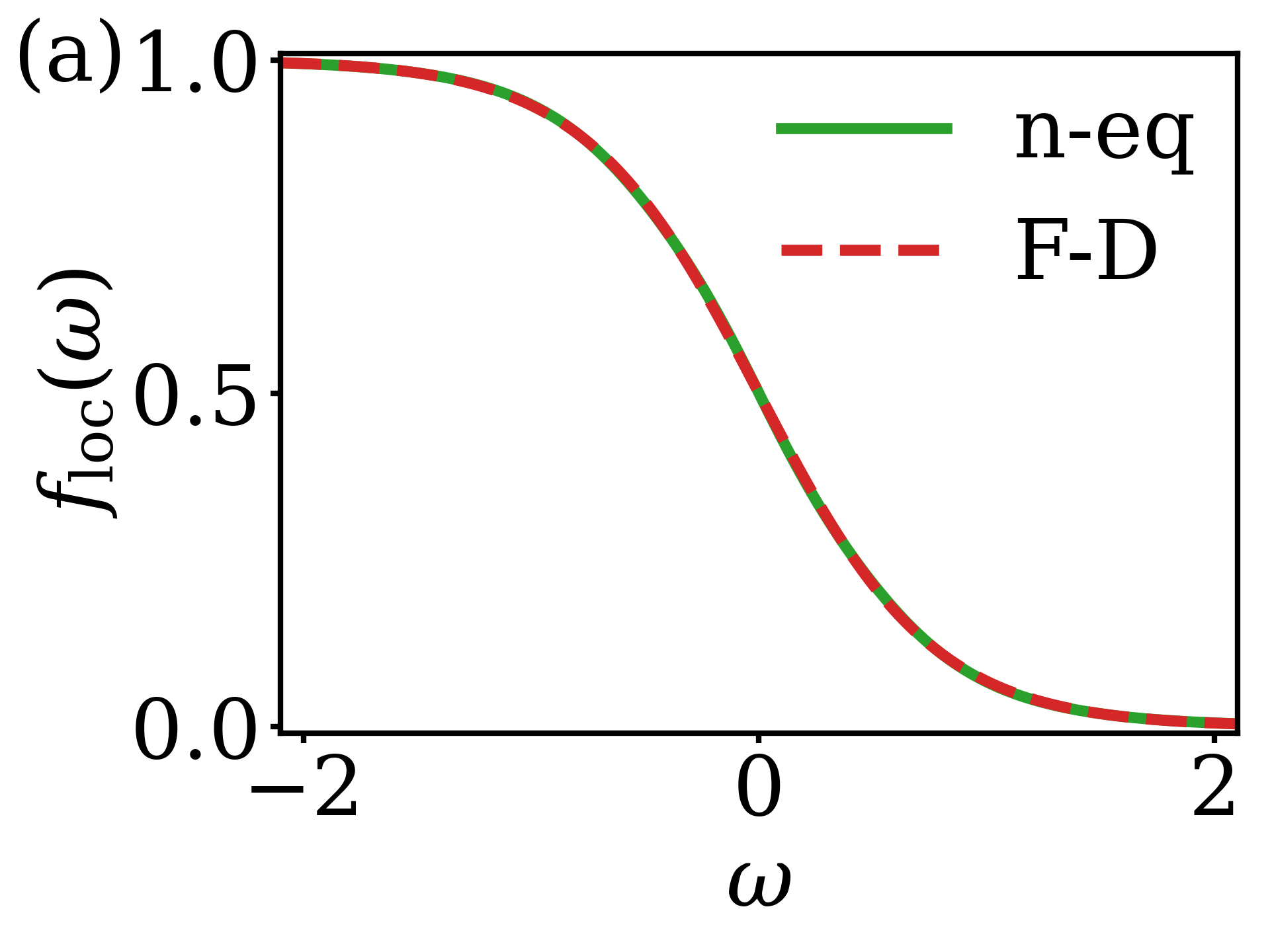}
\includegraphics[width=0.45\columnwidth]{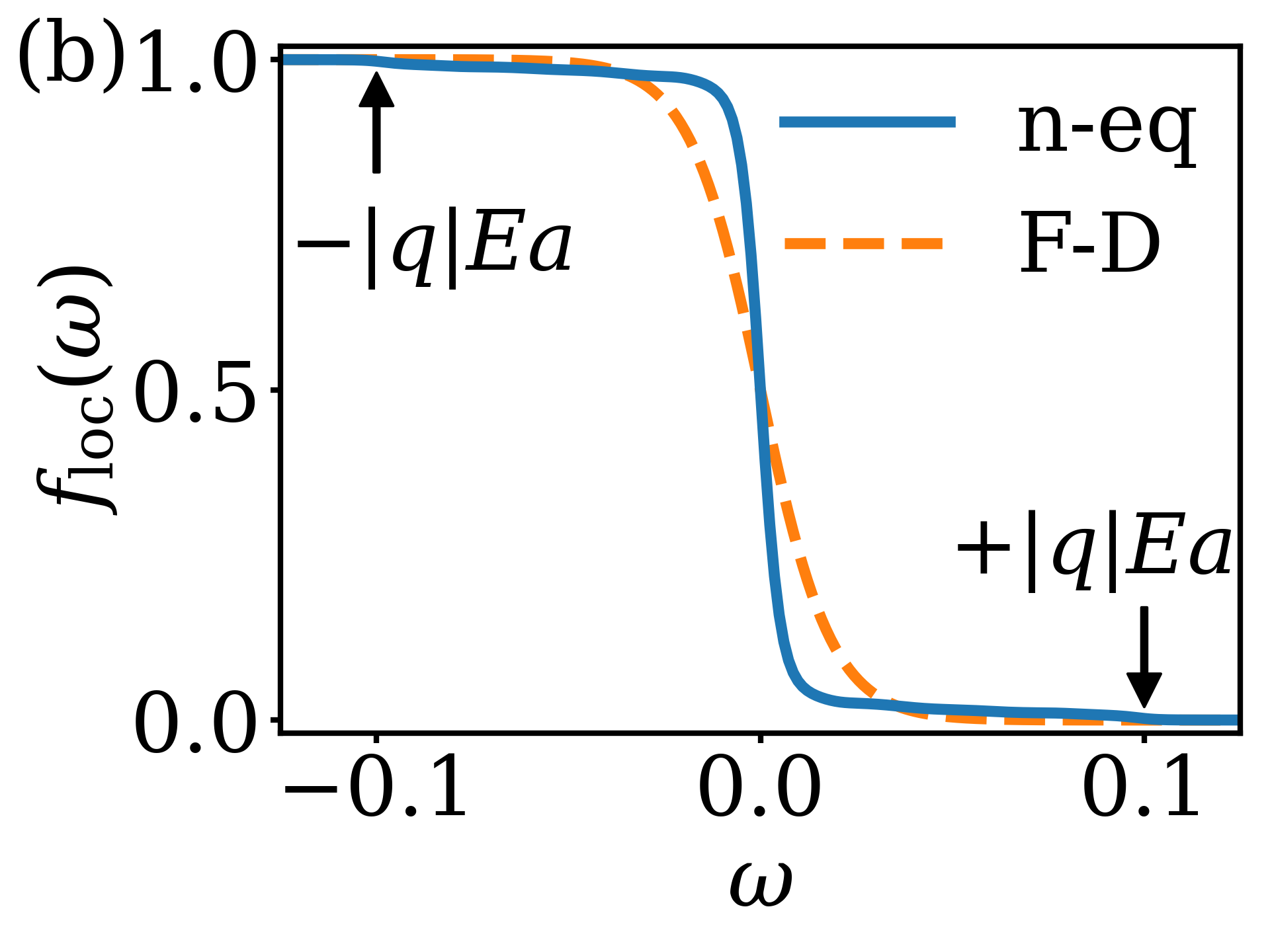}
\caption{\label{fig:dfstep} 
Local distribution function $f_{\rm loc}(\omega)$ of the electric-field-driven DHM, defined in Eq.~(\ref{eq:f_loc}).
(a) In the metallic phase, $f_{\rm loc}(\omega)$ (solid line) cannot be distinguished from a Fermi-Dirac distribution (dashed line) at the temperature {$T_{\rm eff}=0.38$} determined via Eq.~(\ref{eq:teff}).
(b) In the insulating phase, $f_{\rm loc}(\omega)$ {corresponds to $T_{\rm eff}=9.3\times10^{-3}$ but it markedly departs from a Fermi-Dirac distribution}.
{Both examples were taken at the same bath temperature $T=2.5\times10^{-3}$, and parameters $E=0.1$, $t^\perp=0.3$, $U=2.5$, and $\Gamma=7.5\times10^{-4}$}.}
\end{figure}

Let us now discuss the local distribution function of an electronic orbital
\begin{align}
 f_{\rm loc}(\omega)=\frac{1}{2}\left(1-\frac{1}{2}\frac{\sum_m \mathrm{Im}\, G^{\rm K}_m(\omega)}{\sum_m \mathrm{Im}\, G^{\rm R}_m(\omega)}\right),   \label{eq:f_loc}
\end{align}
which informs on the non-equilibrium energy fluctuations. Similarly to the DOS, the electric field has a mild impact on the energy distribution function of the metal: it perfectly matches the equilibrium Fermi-Dirac function albeit at a higher temperature; see Fig.~\ref{fig:dfstep} and the discussion on effective temperature below.
On the other hand, the energy distribution function of the insulator significantly departs from equilibrium: it features structures equally spaced by the energy $|q|Ea$. These structures have already been reported in studies regarding the SOHM. They can be seen as a consequence of the coupling of the impurity site to its neighboring sites in the direction of the field, $\pm \boldsymbol{u}_x$. These sites act as baths with chemical potentials shifted by the potential drop $\pm |q| Ea$~\cite{camille2013}.

\subsection{Non-equilibrium phase diagram}

\begin{figure}
\includegraphics[width=0.8\columnwidth]{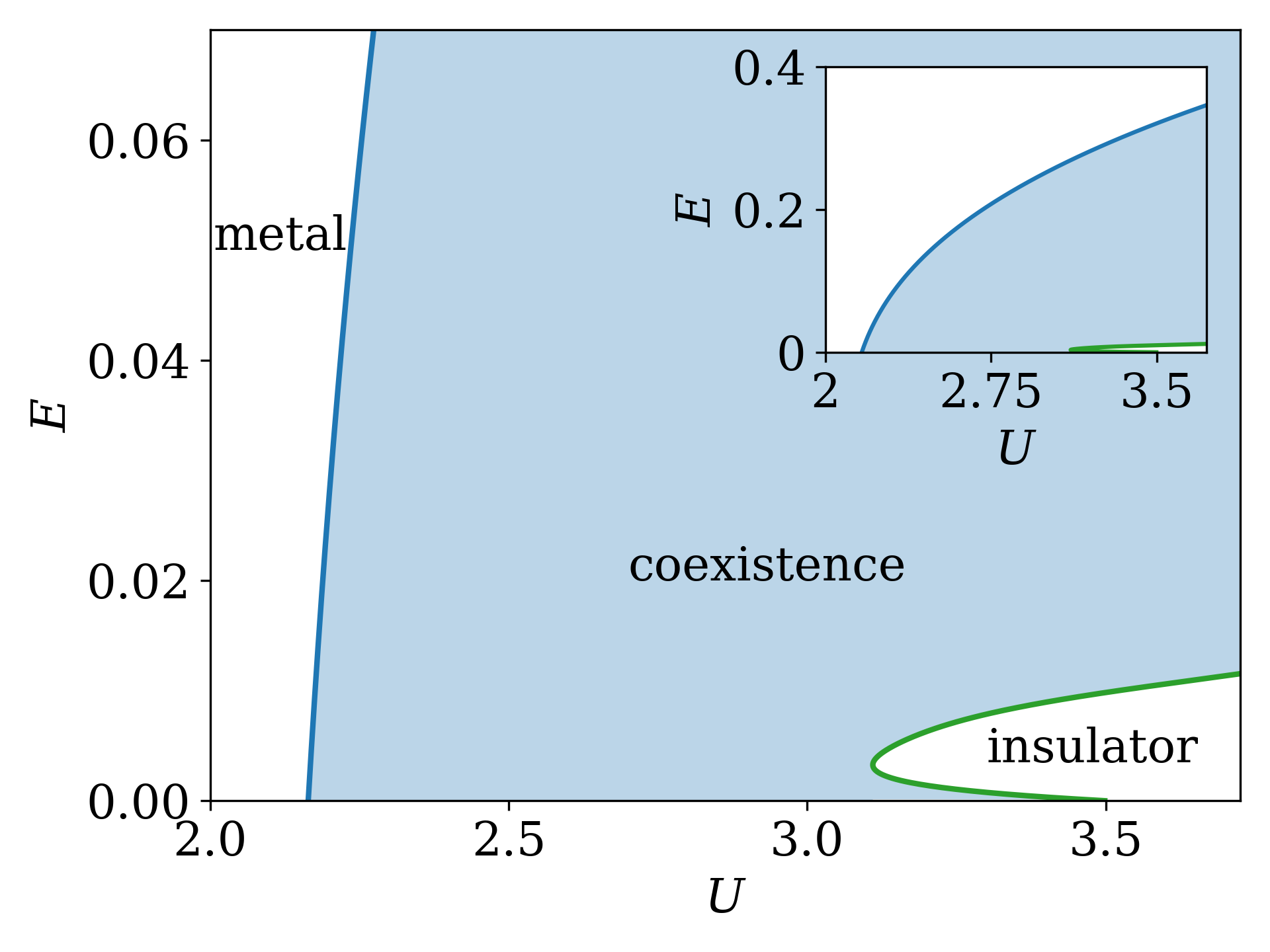}
\caption{\label{fig:t1PD}
Non-equilibrium phase diagram of the electric-field-driven Dimer-Hubbard model (DHM) in the $U-E$ plane. 
The IMT (MIT) line is in blue (green). The inset shows a larger region of the coexistence region. The parameters are $t^\perp=0.3$, $T=2.5\times10^{-3}$, and $\Gamma=7.5\times10^{-4}$.
}
\end{figure}

\begin{figure}[h]
\centering
\includegraphics[width=0.49
\columnwidth]{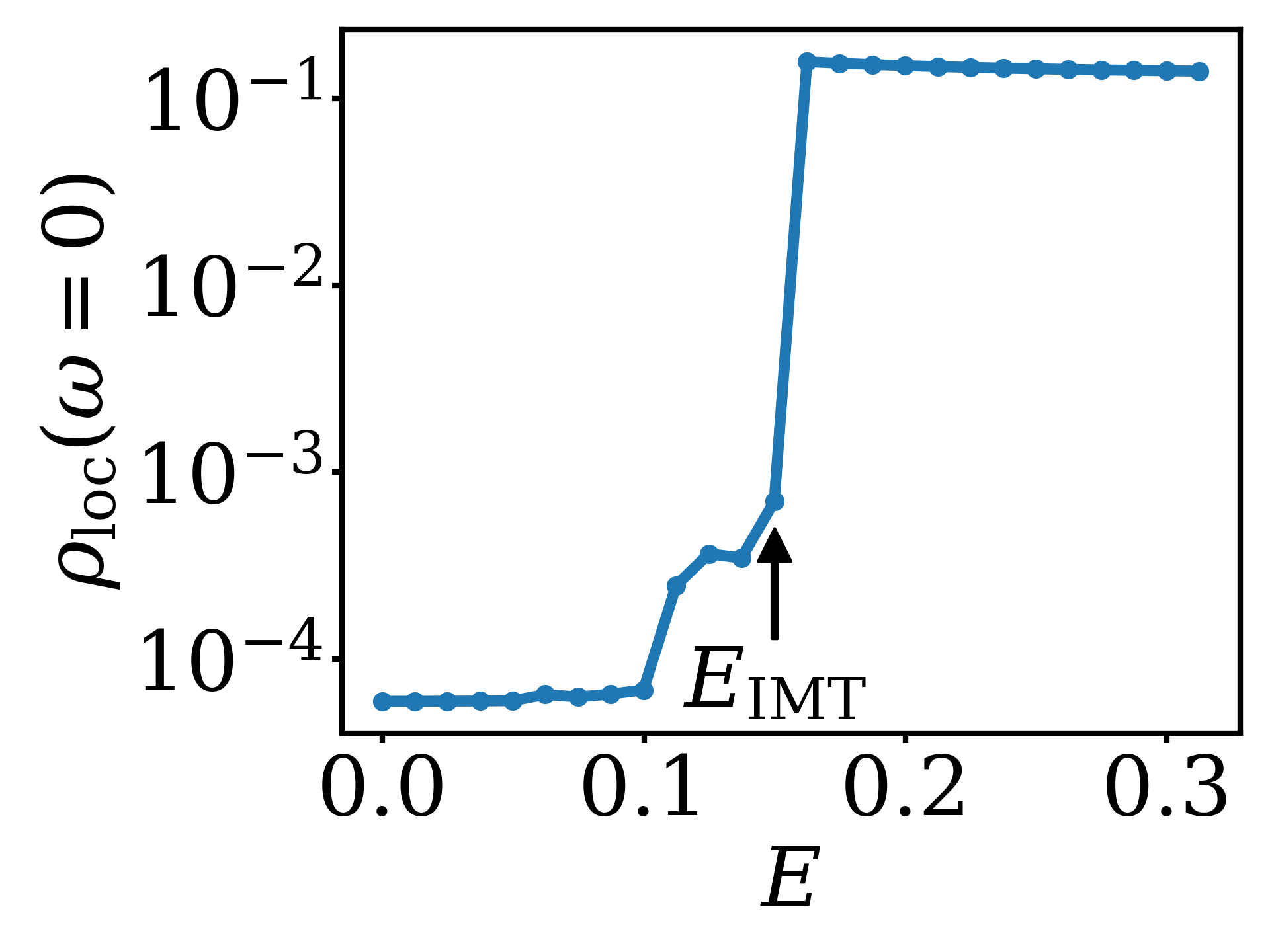}
\caption{\label{fig:E_rho0} {Local density of states at the Fermi level as a function of the electric field. Starting from an insulating solution at $E=0$, the sharp leap of two orders of magnitude is used to locate the insulator-to-metal transition (IMT).
The parameters are $U=2.5$, $t^\perp=0.3$, $T=2.5 \times 10^{-3}$, and $\Gamma=7.5\times10^{-4}$.
}}
\end{figure}

Let us now present the non-equilibrium phase diagram of the electric-field-driven DHM.
It is computed in Fig.~\ref{fig:t1PD} in the $U-E$ plane for a fixed value of $t^\perp$.
The different phases are determined depending on the presence of a gap at the Fermi level in the local density of states, or the lack thereof.
{In practice, we track the value of $\rho_{\rm loc}(\omega=0)$, which presents a sharp variation of at least 2 orders of magnitude between phases, as is shown in Fig.~\ref{fig:E_rho0}.}
Similarly to the equilibrium phase diagram presented in Fig.~\ref{fig:coex}, the solid lines delimit a region (shaded blue) where both the metal and the insulator coexist.
However, while the IMT and MIT lines (blue and green, respectively) have general trends that resemble their equilibrium counterparts, the electric-field scales on which they develop are drastically smaller on the MIT side than on the IMT side: $E_{\rm IMT} \gg E_{\rm MIT}$. This results in a very wide coexistence region that extends to very large values of $U$ and $E$.
For very large electric fields, outside of the ranges of Fig.~\ref{fig:coex}, the IMT and MIT become smooth crossovers between bad metals and dirty insulators.

This is quite in contrast to the case of the electric-field-driven SOHM whose corresponding non-equilibrium phase diagram studied {in Ref.~\cite{camille2015} for $T=5\times10^{-3}$ and $\Gamma=3.4\times10^{-3}$} was found to sport a closed coexistence region qualitatively similar to its equilibrium phase diagram in the $U-T$ plane,  \textit{c.f.} the $t^\perp =0$ case in Fig.~\ref{fig:coex}.

\subsection{Transport properties}

Let us now discuss the far-from-equilibrium transport properties of the electric-field-driven DHM.
The electric field generates an electric current per site and per spin $\boldsymbol{J} = J \boldsymbol{u}_x$ which is computed as,
\begin{align}
    J &= \frac{t^2}{2} \nonumber\sum_m \int\frac{\rmd\omega}{2\pi} \sum_{k_y}\left\{\mathrm{Im}G^{\rm R}_m(\omega,k_y)\mathrm{Im}F^{\rm K}_{m}(\omega,k_y)\right.\nonumber\\
    &\left.\qquad\qquad\qquad -\mathrm{Im}G^{\rm K}_m(\omega,k_y)\mathrm{Im}F^{\rm R}_{m}(\omega,k_y) \right\},
\end{align}
where the first summation is performed over the bonding ($m = {\rm B}$) and antibonding ($m={\rm A}$) bands.

\begin{figure}
\centering
\includegraphics[width=0.49\columnwidth]{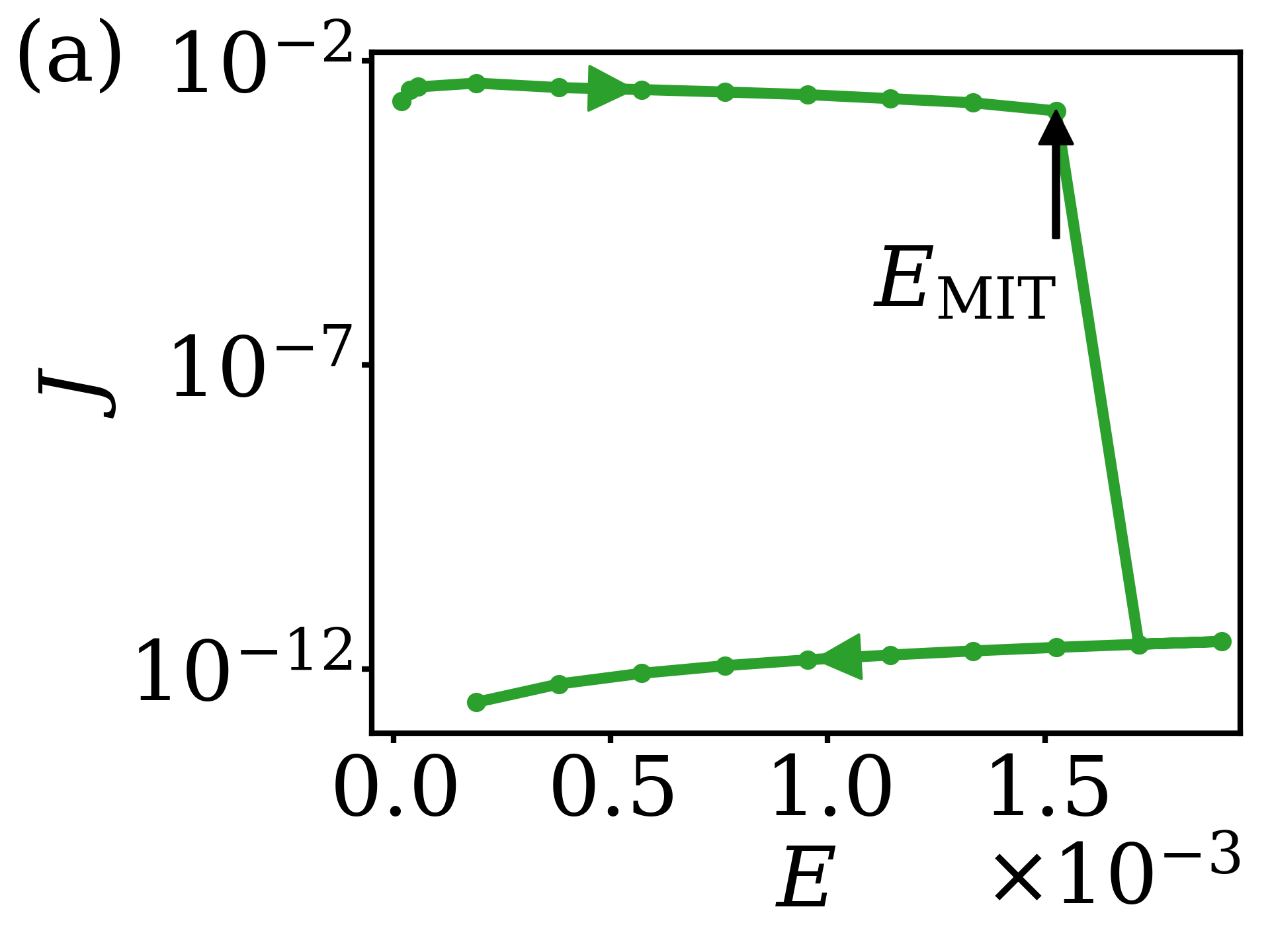}
\includegraphics[width=0.49
\columnwidth]{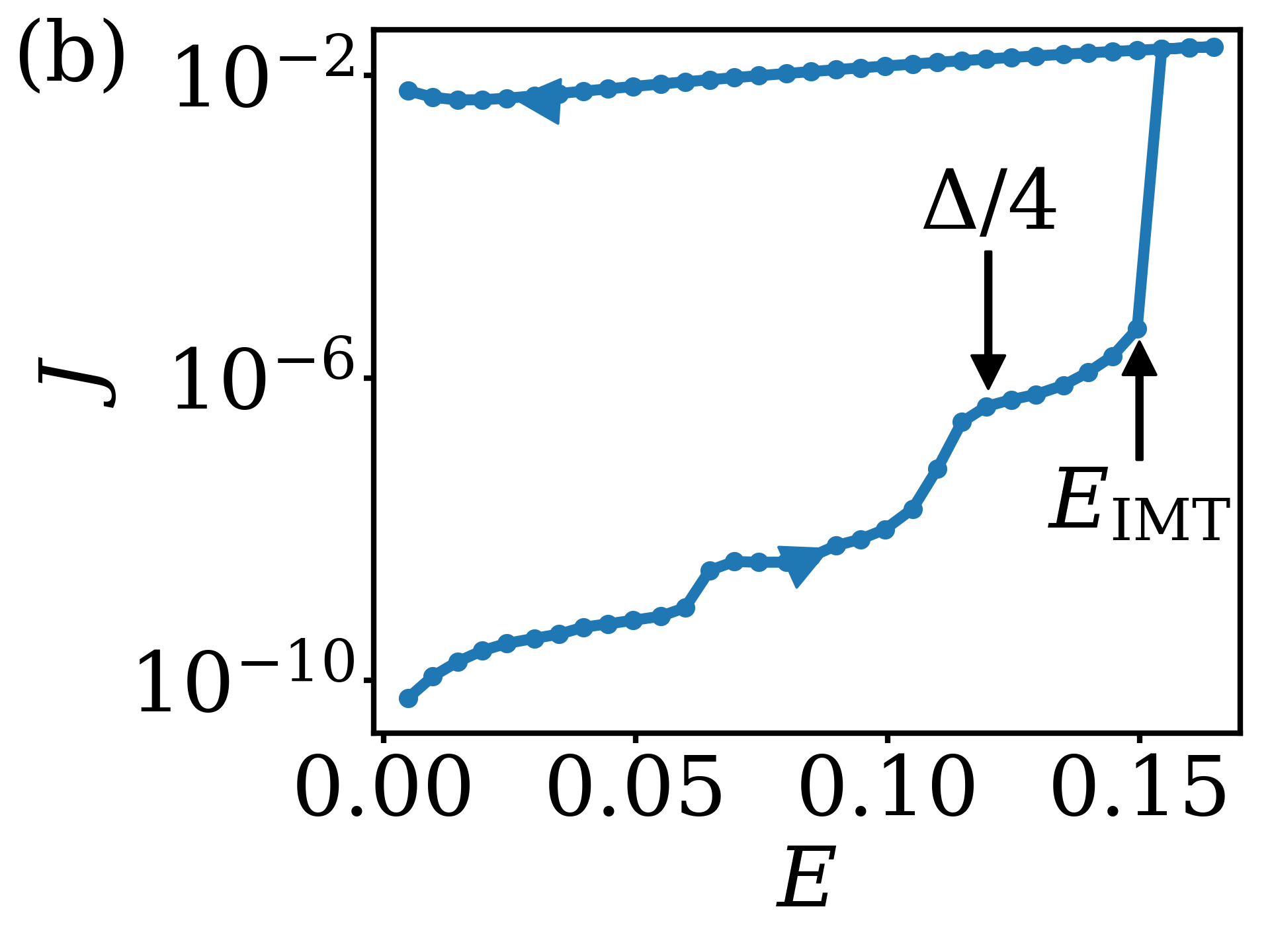}
\caption{\label{fig:J}
Hysteretic $I-V$ characteristics of the DMH (note the semi-log scale). 
(a) Starting from a metallic state in the coexistence region at $U =3.25$, the electric field is increased until the DHM experiences an MIT at $E_{\rm MIT}$. Later, the electric field is decreased and the DHM remains insulating. 
(b) Starting from an insulating state in the coexistence region at $U =2.5$, the electric field is increased until the DMH experiences an IMT at $E_{\rm IMT} \gg E_{\rm MIT}$. 
The parameters are $t^\perp=0.3$ and $\Gamma=7.5\times10^{-4}$.
}
\end{figure}

In Fig.~\ref{fig:J}, we illustrate the $I-V$ characteristics of the DHM starting from low-temperature equilibrium states in the coexistence region and ramping up, and then down, the electric field.
The $I-V$ curves display hysteresis both when starting from the metal or the insulator.  
When starting from a metallic state in the coexistence region, one first expects a linear regime  $J =\sigma_{\rm dc} E$, where $\sigma_{\rm dc}$ is the DC conductivity. The latter quickly renormalizes as the temperature of the sample increases, relegating the linear regime down to extremely small values of electric field which cannot be seen on the scale used here~\cite{camille2015}.
At $E = E_{\rm MIT}$, the DHM experiences an MIT where the current density drops by 9 orders of magnitude. The DHM remains an insulator after a subsequent decrease in the electric field.

When starting from an insulating state in the coexistence region, the applied electric field has little effect on the slowly increasing current. Indeed, the very clean nature of the gap opened by $t^\perp$, as discussed above, makes $J$ much suppressed compared to the case of the SOHM. The Bloch-Zener structures described above have an impact on the current at finite electric fields: the current is enhanced whenever they provide a pathway through the gap. Indeed, states at the lower edge of the gap can be excited by a sequence of electric-field driven transitions of energy $|q|Ea$. 
This happens at values $|q|Ea = \Delta / m$, where $m$ is a small positive integer. In Fig.~\ref{fig:J}, this is seen for $m=4$ (corresponding to $n=2$ in Fig.~\ref{fig:islands}). Smaller values of $m \leq 3$ correspond to field strengths that lay outside the range of the figure and after the IMT. Larger values of $m \geq 5$ correspond to exponentially suppressed BZ in-gap structures and to smaller features in the $I-V$ which are hard to isolate.
Note that these Bloch-Zener effects cannot be seen in the MIT case given the much smaller values of the electric field involved.
As the electric field is furthermore increased, the DHM experiences an IMT at $E = E_{\rm IMT} \gg E_{\rm MIT}$. 
Note the violent drop of resistivity by 4 orders of magnitude. 
The DHM remains metallic after a subsequent decrease in the electric field.

\subsection{Effective temperature}

\begin{figure}
\centering
\includegraphics[width=0.49\columnwidth]{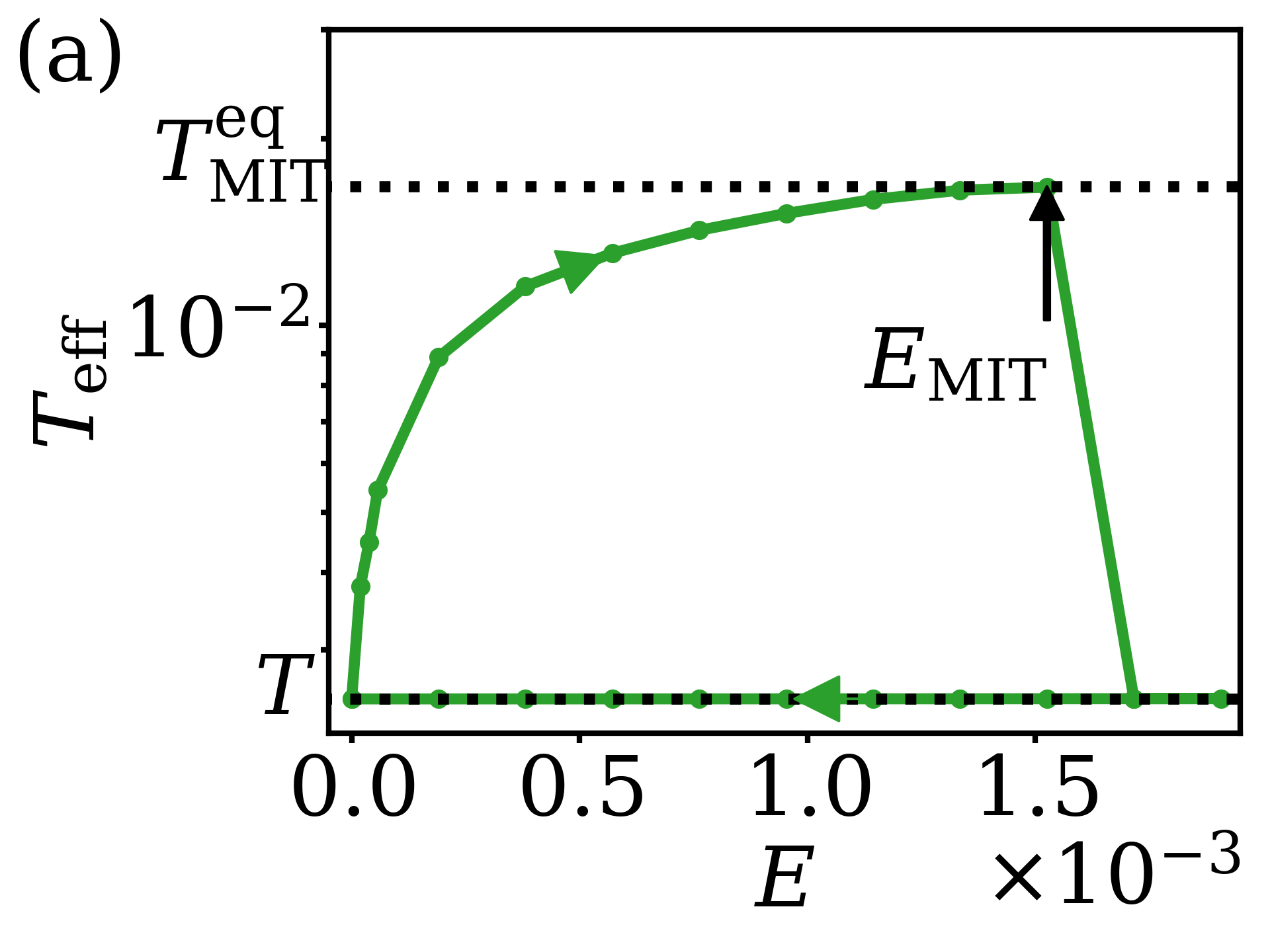}
\includegraphics[width=0.49\columnwidth]{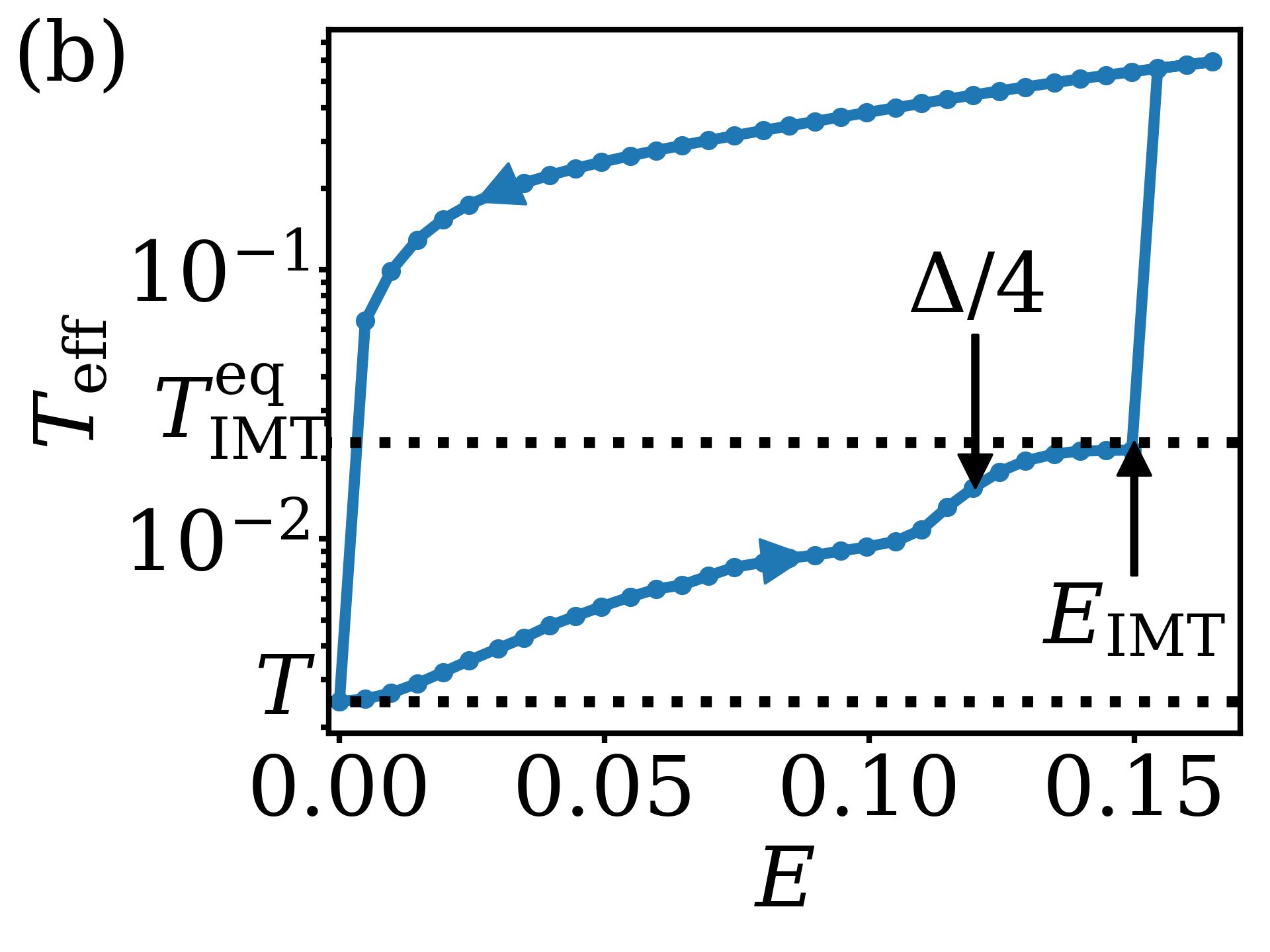}
\caption{\label{fig:Teff}
Effective temperature $T_{\rm eff}$ computed using Eq.~(\ref{eq:teff}) as a function of the increasing and decreasing electric field (note the semi-log scale). (a) Starting from a metal with the same parameters as in Fig.~\ref{fig:J}(a).
(b) Starting from an insulator with the same parameters as in Fig.~\ref{fig:J}(b). 
Both MIT and IMT take place once $T_{\mathrm{eff}}$ reaches the corresponding equilibrium transition temperature, $T_{\rm MIT}^{\rm eq}(U=3.25)=0.017$ and $T_{\rm IMT}^{\rm eq}(U=2.5)=0.023$, marked by horizontal dashed lines.}
\end{figure}

Let us now discuss the thermodynamic properties of the electric-field-driven DHM.
The concept of temperature is \textit{a priori} ill-defined for a non-equilibrium steady state. Indeed, a single scalar quantity cannot in principle inform on the full energy and momentum-dependent non-equilibrium distribution function.
In practice, the notion of effective temperature has however proven useful.
In the long-lasting debate on the microscopic origin of the resistive switching, a strong advocate for a thermally-driven scenario has been a series of experiments where the effective temperature of the sample was monitored \textit{in situ} via the emission of a fluorescent particle acting as a local temperature probe~\cite{zimmers2013}.
Moreover, we have already shown that the non-equilibrium local distribution function of the metallic state is a Fermi-Dirac function with a renormalized temperature (see Fig.~\ref{fig:dfstep}).
In that spirit, we adopt a Sommerfeld-like definition for the local effective temperature~\cite{jongfilament2017}:
\begin{align}
\label{eq:teff} 
    T_\mathrm{eff}^2 = \frac{6}{\pi^2}\int \rmd\omega \, \omega \left[ f_{\rm loc}(\omega)-\Theta(-\omega) \right] \,,
\end{align}
where the local orbital energy distribution function $f_{\rm loc}(\omega)$ has been defined in Eq.~(\ref{eq:f_loc}).
The above definition of $T_{\rm eff}$ can be seen as a simple measure of the ability of the system to create and sustain excitations above the Fermi level. Naturally, $T_{\rm eff}$ boils down to the thermodynamic temperature in equilibrium, when $f_{\rm loc}(\omega)$ is the Fermi-Dirac distribution $f_{\rm FD}(\omega) = [1+\exp(\omega/T)]^{-1}$.
Note that other choices could have been made to extract a $T_{\rm eff}$ from $f_{\rm loc}(\omega)$, such as fitting it to a Fermi-Dirac distribution or using its slope at the Fermi energy, $-\left(\partial_\omega f(\omega)\lvert_{\omega=0}\right)^{-1} /4$. 
{We checked that our results are qualitatively robust with respect to these alternative choices.}
The strength of our definition lies in the fact that it does not require any fitting parameter and it does not require $f_{\rm loc}(\omega)$ to be close to a Fermi-Dirac distribution.

In Fig.~\ref{fig:Teff}, we monitor the effective temperature of the DHM starting from a low-temperature equilibrium state and ramping up, and then down, the electric field. Similarly to Fig.~\ref{fig:J}, we start from both a metallic state and an insulating state in the coexistence region. 
In both cases, $T_\mathrm{eff}$ grows monotonously with $E$, starting from the bath temperature $T$ at $E=0$. 
Naturally, the stronger currents produced in the metallic states are responsible for higher effective temperatures than in the insulating states. This is reflected in the huge discontinuities of $T_\mathrm{eff}$ at the transitions. 
Moreover, the high $T_\mathrm{eff}$ after the IMT induces states known as `bad metals', characterized by a low (but not gapped) density of states around the Fermi energy. 
A first outcome of Fig.~\ref{fig:Teff} is the notion of state-dependent effective temperature: similarly to the hysteretic $I-V$ curves, the effective temperature has a region of bistability tied to the metastability of the state.
In other words, for the same system parameters and the same electric field, the effective temperature greatly depends on the state the system, whether insulating or metallic.
A second  outcome of Fig.~\ref{fig:Teff} is the fact that both non-equilibrium transitions, IMT and MIT, occur whenever the effective temperature matches the corresponding equilibrium transition temperature (represented with dotted lines).

\begin{figure}
\includegraphics[width=1\columnwidth]{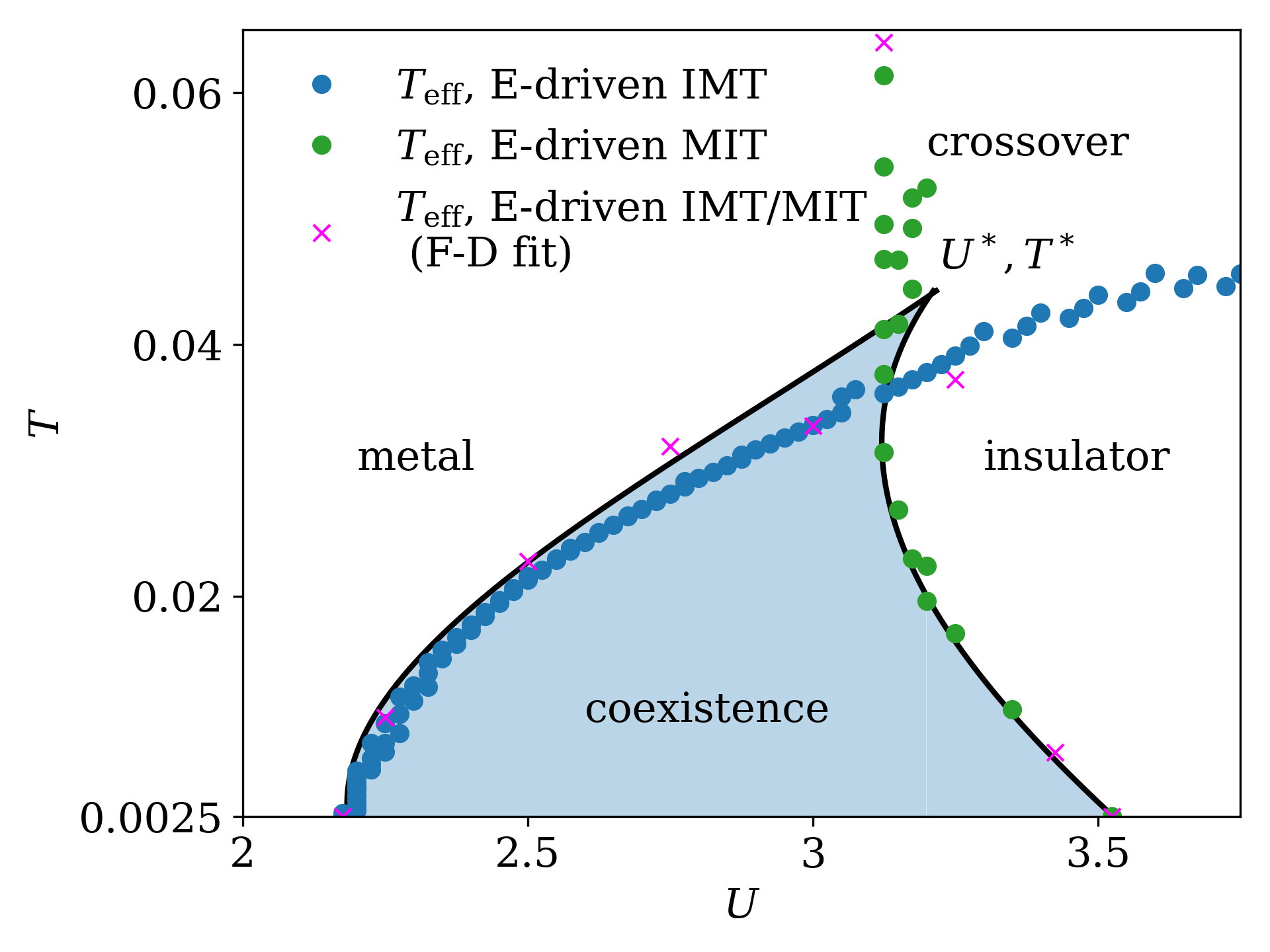}
\caption{\label{fig:pd05} 
Phase diagram of the electric-field-driven DHM in the $U-T_{\rm eff}$ plane.
The effective temperatures $T_{\rm eff}$ were computed using Eq.~(\ref{eq:teff}).
The blue (green) dots correspond to $T_{\rm eff}$ computed just at the IMT (MIT) induced by increasing the electric field. {The magenta crosses indicate instances of $T_{\rm eff}$ at the transition computed by fitting $f_\mathrm{loc}(\omega)$ to a Fermi-Dirac distribution.}
The solid lines correspond to the equilibrium IMT and MIT driven by increasing the temperature.
The parameters are $t^\perp=0.3$, $T=2.5\times10^{-3}$, and $\Gamma = 7.5\times10^{-4}$.
}
\end{figure}

In Fig.~\ref{fig:pd05}, we repeat the previous analysis on the entire non-equilibrium phase diagram. We prepare low-temperature equilibrium metallic and insulating samples at all values of $U$ in the coexistence region, we increase the electric field, and we report the effective temperatures $T^{\rm eq}_{\rm MIT}$ and $T^{\rm eq}_{\rm IMT}$ measured exactly at MIT and the IMT, respectively.
Remarkably, we confirm the previous observations made in Fig.~\ref{fig:Teff}:
\begin{align}
\begin{array}{rl}
 T_{\rm eff}(E_{\rm IMT}; \mathrm{insulator}) &\approx T^{\rm eq}_{\rm IMT}\,, \\
 \vphantom{\biggl( }T_{\rm eff}(E_{\rm MIT}; \mathrm{metal}) &\approx T^{\rm eq}_{\rm MIT}\,.   
\end{array}\label{eq:TT_EQ}
 \end{align}
This matching is one of the main results of this manuscript.
Equations~(\ref{eq:TT_EQ}) account for the large difference in magnitude of the threshold fields $E_{\rm IMT} \gg E_{\rm MIT}$ even though $T^{\rm eq}_{\rm IMT}$ and $T^{\rm eq}_{\rm IMT}$ are on the same order of magnitude.
Notably, this is in agreement with the experimental findings of Refs.~\cite{zimmers2013,marcelo2020} where the resistivity of VO$_2$ driven by a DC bias was found to match the equilibrium resistivity, once the voltage bias is parameterized in terms of the local effective temperature of the sample.
It was already demonstrated numerically for the MIT occurring in the electric-field-driven dissipative SOHM~\cite{camille2015}, and it was given some analytical insight in the context of a simplified one-dimensional model with a correlated gap treated within mean-field theory~\cite{CamilleJongMF}.
This result articulates both the thermal and the electronic scenarios for resistive switching in a unified framework:
once the non-equilibrium electronic mechanisms are accounted for by an effective temperature, the non-equilibrium phase transition reduces to the thermally-driven equilibrium phase transition.

\section{Discussion and conclusion}

In this work, we investigated a driven-dissipative version of the DHM in which both first-order MIT and IMT can be driven by a DC electric field rather than temperature.

The field-driven MIT was found to follow a simple Joule-heating scenario in which the driven metal essentially remains in local thermal equilibrium despite the finite electronic current and heat flow. The balance between the power injected by the electric field and the heat released to the environment brings the sample to its equilibrium transition temperature, triggering the MIT at threshold fields controlled by the dissipation rate, $E_{\rm MIT} \sim \sqrt{\Gamma}$~\cite{camille2015}, and corresponding to meV energy scales.

The field-driven IMT was found to follow a different scenario in which both the inter-orbital hopping and Bloch-Zener (BZ) effect play a crucial role. Indeed, $t^\perp$ is responsible for opening a clean gap $\Delta$ within the pseudogap of the Mott insulator.
Such a clean gap provides a very good insulating ground state where Joule-heating effects are strongly suppressed, auguring very large threshold fields to overcome the gap $\Delta$. However, we found a substantial contribution of the BZ effect to Joule-heating: the electric field is responsible for the formation of in-gap states, located at energies given by multiple values of the electric field, which can form a non-equilibrium pathway to bridge the gap. Altogether, we found threshold fields controlled by fractions of the gap,
$E_{\rm MIT}  \sim \Delta / 4$. This large 100~meV scale is at odds with the available experimental data on threshold fields. This calls for further investigation in the modeling of these driven-dissipative correlated materials. A promising avenue is to explore the influence of a phononic dissipative environment rather than the simple electronic dissipation used here~\cite{millis2018,arrigoni2022}. Notably, it was recently reported that the interplay of the electric field with such a bosonic bath could create a strong non-equilibrium pathway through the gap resulting in electronic avalanches with fields at sub meV scales~\cite{CamilleJongAvalanche}.

 {
 Importantly, the study of the electric-field-driven SOHM and DHM has shown that there are two key factors in building up toward the electric-field-driven transition:  the nature of the insulating state, and its hybridization with a dissipative environment.
The MIT in both models proceeds with a simple Joule heating mechanism involving excitations at the Fermi level. The IMT in the SOHM proceeds similarly but the in-gap excitations are produced by the hybridization with the dissipative bath. For the IMT in the DHM, the low-energy insulating state is impervious to the bath hybridization and the destabilization of the insulator comes from a higher energy mechanism, namely, the Bloch-Zener effect.
The fact that these three distinct non-equilibrium scenarios can be unified in a single equilibrium framework once their effects are measured in terms of effective temperature is a non-trivial observation.
We believe this observation is not limited to the particular model at hand and generalizes to other cases of RS in correlated materials with possibly other microscopic mechanisms at play.
}


{
Most experiments show IMTs with spatially heterogeneous solutions in the form of metallic filaments which were not discussed in this manuscript. Since the insulating phase is spatially homogeneous, it is reasonable to think that both the mechanism driving the onset of the switching and the value threshold field do not involve heterogeneous dynamics. More generally, our homogeneous solutions can be seen as mesoscopic building blocks for a larger sample with a heterogeneous landscape.
Relaxing the assumption of homogeneity would require the use of inhomogeneous DMFT~\cite{bakalov2016,Okamoto2007,okamoto2008,Potthoff1999a,Potthoff1999b} which is computationally challenging. Let us rather highlight the steps for developing an effective field theory description for the electric-field-driven RS which leverages the teachings of our microscopic analysis and operates at a coarse-grained level.
}
With our results in mind, this can simply be achieved by starting from the existing equilibrium field theories and by replacing the thermodynamic temperature by the state-dependent effective temperature $T_{\rm eff}$. In the context of the Hubbard model, we can therefore start from the results of Refs.~\cite{MeanFieldKotliar1,MeanFieldKotliar2,MeanFieldKotliar3,MeanFieldKotliar4} and propose an effective free-energy functional of the form
\begin{align}
    \mathcal{F}_{\rm NESS}[\phi] =& \int \rmd^d x \frac12 (\nabla_x \phi)^2 + \frac12 \vec{\mu} \cdot \left(\vec{K}(\phi) - \vec{K}^* \right) \phi^2 \nonumber \\ 
    & \qquad + \frac14 \lambda \phi^4 -  \vec{h} \cdot \left(\vec{K}(\phi) - \vec{K}^* \right) \phi \; ,
\end{align}
where the field $\phi$ is the scalar order parameter, $d$ is the dimension of the space, $\vec{K}(\phi)-\vec{K}^* = \left(T_{\rm eff}(\phi;E) -T^*,U- U^*\right)$ is a two-component vector which measures the distance to criticality, $\vec{\mu}$ and $\vec{h}$ are two-component parameters, and $\lambda > 0$. In practice, the free energy has to be supplemented with kinematic and geometric constraints on the fields such as the Maxwell and Kirchhoff laws.
Such a low-energy description, to be presented in Ref.~\cite{futureus}, promises to bridge the gap between the {microscopic} computations discussed in this manuscript and the heuristic models of resistor networks that have successfully been used to account for the filament formation observed in experiments~\cite{MarceloResistorNetworks1,MarceloResistorNetworks2}.

\medskip

\paragraph*{Acknowledgements.}
MD and CA acknowledge the support from the French ANR ``MoMA'' project ANR-19-CE30-0020.
CA is grateful for the support from project 6004-1 of the Indo-French Centre for the Promotion of Advanced Research (IFCPAR). We warmly thank Marcelo Rozenberg and Marcello Civelli for inspiring discussions. MD wishes to thank Lorenzo Fratino and Soumen Bag for many helpful conversations and advice.

\appendix

\section{Estimation of $\Gamma$ from experimental data}
\label{app:gamma}
Heat transfer to the electronic baths, controlled by the single rate $\Gamma$, is the only dissipative mechanism included in the driven-dissipative DHM given in Eq.~(\ref{eq:DHM}). In reality, the dissipative environment of the active electronic degrees of freedom is more complex with, in particular, the coupling to lattice phonons. Consequently, the heuristic parameter $\Gamma$ is to be understood as an effective dissipative rate that aims to encompass the different dissipative processes that are not explicitly included in our model.
In this Section, we discuss how we estimated from experimental data the value $\Gamma=7.5\times10^{-4}$~eV used throughout this manuscript. 

In a steady state, the electric field is constantly pumping energy into the system, which is dissipated into the reservoirs. The rate of heat exchange with the reservoir per unit volume is computed as~\cite{jong2013}

\begin{equation}
\label{eq:Q}
    \mathcal{Q}=2\Gamma\sum_m\int \rmd\omega\rho_m(\omega)\left[f_m(\omega)-f_\mathrm{th}(\omega)\right]\omega\,.
\end{equation}

\noindent Index $m$ labels the bonding and antibonding bands ${\rm B}$ and ${\rm A}$, and $\rho_m(\omega)$, $f_m(\omega)$ are the band-specific DOS and local energy distribution functions. $f_\textrm{th}(\omega)$ is the energy distribution function of the bath, which is a Fermi-Dirac function at temperature $T$.

According to the first law of thermodynamics, the heat and work rates compensate in a steady state, $\mathcal{W+Q}=0$, and we therefore can determine $\mathcal{W}$ through Eq.~(\ref{eq:Q}).

In a metallic state, we can reasonably approximate $f_m(\omega)\approx f_{\rm loc}(\omega)$, and $\rho_m(\omega)\approx1/8t$ for $\omega\sim0$, where the integrand in the {\sc r.h.s.} of~(\ref{eq:Q}) is nonzero. Furthermore, using the definition of $T_{\rm eff}$ in Eq.~(\ref{eq:teff}), and reinstating the universal constants, we obtain an expression for the dissipative rate $\Gamma$ that can be used to match to experimental data:

\begin{equation}
\label{eq:gammaexp}
    \Gamma = \frac{12}{\pi^2} \frac{\hbar}{k_{\rm B}^2}\frac{t\mathcal{W}}{T_\textrm{eff}^2-T^2}.
\end{equation}

We extract the corresponding experimental values from Ref.~\cite{jouleh2016}. In this experiment, a thin film of VO$_2$ of volume $6,0\times10^{-21}$~$\mathrm{m^3}$, placed on a TiO$_2$ substrate (which plays the role of the bath), is driven through the transition by an external electric field. In the presence of this substrate, the critical temperature is brought down to $T^{\rm eq}_{\rm IMT}=270$~K. $T_{\rm eff}$ is shown to reach this value at the electrically-driven transition, which is where we choose to evaluate the {\sc r.h.s.} of Eq.~(\ref{eq:gammaexp}). The total power needed to bring the system to the transition is measured to be $\mathcal{W}\times N_{\rm s}=6$~mW for the bath temperature $T=230$~K. $N_{\rm s}$ denotes the number of vanadium atoms in the sample, which we estimated from the volume of the sample and the lattice constant $a=4.5$~\AA (caption, Table~\ref{tab:table1}).

Replacing the experimental values above in Eq.~(\ref{eq:gammaexp}), we compute an estimate for the dissipative rate $\Gamma \sim 10^{-3} \textrm{eV}$. In the main text, we use a slightly smaller value of $7.5\times10^{-4}$~eV. This ensures that $\Gamma$ remains the smallest energy scale in the system (below the bath temperature $T$) and we stay within the bounds of the IPT approximation (see App.~\ref{app:IPT}).

\section{Iterated Perturbation Theory (IPT) for the DHM}
\label{app:IPT}

\subsection{Practical expressions}

\begin{figure}
\includegraphics[width=\columnwidth]{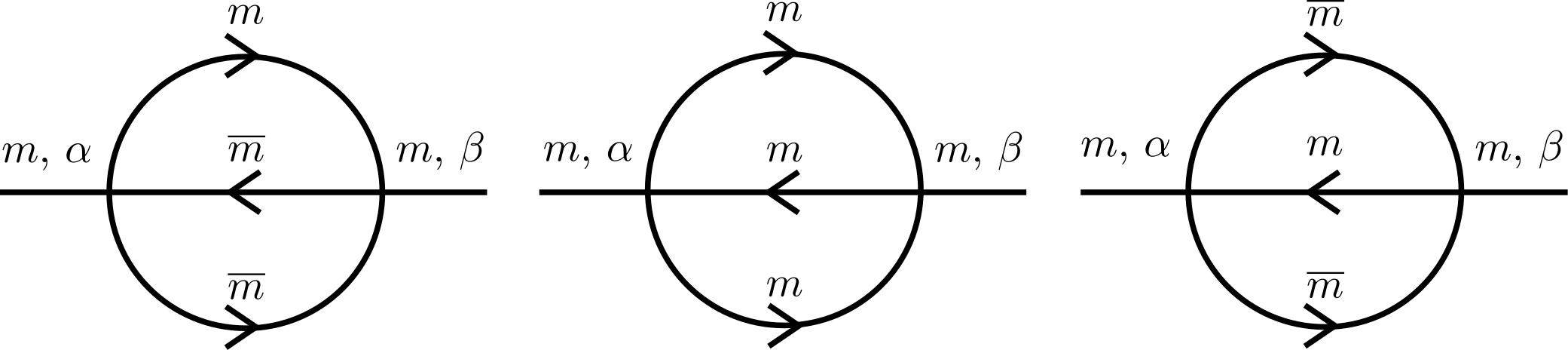}
\caption{Second-order Feynman diagrams contributing to the IPT self-energy $\Sigma_{U \, m}^{\alpha\beta}$ in the bonding - antibonding basis ($m=$~A, B ; $\overline{\mathrm{A}}=$~B, $\overline{\mathrm{B}}=$~A) and in the Kadanoff-Baym-Keldysh contour basis ($\alpha, \beta =+,-$). The lines are the non-interacting impurity Green's function $\mathcal{G}^{\alpha\beta}_m$. Spin indices were dropped for the sake of compactness. \label{fig:ipt_diag}}
\end{figure} 
 
In this Section, we provide an alternative expression for the IPT self-energy of the DHM, useful for its numerical implementation.

Figure~\ref{fig:ipt_diag} displays the Feynman diagrams contributing to the IPT self-energy.
In Eq.~(\ref{IPTc}), we reported its expression at half-filling in the basis of the original Kadanoff-Baym-Keldysh contour ($\alpha, \beta =+,-$) and in the real time domain. In practice, we employ a more practical formulation, and we work with the retarded and Keldysh components.
Moreover, it is most efficient to compute only the imaginary parts, since they are compactly supported functions in Fourier space (frequency domain). 
The real part of the retarded component can eventually be recovered through the Kramers-Kronig relations, and the Keldysh component as defined in Eq.~(\ref{eq:keldysh}) is purely imaginary. 
Applying the particle-hole relations at half-filling, given by

\begin{align}
\label{eq:hf}
    \mathrm{Im}\mathcal{G}^{\rm R}_{ab}(\omega)&=\;\;\,(2\delta_{ab}-1)\mathrm{Im}\mathcal{G}^{\rm R}_{ab}(-\omega)\;,\nonumber\\
    \mathrm{Im}\mathcal{G}^{\rm K}_{ab}(\omega)&=-(2\delta_{ab}-1)\mathrm{Im}\mathcal{G}^{\rm K}_{ab}(-\omega)\;,
\end{align}
we come to
\begin{equation}
\begin{split}
    \mathrm{Im}\Sigma_{U\,ab}^{\rm R}(\omega)=\frac{U^2}{4} \{-3\mathcal{G}_{ab}^{\rm K}(\omega)\circ\mathcal{G}_{ab}^{\rm K}(\omega)\circ\mathrm{Im}\mathcal{G}_{ab}^{\rm R}(\omega)&  \\
    +  4\mathrm{Im}\mathcal{G}_{ab}^{\rm R}(\omega)\circ\mathrm{Im}\mathcal{G}_{ab}^{\rm R}(\omega)\circ\mathrm{Im}\mathcal{G}_{ab}^{\rm R}(\omega)\} (2\delta_{ab}-1)\;,&
\label{eq:IPT_R}
\end{split}
\end{equation}
and
\begin{equation}
\begin{split}
    \Sigma_{U\,ab}^{\rm K}(\omega)=\frac{U^2}{4} \{12\mathcal{G}_{ab}^{\rm K}(\omega)\circ\mathrm{Im}\mathcal{G}_{ab}^{\rm R}(\omega)\circ\mathrm{Im}\mathcal{G}_{ab}^{\rm R}(\omega)&  \\
    -  \mathcal{G}_{ab}^{\rm K}(\omega)\circ\mathcal{G}_{ab}^{\rm K}(\omega)\circ\mathcal{G}_{ab}^{\rm K}(\omega)\} (2\delta_{ab}-1)\;,&
\label{eq:IPT_K}
\end{split}
\end{equation}
where we have denoted the convolution between two functions as $(f \circ g)(\omega) \equiv \int \frac{\rmd\omega'}{2\pi}f(\omega-\omega')g(\omega')$, and $a,b=1,2$ are the orbital basis indices. Rotation to the bonding - antibonding basis ($m=\mathrm{B}$, $\mathrm{A}$) used throughout the manuscript can be made via the following dictionary:
\begin{align}
    \label{eq:dict}
    G_{\rm B}=G_{11}+G_{12} \; , \\
    G_{\rm A}=G_{11}-G_{12} \; , \nonumber
\end{align}
which applies as well for the self-energies.

\medskip

\subsection{Dimer limit}
\begin{table*}
\centering
\begin{tabular*}{0.95\linewidth}{@{\extracolsep{\fill}}|c|c|c|}
\hline\hline
Energy level     &  Eigenstate  & $N$ \\
\hline 
    $\vphantom{\Biggl(}$ $-\frac{U}{2} \left(1-\sqrt{1+\left(\frac{4t^\perp}{U}\right)^2}\right)$ &  $\left(1-\sqrt{1+\left(\frac{4t^\perp}{U}\right)^2}\right)(\ket{\uparrow,\downarrow}-\ket{\downarrow,\uparrow})+\frac{4t^\perp}{U}\left(\ket{\uparrow\downarrow,0}+\ket{0,\uparrow\downarrow}\right)$  &  $2$ \\[10pt] 
     \hline
       $\vphantom{\Bigl(}$ & $\ket{\uparrow\downarrow,0}-\ket{0,\uparrow\downarrow}$ & $2$ \\
   $\vphantom{\Bigl(}$ $0$    & $\ket{\uparrow\downarrow,\uparrow\downarrow}$           & 4\\
   $\vphantom{\Bigl(}$      &$\ket{0}$&0 \\
        \hline
  $\vphantom{\Bigl(}$  $t^\perp-U/2$   &  $\ket{\uparrow,\uparrow\downarrow}+\ket{\uparrow\downarrow,\uparrow},\;\;\ket{\downarrow,\uparrow\downarrow}+\ket{\uparrow\downarrow,\downarrow}$  & $3$ \\
      $\vphantom{\Bigl(}$            &  $\ket{\uparrow,0}-\ket{0,\uparrow},\;\;\ket{\downarrow,0}-\ket{0,\downarrow}$  & $1$ \\
                  \hline
   $\vphantom{\Bigl(}$ $-t^\perp-U/2$   &  $\ket{\uparrow,\uparrow\downarrow}-\ket{\uparrow\downarrow,\uparrow},\;\;\ket{\downarrow,\uparrow\downarrow}-\ket{\uparrow\downarrow,\downarrow}$  & $3$ \\
      $\vphantom{\Bigl(}$            &  $\ket{\uparrow,0}+\ket{0,\uparrow},\;\;\ket{\downarrow,0}+\ket{0,\downarrow}$  & $1$ \\
        \hline
   $\vphantom{\Bigl(}$ $-U$      &  $\ket{\uparrow,\uparrow}, \;\;\ket{\downarrow,\downarrow},\;\; \ket{\uparrow,\downarrow}+\ket{\downarrow,\uparrow}$ & $2$\\
    \hline
   $\vphantom{\Biggl(}$ $-\frac{U}{2} \left(1+\sqrt{1+\left(\frac{4t^\perp}{U}\right)^2}\right)$   &   $\ket{\phi_\mathrm{GS}} = \left(1+\sqrt{1+\left(\frac{4t^\perp}{U}\right)^2}\right)(\ket{\uparrow,\downarrow}-\ket{\downarrow,\uparrow})+\frac{4t^\perp}{U}\left(\ket{\uparrow\downarrow,0}+\ket{0,\uparrow\downarrow}\right)$ & $2$\\
    \hline\hline
\end{tabular*}
\caption{
Eigensystem of the DHM Hamiltonian in the dimer limit. We use the shorthand notation $\ket{\psi,\phi}\equiv c^\dagger_{1\psi}c^\dagger_{2\phi}\ket{0}$, with  $\psi,\phi\in\{\uparrow,\downarrow,\uparrow\downarrow\}$,  $c^\dagger_{a\uparrow\downarrow}\equiv c^\dagger_{a\uparrow}c^\dagger_{a\downarrow}$. The states are ordered from the highest excited state (top) to the ground state (bottom), and they are not normalized. $N$ denotes the number of particles in each state, which is a well-defined quantum number for all eigenstates since $[H,N]=0$. The other good quantum numbers are the total spin $\boldsymbol{S}^2$, the spin projection $S_z$, and the orbital permutation parity $P_{12}$. \label{tab:table2}
}
\end{table*}

Let us now discuss the range of validity of the IPT approximation. To this end, we work in a limit where the system is diagonalizable and we can derive exact expressions, in order to compare them to the IPT scheme.

We consider the equilibrium DHM at half-filling, in the dimer limit ($t/U \rightarrow 0$, keeping $t^\perp$ finite) and in the absence of dissipation ($\Gamma=0$). This defines the two-orbital, single-site Hamiltonian
\begin{align}
    \label{eq_Hat}
    H_{\mathrm{dim}}=&-t^\perp\sum_{\sigma}(c^\dagger_{1\sigma}c_{2\sigma}+c^\dagger_{2\sigma}c_{1\sigma}) + U (n_{1\uparrow}n_{1\downarrow}+n_{2\uparrow}n_{2\downarrow})\nonumber \\ 
    &\qquad- \frac{U}{2}\sum_{\sigma}(n_{1\sigma}+n_{2\sigma})\,,
\end{align}
which can be diagonalized exactly. This limit is usually referred to as the atomic limit when $t^\perp=0$ (SOHM). The energy spectrum and eigenstates are shown in Table~\ref{tab:table2}. We can subsequently write the bonding retarded Green's function of the dimer, $G^{\rm R\, ({\rm dim})}_{\rm B}(\omega)$, in the spectral representation:
\begin{align}
    G^{\rm R \, ({\rm dim})}_{\rm B}(\omega) = \sum_{\sigma,j,k} &\frac{|\bra{\phi_j}c^\dagger_{{\rm B}\sigma}\ket{\phi_k}|^2}{\omega-(E_j-E_k)+\rmi0^+}\label{eq:spec}\\
    &\times(\mathrm{e}^{(E_j-E_\mathrm{GS})/T}+\mathrm{e}^{(E_k-E_\mathrm{GS})/T})\nonumber\\
    \stackrel{\left(T\to0\right)}{=}
    \sum_{\sigma,j} &\frac{|\bra{\phi_j}c^\dagger_{{\rm B}\sigma}\ket{\phi_\mathrm{GS}}|^2}{\omega-(E_j-E_\mathrm{GS})+\rmi0^+}\nonumber\\
    +&\frac{|\bra{\phi_\mathrm{GS}}c_{{\rm B}\sigma}\ket{\phi_j}|^2}{\omega-(E_\mathrm{GS}-E_j)+\rmi0^+}\\
    =\;\;\;\;\;\;\;&\frac{1}{\omega+t^\perp-\frac{U^2}{4}\frac{1}{\omega-3t^\perp+\rmi0^+}}.
\end{align}
$\ket{\phi_j}$ are the eigenstates of $H_{\rm at}$, $E_j$ are the corresponding energies, and `GS' labels the ground state of the system. From the last line we extract the bonding self-energy at zero temperature,
\begin{equation}
\label{eq:self_dim}
    \Sigma^{\rm R\, ({\rm dim})}_{U\,\rm{B}}(\omega)=\frac{U^2}{4}\frac{1}{\omega-3t^\perp+\rmi0^+}\;.
\end{equation}

Let us now compare this expression to the self-energy computed with IPT in the dimer limit. In this setting, the non-interacting Green's functions read \begin{align}
    &\mathcal{G}^{\rm R}_{\rm B/A}(\omega)=\frac{1}{\omega \pm t^\perp + \rmi 0^+}\;, \\
    &\mathcal{G}^{{\rm K}}_{\rm B/A}=2 \rmi \tanh{\left(\frac{\omega}{2T}\right)}\mathrm{Im}\mathcal{G}^{\rm R}_{\rm B/A}(\omega)\;.\nonumber
\end{align}
Replacing them in equation~(\ref{eq:IPT_R}), we obtain
\begin{align}
    \Sigma^{\mathrm{R}}_{U\,\rm{B}}(\omega)=&\frac{U^2}{16}\biggl( \left[ 1+3\mathrm{tanh}^2\left(\frac{t^\perp}{2T}\right) \right]\frac{1}{\omega-3t^\perp+\rmi0^+} \nonumber\\
    +&3\left[ 1-\mathrm{tanh}^2\left(\frac{t^\perp}{2T}\right) \right]\frac{1}{\omega+t^\perp+\rmi0^+} \biggr) \label{eq:IPTf}\\
    \stackrel{\left(T\to0\right)}{=}&\;\frac{U^2}{4}\frac{1}{\omega-3t^\perp+\rmi0^+}\;.\label{eq:IPT0}
\end{align}
Note that the terms with a pole at $3t^\perp$ originate from the third Feynman diagram in Fig.~\ref{fig:ipt_diag}.
The bonding retarded Green's function is 
\begin{equation}
\label{eq:GR_IPT}
    G^\mathrm{R}_{\rm B}(\omega)=\frac{1}{\omega+t^\perp-\Sigma^{\mathrm{R}}_{U\,{\rm B}}(\omega)}\;.
\end{equation}
Expression~(\ref{eq:IPT0}) perfectly matches $\Sigma^{\rm R\, ({\rm dim})}_{U\,\rm B}$ in Eq.~(\ref{eq:self_dim}), which implies that the IPT approximation at $T=0$ is exact in the limit $t/U\to0$. 
This is remarkable for a perturbative approach, which is expected to be valid \textit{a priori} only around the opposite limit $U/t\to0$.

It is important to remark that the comparisons in this Section are performed in the absence of a dissipative environment (meaning $\Gamma=0$). Once the bath degrees of freedom are integrated out (see Section~\ref{sec:methods}), it is not possible to perform exact spectral diagonalization. By keeping $\Gamma$ as the smallest scale of the system, we minimize any effect it may have on the validity of the approximation.

The picture is different at finite temperature, where the second term in Eq.~(\ref{eq:IPTf}) becomes finite. 
The presence of this term produces a second root in the denominator of $G^\mathrm{R}_{\rm B}$ in Eq.~(\ref{eq:GR_IPT}), yielding a secondary peak in the DOS located at $\omega \sim -t^\perp$. Comparing to the exact expression in Eq.~(\ref{eq:spec}), there is no peak located at a similar frequency: inspecting Table~\ref{tab:table2} shows that $|E_j-E_k| \neq~t^\perp$~$\forall \,j,\,k$. Analogous reasoning applies for the antibonding component, producing a secondary peak at $\omega \sim t^\perp$. We note that these spurious peaks are relatively small compared to the primary structures. In order to keep this spurious effect under control, when $t^\perp$ is finite we make sure to work in regimes $ T\ll t^\perp$, where the second term is suppressed. 

This range of validity extends to all temperatures when $t^\perp$ is zero or very small. 
Indeed, in the lattice model, the width of the spectral features is controlled by $t$. If $t^\perp \ll t$, the spurious peaks described above cannot be distinguished from the other structures. This means that in this case, the approximation is still valid at high temperatures. Furthermore, in the case of the SOHM ($t^\perp=0$) the matching is exact for all temperatures: Eq.~(\ref{eq:IPTf}) has only one term centered around $\omega \sim 0$, and extracting a self-energy from Eq.~(\ref{eq:spec}) yields the same result.

\section{Estimation of the in-gap density of states}
\label{app:rho0}

In the following, we derive semi-analytical expressions for the in-gap density of states of the SOHM and the DHM. The objective is to account for the large difference observed in Fig.~\ref{fig:gap}.

The density of states of each band $m=$~B, A can be written, in terms of the non-interacting Green's functions of the impurity $\mathcal{G}_m(\omega)$, as
\begin{widetext}
\begin{equation}
\label{eq:dos_im}
    \rho_m(\omega)=\frac{1}{\pi}\frac{\mathrm{Im}({\mathcal{G}_m^{\rm R}}^{-1})(\omega)-\mathrm{Im}\Sigma^{\rm R}_{U\,m}(\omega)}{\left[ \mathrm{Re}({\mathcal{G}_m^{\rm R}}^{-1})(\omega)-\mathrm{Re}\Sigma^{\rm R}_{U\,m}(\omega) \right]^2 + \left[ (\mathrm{Im}{\mathcal{G}_m^{\rm R}}^{-1})(\omega)-\mathrm{Im}\Sigma^{\rm R}_{U\,m}(\omega) \right]^2}\,, 
\end{equation}
\end{widetext}
and the local DOS as $\rho_{\rm  loc}(\omega)=\frac{1}{2}\sum_m\rho_m(\omega)$. 

In the SOHM, this simplifies as $\rho(\omega)=\rho_B(\omega)=\rho_A(\omega)$, which holds in the DHM at $\omega=0$ due to the orbital permutation and particle-hole symmetries in Eq.~(\ref{eq:hf}).

In the dimer limit and at $T=0$, we can estimate the density of states at $\omega=0$ by working with the expression~(\ref{eq:dos_im}). 
In this limit, we can explicitly write $\mathcal{G}_m^{\rm R}$ and $\Sigma^{\rm R}_{U\,m}$, using the expressions given in App.~\ref{app:IPT}. The smallest energy scale of the system is always $\Gamma$ (see the typical values in Table~\ref{tab:table1}), and we therefore include it as the regularizing term (replacing $\rm{i}0^{+}\to\rm{i}\Gamma$). In the SOHM (atomic limit), we have:
\begin{align}
    \mathrm{Re}\Sigma^{\rm R}_{U\,m}|_{\omega=0}=\mathrm{Re}(&\mathcal{G}_m^{\rm R\,-1})|_{\omega=0}=0\,,\\\nonumber
    |\mathrm{Im}\Sigma^{\rm R}_{U\,m}|_{\omega=0}\sim\frac{U^2}{\Gamma}\gg\mathrm{Im}(&\mathcal{G}_m^{\rm R\,-1})|_{\omega=0}=\Gamma\,,
\end{align}
and therefore the gap density is given by
\begin{align}
\label{eq:r0sohm}
    \rho(\omega=0)\sim\Gamma/U^2.
\end{align}

For the DHM, the typical scales given in Table~\ref{tab:table1} lead us to work in the regime $U \gg t^\perp \gg \Gamma$.
In the dimer limit, $\mathrm{Re}({\mathcal{G}_B^{\rm R}}^{-1})|_{\omega=0} = t^\perp$ and $\mathrm{Im}({\mathcal{G}_B^{\rm R}}^{-1})|_{\omega=0} = \Gamma$.
The real part of the self-energy is given by $\mathrm{Re}\Sigma^{\rm R}_{U\,B}|_{\omega=0}=-\frac{U^2}{4}\frac{3t^\perp}{(3t^\perp)^2+\Gamma^2}\sim~-U^2/12t^\perp$, and numerical computations of $\mathrm{Im}\Sigma^{\rm R}_{U\,B}|_{\omega=0}$ at finite $\Gamma$ show that $\mathrm{Im}\Sigma^{\rm R}_{U\,B}|_{\omega=0} \ll \Gamma$. The latter term can therefore be neglected from Eq.~(\ref{eq:dos_im}). Consequently, the density of states in the gap is given by
\begin{align}
\label{eq:r0dhm}
    \rho(\omega=0)\sim\frac{{t^\perp}^2\Gamma}{U^4}.
\end{align}
Comparing $\rho(\omega=0)$ in the two models, we see that they differ by a factor of $(U/t^\perp)^2$, which accounts for the difference observed in Fig.~\ref{fig:gap}.

\bibliography{bilbio}

\providecommand{\noopsort}[1]{}\providecommand{\singleletter}[1]{#1}%
\begin{thebibliography}{86}%
\makeatletter
\providecommand \@ifxundefined [1]{%
 \@ifx{#1\undefined}
}%
\providecommand \@ifnum [1]{%
 \ifnum #1\expandafter \@firstoftwo
 \else \expandafter \@secondoftwo
 \fi
}%
\providecommand \@ifx [1]{%
 \ifx #1\expandafter \@firstoftwo
 \else \expandafter \@secondoftwo
 \fi
}%
\providecommand \natexlab [1]{#1}%
\providecommand \enquote  [1]{``#1''}%
\providecommand \bibnamefont  [1]{#1}%
\providecommand \bibfnamefont [1]{#1}%
\providecommand \citenamefont [1]{#1}%
\providecommand \href@noop [0]{\@secondoftwo}%
\providecommand \href [0]{\begingroup \@sanitize@url \@href}%
\providecommand \@href[1]{\@@startlink{#1}\@@href}%
\providecommand \@@href[1]{\endgroup#1\@@endlink}%
\providecommand \@sanitize@url [0]{\catcode `\\12\catcode `\$12\catcode
  `\&12\catcode `\#12\catcode `\^12\catcode `\_12\catcode `\%12\relax}%
\providecommand \@@startlink[1]{}%
\providecommand \@@endlink[0]{}%
\providecommand \url  [0]{\begingroup\@sanitize@url \@url }%
\providecommand \@url [1]{\endgroup\@href {#1}{\urlprefix }}%
\providecommand \urlprefix  [0]{URL }%
\providecommand \Eprint [0]{\href }%
\providecommand \doibase [0]{https://doi.org/}%
\providecommand \selectlanguage [0]{\@gobble}%
\providecommand \bibinfo  [0]{\@secondoftwo}%
\providecommand \bibfield  [0]{\@secondoftwo}%
\providecommand \translation [1]{[#1]}%
\providecommand \BibitemOpen [0]{}%
\providecommand \bibitemStop [0]{}%
\providecommand \bibitemNoStop [0]{.\EOS\space}%
\providecommand \EOS [0]{\spacefactor3000\relax}%
\providecommand \BibitemShut  [1]{\csname bibitem#1\endcsname}%
\let\auto@bib@innerbib\@empty
\bibitem [{\citenamefont {Cox}(2010)}]{cox2010}%
  \BibitemOpen
  \bibfield  {author} {\bibinfo {author} {\bibfnamefont {P.~A.}\ \bibnamefont
  {Cox}},\ }\bibfield  {title} {\bibinfo {title} {Transition metal oxides: an
  introduction to their electronic structure and properties},\ }\href@noop {}
  {\bibfield  {journal} {\bibinfo  {journal} {Oxford University Press, Oxford,
  UK}\ } (\bibinfo {year} {2010})}\BibitemShut {NoStop}%
\bibitem [{\citenamefont {Maekawa}\ \emph {et~al.}(2004)\citenamefont
  {Maekawa}, \citenamefont {Tohyama}, \citenamefont {Barnes}, \citenamefont
  {Ishihara}, \citenamefont {Koshibae},\ and\ \citenamefont
  {Khaliullin}}]{maekawa2004}%
  \BibitemOpen
  \bibfield  {author} {\bibinfo {author} {\bibfnamefont {S.}~\bibnamefont
  {Maekawa}}, \bibinfo {author} {\bibfnamefont {T.}~\bibnamefont {Tohyama}},
  \bibinfo {author} {\bibfnamefont {S.~E.}\ \bibnamefont {Barnes}}, \bibinfo
  {author} {\bibfnamefont {S.}~\bibnamefont {Ishihara}}, \bibinfo {author}
  {\bibfnamefont {W.}~\bibnamefont {Koshibae}},\ and\ \bibinfo {author}
  {\bibfnamefont {G.}~\bibnamefont {Khaliullin}},\ }\bibfield  {title}
  {\bibinfo {title} {Physics of transition metal oxides},\ }\href@noop {}
  {\bibfield  {journal} {\bibinfo  {journal} {Springer Science \& Business
  Media, Berlin, Germany}\ } (\bibinfo {year} {2004})}\BibitemShut {NoStop}%
\bibitem [{\citenamefont {Khomskii}(2014)}]{khomskii2014}%
  \BibitemOpen
  \bibfield  {author} {\bibinfo {author} {\bibfnamefont {D.}~\bibnamefont
  {Khomskii}},\ }\bibfield  {title} {\bibinfo {title} {Transition metal
  compounds},\ }\href@noop {} {\bibfield  {journal} {\bibinfo  {journal}
  {Cambridge University Press, Cambridge, UK}\ } (\bibinfo {year}
  {2014})}\BibitemShut {NoStop}%
\bibitem [{\citenamefont {Sawa}(2008)}]{sawa2008}%
  \BibitemOpen
  \bibfield  {author} {\bibinfo {author} {\bibfnamefont {A.}~\bibnamefont
  {Sawa}},\ }\bibfield  {title} {\bibinfo {title} {Resistive switching in
  transition metal oxides},\ }\href
  {https://doi.org/https://doi.org/10.1016/S1369-7021(08)70119-6} {\bibfield
  {journal} {\bibinfo  {journal} {Mater. Today}\ }\textbf {\bibinfo {volume}
  {11}},\ \bibinfo {pages} {28} (\bibinfo {year} {2008})}\BibitemShut {NoStop}%
\bibitem [{\citenamefont {Lee}\ \emph {et~al.}(2015)\citenamefont {Lee},
  \citenamefont {Lee},\ and\ \citenamefont {Noh}}]{sung2015}%
  \BibitemOpen
  \bibfield  {author} {\bibinfo {author} {\bibfnamefont {J.~S.}\ \bibnamefont
  {Lee}}, \bibinfo {author} {\bibfnamefont {S.}~\bibnamefont {Lee}},\ and\
  \bibinfo {author} {\bibfnamefont {T.~W.}\ \bibnamefont {Noh}},\ }\bibfield
  {title} {\bibinfo {title} {Resistive switching phenomena: A review of
  statistical physics approaches},\ }\href {https://doi.org/10.1063/1.4929512}
  {\bibfield  {journal} {\bibinfo  {journal} {Appl. Phys. Rev.}\ }\textbf
  {\bibinfo {volume} {2}},\ \bibinfo {pages} {031303} (\bibinfo {year}
  {2015})}\BibitemShut {NoStop}%
\bibitem [{\citenamefont {Cope}\ and\ \citenamefont {Penn}(1968)}]{cope1968}%
  \BibitemOpen
  \bibfield  {author} {\bibinfo {author} {\bibfnamefont {R.}~\bibnamefont
  {Cope}}\ and\ \bibinfo {author} {\bibfnamefont {A.}~\bibnamefont {Penn}},\
  }\bibfield  {title} {\bibinfo {title} {High-speed solid-state thermal
  switches based on vanadium dioxide},\ }\href@noop {} {\bibfield  {journal}
  {\bibinfo  {journal} {J. Phys. D: Appl. Phys.}\ }\textbf {\bibinfo {volume}
  {1}},\ \bibinfo {pages} {161} (\bibinfo {year} {1968})}\BibitemShut {NoStop}%
\bibitem [{\citenamefont {Lin}\ \emph {et~al.}(2018)\citenamefont {Lin},
  \citenamefont {Ramanathan},\ and\ \citenamefont {Guha}}]{lin2018}%
  \BibitemOpen
  \bibfield  {author} {\bibinfo {author} {\bibfnamefont {J.}~\bibnamefont
  {Lin}}, \bibinfo {author} {\bibfnamefont {S.}~\bibnamefont {Ramanathan}},\
  and\ \bibinfo {author} {\bibfnamefont {S.}~\bibnamefont {Guha}},\ }\bibfield
  {title} {\bibinfo {title} {Electrically driven insulator–metal
  transition-based devices—part ii: Transient characteristics},\ }\href
  {https://doi.org/10.1109/TED.2018.2859188} {\bibfield  {journal} {\bibinfo
  {journal} {IEEE Trans. Electron Devices}\ }\textbf {\bibinfo {volume} {65}},\
  \bibinfo {pages} {3989} (\bibinfo {year} {2018})}\BibitemShut {NoStop}%
\bibitem [{\citenamefont {del Valle}\ \emph {et~al.}(2018)\citenamefont {del
  Valle}, \citenamefont {Ramírez}, \citenamefont {Rozenberg},\ and\
  \citenamefont {Schuller}}]{marcelo2018}%
  \BibitemOpen
  \bibfield  {author} {\bibinfo {author} {\bibfnamefont {J.}~\bibnamefont {del
  Valle}}, \bibinfo {author} {\bibfnamefont {J.~G.}\ \bibnamefont {Ramírez}},
  \bibinfo {author} {\bibfnamefont {M.~J.}\ \bibnamefont {Rozenberg}},\ and\
  \bibinfo {author} {\bibfnamefont {I.~K.}\ \bibnamefont {Schuller}},\
  }\bibfield  {title} {\bibinfo {title} {Challenges in materials and devices
  for resistive-switching-based neuromorphic computing},\ }\href
  {https://doi.org/10.1063/1.5047800} {\bibfield  {journal} {\bibinfo
  {journal} {J. Appl. Phys.}\ }\textbf {\bibinfo {volume} {124}},\ \bibinfo
  {pages} {211101} (\bibinfo {year} {2018})}\BibitemShut {NoStop}%
\bibitem [{\citenamefont {Hoffmann}\ \emph {et~al.}(2022)\citenamefont
  {Hoffmann}, \citenamefont {Ramanathan}, \citenamefont {Grollier},
  \citenamefont {Kent}, \citenamefont {Rozenberg}, \citenamefont {Schuller},
  \citenamefont {Shpyrko}, \citenamefont {Dynes}, \citenamefont {Fainman},
  \citenamefont {Frano} \emph {et~al.}}]{marcelo2022}%
  \BibitemOpen
  \bibfield  {author} {\bibinfo {author} {\bibfnamefont {A.}~\bibnamefont
  {Hoffmann}}, \bibinfo {author} {\bibfnamefont {S.}~\bibnamefont
  {Ramanathan}}, \bibinfo {author} {\bibfnamefont {J.}~\bibnamefont
  {Grollier}}, \bibinfo {author} {\bibfnamefont {A.~D.}\ \bibnamefont {Kent}},
  \bibinfo {author} {\bibfnamefont {M.~J.}\ \bibnamefont {Rozenberg}}, \bibinfo
  {author} {\bibfnamefont {I.~K.}\ \bibnamefont {Schuller}}, \bibinfo {author}
  {\bibfnamefont {O.~G.}\ \bibnamefont {Shpyrko}}, \bibinfo {author}
  {\bibfnamefont {R.~C.}\ \bibnamefont {Dynes}}, \bibinfo {author}
  {\bibfnamefont {Y.}~\bibnamefont {Fainman}}, \bibinfo {author} {\bibfnamefont
  {A.}~\bibnamefont {Frano}}, \emph {et~al.},\ }\bibfield  {title} {\bibinfo
  {title} {Quantum materials for energy-efficient neuromorphic computing:
  Opportunities and challenges},\ }\href@noop {} {\bibfield  {journal}
  {\bibinfo  {journal} {APL Mater.}\ }\textbf {\bibinfo {volume} {10}},\
  \bibinfo {pages} {070904} (\bibinfo {year} {2022})}\BibitemShut {NoStop}%
\bibitem [{\citenamefont {Ridley}(1963)}]{Ridley1963}%
  \BibitemOpen
  \bibfield  {author} {\bibinfo {author} {\bibfnamefont {B.~K.}\ \bibnamefont
  {Ridley}},\ }\bibfield  {title} {\bibinfo {title} {Specific negative
  resistance in solids},\ }\href {https://doi.org/10.1088/0370-1328/82/6/315}
  {\bibfield  {journal} {\bibinfo  {journal} {Proc. Phys. Soc.}\ }\textbf
  {\bibinfo {volume} {82}},\ \bibinfo {pages} {954} (\bibinfo {year}
  {1963})}\BibitemShut {NoStop}%
\bibitem [{\citenamefont {Volkov}\ and\ \citenamefont
  {Kogan}(1969)}]{kogan1969}%
  \BibitemOpen
  \bibfield  {author} {\bibinfo {author} {\bibfnamefont {A.~F.}\ \bibnamefont
  {Volkov}}\ and\ \bibinfo {author} {\bibfnamefont {S.~M.}\ \bibnamefont
  {Kogan}},\ }\bibfield  {title} {\bibinfo {title} {Physical phenomena in
  semiconductors with negative differential conductivity},\ }\href
  {https://doi.org/10.1070/pu1969v011n06abeh003780} {\bibfield  {journal}
  {\bibinfo  {journal} {Phys.-Usp.}\ }\textbf {\bibinfo {volume} {11}},\
  \bibinfo {pages} {881} (\bibinfo {year} {1969})}\BibitemShut {NoStop}%
\bibitem [{\citenamefont {Duchene}\ \emph {et~al.}(1971)\citenamefont
  {Duchene}, \citenamefont {Terraillon}, \citenamefont {Pailly},\ and\
  \citenamefont {Adam}}]{duchene1971}%
  \BibitemOpen
  \bibfield  {author} {\bibinfo {author} {\bibfnamefont {J.}~\bibnamefont
  {Duchene}}, \bibinfo {author} {\bibfnamefont {M.}~\bibnamefont {Terraillon}},
  \bibinfo {author} {\bibfnamefont {P.}~\bibnamefont {Pailly}},\ and\ \bibinfo
  {author} {\bibfnamefont {G.}~\bibnamefont {Adam}},\ }\bibfield  {title}
  {\bibinfo {title} {Filamentary conduction in vo2 coplanar thin‐film
  devices},\ }\href {https://doi.org/10.1063/1.1653835} {\bibfield  {journal}
  {\bibinfo  {journal} {Appl. Phys. Lett.}\ }\textbf {\bibinfo {volume} {19}},\
  \bibinfo {pages} {115} (\bibinfo {year} {1971})}\BibitemShut {NoStop}%
\bibitem [{\citenamefont {Stefanovich}\ \emph {et~al.}(2000)\citenamefont
  {Stefanovich}, \citenamefont {Pergament},\ and\ \citenamefont
  {Stefanovich}}]{stefanovich2000}%
  \BibitemOpen
  \bibfield  {author} {\bibinfo {author} {\bibfnamefont {G.}~\bibnamefont
  {Stefanovich}}, \bibinfo {author} {\bibfnamefont {A.}~\bibnamefont
  {Pergament}},\ and\ \bibinfo {author} {\bibfnamefont {D.}~\bibnamefont
  {Stefanovich}},\ }\bibfield  {title} {\bibinfo {title} {Electrical switching
  and {M}ott transition in {VO}$_2$},\ }\href
  {https://doi.org/10.1088/0953-8984/12/41/310} {\bibfield  {journal} {\bibinfo
   {journal} {J. Phys.: Condens. Matter}\ }\textbf {\bibinfo {volume} {12}},\
  \bibinfo {pages} {8837} (\bibinfo {year} {2000})}\BibitemShut {NoStop}%
\bibitem [{\citenamefont {Gopalakrishnan}\ \emph {et~al.}(2009)\citenamefont
  {Gopalakrishnan}, \citenamefont {Ruzmetov},\ and\ \citenamefont
  {Ramanathan}}]{debate2009}%
  \BibitemOpen
  \bibfield  {author} {\bibinfo {author} {\bibfnamefont {G.}~\bibnamefont
  {Gopalakrishnan}}, \bibinfo {author} {\bibfnamefont {D.}~\bibnamefont
  {Ruzmetov}},\ and\ \bibinfo {author} {\bibfnamefont {S.}~\bibnamefont
  {Ramanathan}},\ }\bibfield  {title} {\bibinfo {title} {On the triggering
  mechanism for the metal--insulator transition in thin film vo2 devices:
  electric field versus thermal effects},\ }\href@noop {} {\bibfield  {journal}
  {\bibinfo  {journal} {J. Mater. Sci.}\ }\textbf {\bibinfo {volume} {44}},\
  \bibinfo {pages} {5345} (\bibinfo {year} {2009})}\BibitemShut {NoStop}%
\bibitem [{\citenamefont {Janod}\ \emph {et~al.}(2015)\citenamefont {Janod},
  \citenamefont {Tranchant}, \citenamefont {Corraze}, \citenamefont {Querré},
  \citenamefont {Stoliar}, \citenamefont {Rozenberg}, \citenamefont {Cren},
  \citenamefont {Roditchev}, \citenamefont {Phuoc}, \citenamefont {Besland},\
  and\ \citenamefont {Cario}}]{janod2015}%
  \BibitemOpen
  \bibfield  {author} {\bibinfo {author} {\bibfnamefont {E.}~\bibnamefont
  {Janod}}, \bibinfo {author} {\bibfnamefont {J.}~\bibnamefont {Tranchant}},
  \bibinfo {author} {\bibfnamefont {B.}~\bibnamefont {Corraze}}, \bibinfo
  {author} {\bibfnamefont {M.}~\bibnamefont {Querré}}, \bibinfo {author}
  {\bibfnamefont {P.}~\bibnamefont {Stoliar}}, \bibinfo {author} {\bibfnamefont
  {M.}~\bibnamefont {Rozenberg}}, \bibinfo {author} {\bibfnamefont
  {T.}~\bibnamefont {Cren}}, \bibinfo {author} {\bibfnamefont {D.}~\bibnamefont
  {Roditchev}}, \bibinfo {author} {\bibfnamefont {V.~T.}\ \bibnamefont
  {Phuoc}}, \bibinfo {author} {\bibfnamefont {M.-P.}\ \bibnamefont {Besland}},\
  and\ \bibinfo {author} {\bibfnamefont {L.}~\bibnamefont {Cario}},\ }\bibfield
   {title} {\bibinfo {title} {Resistive switching in {M}ott insulators and
  correlated systems},\ }\href
  {https://doi.org/https://doi.org/10.1002/adfm.201500823} {\bibfield
  {journal} {\bibinfo  {journal} {Adv. Funct. Mater.}\ }\textbf {\bibinfo
  {volume} {25}},\ \bibinfo {pages} {6287} (\bibinfo {year}
  {2015})}\BibitemShut {NoStop}%
\bibitem [{\citenamefont {Higgins}\ \emph {et~al.}(1977)\citenamefont
  {Higgins}, \citenamefont {Temple},\ and\ \citenamefont
  {Lewis}}]{higgins1977}%
  \BibitemOpen
  \bibfield  {author} {\bibinfo {author} {\bibfnamefont {J.}~\bibnamefont
  {Higgins}}, \bibinfo {author} {\bibfnamefont {B.}~\bibnamefont {Temple}},\
  and\ \bibinfo {author} {\bibfnamefont {J.}~\bibnamefont {Lewis}},\ }\bibfield
   {title} {\bibinfo {title} {Electrical properties of vanadate-glass threshold
  switches},\ }\href
  {https://doi.org/https://doi.org/10.1016/0022-3093(77)90005-9} {\bibfield
  {journal} {\bibinfo  {journal} {J. Non-Cryst. Solids}\ }\textbf {\bibinfo
  {volume} {23}},\ \bibinfo {pages} {187} (\bibinfo {year} {1977})}\BibitemShut
  {NoStop}%
\bibitem [{\citenamefont {Wu}\ \emph {et~al.}(2011)\citenamefont {Wu},
  \citenamefont {Zimmers}, \citenamefont {Aubin}, \citenamefont {Ghosh},
  \citenamefont {Liu},\ and\ \citenamefont {Lopez}}]{zimmers2011}%
  \BibitemOpen
  \bibfield  {author} {\bibinfo {author} {\bibfnamefont {B.}~\bibnamefont
  {Wu}}, \bibinfo {author} {\bibfnamefont {A.}~\bibnamefont {Zimmers}},
  \bibinfo {author} {\bibfnamefont {H.}~\bibnamefont {Aubin}}, \bibinfo
  {author} {\bibfnamefont {R.}~\bibnamefont {Ghosh}}, \bibinfo {author}
  {\bibfnamefont {Y.}~\bibnamefont {Liu}},\ and\ \bibinfo {author}
  {\bibfnamefont {R.}~\bibnamefont {Lopez}},\ }\bibfield  {title} {\bibinfo
  {title} {Electric-field-driven phase transition in vanadium dioxide},\ }\href
  {https://doi.org/10.1103/PhysRevB.84.241410} {\bibfield  {journal} {\bibinfo
  {journal} {Phys. Rev. B}\ }\textbf {\bibinfo {volume} {84}},\ \bibinfo
  {pages} {241410} (\bibinfo {year} {2011})}\BibitemShut {NoStop}%
\bibitem [{\citenamefont {Zimmers}\ \emph {et~al.}(2013)\citenamefont
  {Zimmers}, \citenamefont {Aigouy}, \citenamefont {Mortier}, \citenamefont
  {Sharoni}, \citenamefont {Wang}, \citenamefont {West}, \citenamefont
  {Ramirez},\ and\ \citenamefont {Schuller}}]{zimmers2013}%
  \BibitemOpen
  \bibfield  {author} {\bibinfo {author} {\bibfnamefont {A.}~\bibnamefont
  {Zimmers}}, \bibinfo {author} {\bibfnamefont {L.}~\bibnamefont {Aigouy}},
  \bibinfo {author} {\bibfnamefont {M.}~\bibnamefont {Mortier}}, \bibinfo
  {author} {\bibfnamefont {A.}~\bibnamefont {Sharoni}}, \bibinfo {author}
  {\bibfnamefont {S.}~\bibnamefont {Wang}}, \bibinfo {author} {\bibfnamefont
  {K.~G.}\ \bibnamefont {West}}, \bibinfo {author} {\bibfnamefont {J.~G.}\
  \bibnamefont {Ramirez}},\ and\ \bibinfo {author} {\bibfnamefont {I.~K.}\
  \bibnamefont {Schuller}},\ }\bibfield  {title} {\bibinfo {title} {Role of
  thermal heating on the voltage induced insulator-metal transition in
  {VO}$_{2}$},\ }\href {https://doi.org/10.1103/PhysRevLett.110.056601}
  {\bibfield  {journal} {\bibinfo  {journal} {Phys. Rev. Lett.}\ }\textbf
  {\bibinfo {volume} {110}},\ \bibinfo {pages} {056601} (\bibinfo {year}
  {2013})}\BibitemShut {NoStop}%
\bibitem [{\citenamefont {Li}\ \emph {et~al.}(2016)\citenamefont {Li},
  \citenamefont {Sharma}, \citenamefont {Gala}, \citenamefont {Shukla},
  \citenamefont {Paik}, \citenamefont {Datta}, \citenamefont {Schlom},
  \citenamefont {Bain},\ and\ \citenamefont {Skowronski}}]{jouleh2016}%
  \BibitemOpen
  \bibfield  {author} {\bibinfo {author} {\bibfnamefont {D.}~\bibnamefont
  {Li}}, \bibinfo {author} {\bibfnamefont {A.~A.}\ \bibnamefont {Sharma}},
  \bibinfo {author} {\bibfnamefont {D.~K.}\ \bibnamefont {Gala}}, \bibinfo
  {author} {\bibfnamefont {N.}~\bibnamefont {Shukla}}, \bibinfo {author}
  {\bibfnamefont {H.}~\bibnamefont {Paik}}, \bibinfo {author} {\bibfnamefont
  {S.}~\bibnamefont {Datta}}, \bibinfo {author} {\bibfnamefont {D.~G.}\
  \bibnamefont {Schlom}}, \bibinfo {author} {\bibfnamefont {J.~A.}\
  \bibnamefont {Bain}},\ and\ \bibinfo {author} {\bibfnamefont
  {M.}~\bibnamefont {Skowronski}},\ }\bibfield  {title} {\bibinfo {title}
  {Joule heating-induced metal–insulator transition in epitaxial
  {VO}$_2$/{T}i{O}$_2$ devices},\ }\href
  {https://doi.org/10.1021/acsami.6b03501} {\bibfield  {journal} {\bibinfo
  {journal} {ACS Appl. Mater. Interfaces}\ }\textbf {\bibinfo {volume} {8}},\
  \bibinfo {pages} {12908} (\bibinfo {year} {2016})},\ \bibinfo {note} {pMID:
  27136956}\BibitemShut {NoStop}%
\bibitem [{\citenamefont {del Valle}\ \emph {et~al.}(2017)\citenamefont {del
  Valle}, \citenamefont {Kalcheim}, \citenamefont {Trastoy}, \citenamefont
  {Charnukha}, \citenamefont {Basov},\ and\ \citenamefont
  {Schuller}}]{schuller2017}%
  \BibitemOpen
  \bibfield  {author} {\bibinfo {author} {\bibfnamefont {J.}~\bibnamefont {del
  Valle}}, \bibinfo {author} {\bibfnamefont {Y.}~\bibnamefont {Kalcheim}},
  \bibinfo {author} {\bibfnamefont {J.}~\bibnamefont {Trastoy}}, \bibinfo
  {author} {\bibfnamefont {A.}~\bibnamefont {Charnukha}}, \bibinfo {author}
  {\bibfnamefont {D.~N.}\ \bibnamefont {Basov}},\ and\ \bibinfo {author}
  {\bibfnamefont {I.~K.}\ \bibnamefont {Schuller}},\ }\bibfield  {title}
  {\bibinfo {title} {Electrically induced multiple metal-insulator transitions
  in oxide nanodevices},\ }\href
  {https://doi.org/10.1103/PhysRevApplied.8.054041} {\bibfield  {journal}
  {\bibinfo  {journal} {Phys. Rev. Applied}\ }\textbf {\bibinfo {volume} {8}},\
  \bibinfo {pages} {054041} (\bibinfo {year} {2017})}\BibitemShut {NoStop}%
\bibitem [{\citenamefont {Polozov}\ \emph {et~al.}(2020)\citenamefont
  {Polozov}, \citenamefont {Maklakov}, \citenamefont {Rakhmanov}, \citenamefont
  {Maklakov},\ and\ \citenamefont {Kisel}}]{overheating2020}%
  \BibitemOpen
  \bibfield  {author} {\bibinfo {author} {\bibfnamefont {V.~I.}\ \bibnamefont
  {Polozov}}, \bibinfo {author} {\bibfnamefont {S.~S.}\ \bibnamefont
  {Maklakov}}, \bibinfo {author} {\bibfnamefont {A.~L.}\ \bibnamefont
  {Rakhmanov}}, \bibinfo {author} {\bibfnamefont {S.~A.}\ \bibnamefont
  {Maklakov}},\ and\ \bibinfo {author} {\bibfnamefont {V.~N.}\ \bibnamefont
  {Kisel}},\ }\bibfield  {title} {\bibinfo {title} {Blow-up overheating
  instability in vanadium dioxide thin films},\ }\href
  {https://doi.org/10.1103/PhysRevB.101.214310} {\bibfield  {journal} {\bibinfo
   {journal} {Phys. Rev. B}\ }\textbf {\bibinfo {volume} {101}},\ \bibinfo
  {pages} {214310} (\bibinfo {year} {2020})}\BibitemShut {NoStop}%
\bibitem [{\citenamefont {Kalcheim}\ \emph {et~al.}(2020)\citenamefont
  {Kalcheim}, \citenamefont {Camjayi}, \citenamefont {del Valle}, \citenamefont
  {Salev}, \citenamefont {Rozenberg},\ and\ \citenamefont
  {Schuller}}]{marcelo2020}%
  \BibitemOpen
  \bibfield  {author} {\bibinfo {author} {\bibfnamefont {Y.}~\bibnamefont
  {Kalcheim}}, \bibinfo {author} {\bibfnamefont {A.}~\bibnamefont {Camjayi}},
  \bibinfo {author} {\bibfnamefont {J.}~\bibnamefont {del Valle}}, \bibinfo
  {author} {\bibfnamefont {P.}~\bibnamefont {Salev}}, \bibinfo {author}
  {\bibfnamefont {M.}~\bibnamefont {Rozenberg}},\ and\ \bibinfo {author}
  {\bibfnamefont {I.~K.}\ \bibnamefont {Schuller}},\ }\bibfield  {title}
  {\bibinfo {title} {Non-thermal resistive switching in mott insulator
  nanowires},\ }\href {https://doi.org/10.1038/s41467-020-16752-1} {\bibfield
  {journal} {\bibinfo  {journal} {Nature Communications}\ }\textbf {\bibinfo
  {volume} {11}},\ \bibinfo {pages} {2985} (\bibinfo {year}
  {2020})}\BibitemShut {NoStop}%
\bibitem [{\citenamefont {del Valle}\ \emph {et~al.}(2021)\citenamefont {del
  Valle}, \citenamefont {Vargas}, \citenamefont {Rocco}, \citenamefont {Salev},
  \citenamefont {Kalcheim}, \citenamefont {Lapa}, \citenamefont {Adda},
  \citenamefont {Lee}, \citenamefont {Wang}, \citenamefont {Fratino},
  \citenamefont {Rozenberg},\ and\ \citenamefont {Schuller}}]{schuller2021}%
  \BibitemOpen
  \bibfield  {author} {\bibinfo {author} {\bibfnamefont {J.}~\bibnamefont {del
  Valle}}, \bibinfo {author} {\bibfnamefont {N.~M.}\ \bibnamefont {Vargas}},
  \bibinfo {author} {\bibfnamefont {R.}~\bibnamefont {Rocco}}, \bibinfo
  {author} {\bibfnamefont {P.}~\bibnamefont {Salev}}, \bibinfo {author}
  {\bibfnamefont {Y.}~\bibnamefont {Kalcheim}}, \bibinfo {author}
  {\bibfnamefont {P.~N.}\ \bibnamefont {Lapa}}, \bibinfo {author}
  {\bibfnamefont {C.}~\bibnamefont {Adda}}, \bibinfo {author} {\bibfnamefont
  {M.-H.}\ \bibnamefont {Lee}}, \bibinfo {author} {\bibfnamefont {P.~Y.}\
  \bibnamefont {Wang}}, \bibinfo {author} {\bibfnamefont {L.}~\bibnamefont
  {Fratino}}, \bibinfo {author} {\bibfnamefont {M.~J.}\ \bibnamefont
  {Rozenberg}},\ and\ \bibinfo {author} {\bibfnamefont {I.~K.}\ \bibnamefont
  {Schuller}},\ }\bibfield  {title} {\bibinfo {title} {Spatiotemporal
  characterization of the field-induced insulator-to-metal transition},\ }\href
  {https://doi.org/10.1126/science.abd9088} {\bibfield  {journal} {\bibinfo
  {journal} {Science}\ }\textbf {\bibinfo {volume} {373}},\ \bibinfo {pages}
  {907} (\bibinfo {year} {2021})}\BibitemShut {NoStop}%
\bibitem [{\citenamefont {Sugimoto}\ \emph {et~al.}(2008)\citenamefont
  {Sugimoto}, \citenamefont {Onoda},\ and\ \citenamefont
  {Nagaosa}}]{sugimoto2008}%
  \BibitemOpen
  \bibfield  {author} {\bibinfo {author} {\bibfnamefont {N.}~\bibnamefont
  {Sugimoto}}, \bibinfo {author} {\bibfnamefont {S.}~\bibnamefont {Onoda}},\
  and\ \bibinfo {author} {\bibfnamefont {N.}~\bibnamefont {Nagaosa}},\
  }\bibfield  {title} {\bibinfo {title} {Field-induced metal-insulator
  transition and switching phenomenon in correlated insulators},\ }\href
  {https://doi.org/10.1103/PhysRevB.78.155104} {\bibfield  {journal} {\bibinfo
  {journal} {Phys. Rev. B}\ }\textbf {\bibinfo {volume} {78}},\ \bibinfo
  {pages} {155104} (\bibinfo {year} {2008})}\BibitemShut {NoStop}%
\bibitem [{\citenamefont {Han}\ \emph {et~al.}(2018)\citenamefont {Han},
  \citenamefont {Li}, \citenamefont {Aron},\ and\ \citenamefont
  {Kotliar}}]{CamilleJongMF}%
  \BibitemOpen
  \bibfield  {author} {\bibinfo {author} {\bibfnamefont {J.~E.}\ \bibnamefont
  {Han}}, \bibinfo {author} {\bibfnamefont {J.}~\bibnamefont {Li}}, \bibinfo
  {author} {\bibfnamefont {C.}~\bibnamefont {Aron}},\ and\ \bibinfo {author}
  {\bibfnamefont {G.}~\bibnamefont {Kotliar}},\ }\bibfield  {title} {\bibinfo
  {title} {Nonequilibrium mean-field theory of resistive phase transitions},\
  }\href {https://doi.org/10.1103/PhysRevB.98.035145} {\bibfield  {journal}
  {\bibinfo  {journal} {Phys. Rev. B}\ }\textbf {\bibinfo {volume} {98}},\
  \bibinfo {pages} {035145} (\bibinfo {year} {2018})}\BibitemShut {NoStop}%
\bibitem [{\citenamefont {Mitra}\ and\ \citenamefont
  {Millis}(2008)}]{millis2008}%
  \BibitemOpen
  \bibfield  {author} {\bibinfo {author} {\bibfnamefont {A.}~\bibnamefont
  {Mitra}}\ and\ \bibinfo {author} {\bibfnamefont {A.~J.}\ \bibnamefont
  {Millis}},\ }\bibfield  {title} {\bibinfo {title} {Current-driven quantum
  criticality in itinerant electron ferromagnets},\ }\href
  {https://doi.org/10.1103/PhysRevB.77.220404} {\bibfield  {journal} {\bibinfo
  {journal} {Phys. Rev. B}\ }\textbf {\bibinfo {volume} {77}},\ \bibinfo
  {pages} {220404} (\bibinfo {year} {2008})}\BibitemShut {NoStop}%
\bibitem [{\citenamefont {Georges}\ \emph {et~al.}(1996)\citenamefont
  {Georges}, \citenamefont {Kotliar}, \citenamefont {Krauth},\ and\
  \citenamefont {Rozenberg}}]{review1996}%
  \BibitemOpen
  \bibfield  {author} {\bibinfo {author} {\bibfnamefont {A.}~\bibnamefont
  {Georges}}, \bibinfo {author} {\bibfnamefont {G.}~\bibnamefont {Kotliar}},
  \bibinfo {author} {\bibfnamefont {W.}~\bibnamefont {Krauth}},\ and\ \bibinfo
  {author} {\bibfnamefont {M.~J.}\ \bibnamefont {Rozenberg}},\ }\bibfield
  {title} {\bibinfo {title} {Dynamical mean-field theory of strongly correlated
  fermion systems and the limit of infinite dimensions},\ }\href
  {https://doi.org/10.1103/RevModPhys.68.13} {\bibfield  {journal} {\bibinfo
  {journal} {Rev. Mod. Phys.}\ }\textbf {\bibinfo {volume} {68}},\ \bibinfo
  {pages} {13} (\bibinfo {year} {1996})}\BibitemShut {NoStop}%
\bibitem [{\citenamefont {Freericks}\ \emph {et~al.}(2006)\citenamefont
  {Freericks}, \citenamefont {Turkowski},\ and\ \citenamefont
  {Zlati\ifmmode~\acute{c}\else \'{c}\fi{}}}]{Freericks2006}%
  \BibitemOpen
  \bibfield  {author} {\bibinfo {author} {\bibfnamefont {J.~K.}\ \bibnamefont
  {Freericks}}, \bibinfo {author} {\bibfnamefont {V.~M.}\ \bibnamefont
  {Turkowski}},\ and\ \bibinfo {author} {\bibfnamefont {V.}~\bibnamefont
  {Zlati\ifmmode~\acute{c}\else \'{c}\fi{}}},\ }\bibfield  {title} {\bibinfo
  {title} {Nonequilibrium dynamical mean-field theory},\ }\href
  {https://doi.org/10.1103/PhysRevLett.97.266408} {\bibfield  {journal}
  {\bibinfo  {journal} {Phys. Rev. Lett.}\ }\textbf {\bibinfo {volume} {97}},\
  \bibinfo {pages} {266408} (\bibinfo {year} {2006})}\BibitemShut {NoStop}%
\bibitem [{\citenamefont {Okamoto}(2008)}]{okamoto2008}%
  \BibitemOpen
  \bibfield  {author} {\bibinfo {author} {\bibfnamefont {S.}~\bibnamefont
  {Okamoto}},\ }\bibfield  {title} {\bibinfo {title} {Nonlinear transport
  through strongly correlated two-terminal heterostructures: A dynamical
  mean-field approach},\ }\href
  {https://doi.org/10.1103/PhysRevLett.101.116807} {\bibfield  {journal}
  {\bibinfo  {journal} {Phys. Rev. Lett.}\ }\textbf {\bibinfo {volume} {101}},\
  \bibinfo {pages} {116807} (\bibinfo {year} {2008})}\BibitemShut {NoStop}%
\bibitem [{\citenamefont {Aoki}\ \emph {et~al.}(2014)\citenamefont {Aoki},
  \citenamefont {Tsuji}, \citenamefont {Eckstein}, \citenamefont {Kollar},
  \citenamefont {Oka},\ and\ \citenamefont {Werner}}]{nessdmft2014}%
  \BibitemOpen
  \bibfield  {author} {\bibinfo {author} {\bibfnamefont {H.}~\bibnamefont
  {Aoki}}, \bibinfo {author} {\bibfnamefont {N.}~\bibnamefont {Tsuji}},
  \bibinfo {author} {\bibfnamefont {M.}~\bibnamefont {Eckstein}}, \bibinfo
  {author} {\bibfnamefont {M.}~\bibnamefont {Kollar}}, \bibinfo {author}
  {\bibfnamefont {T.}~\bibnamefont {Oka}},\ and\ \bibinfo {author}
  {\bibfnamefont {P.}~\bibnamefont {Werner}},\ }\bibfield  {title} {\bibinfo
  {title} {Nonequilibrium dynamical mean-field theory and its applications},\
  }\href {https://doi.org/10.1103/RevModPhys.86.779} {\bibfield  {journal}
  {\bibinfo  {journal} {Rev. Mod. Phys.}\ }\textbf {\bibinfo {volume} {86}},\
  \bibinfo {pages} {779} (\bibinfo {year} {2014})}\BibitemShut {NoStop}%
\bibitem [{\citenamefont {Joura}\ \emph {et~al.}(2008)\citenamefont {Joura},
  \citenamefont {Freericks},\ and\ \citenamefont {Pruschke}}]{freericks2008}%
  \BibitemOpen
  \bibfield  {author} {\bibinfo {author} {\bibfnamefont {A.~V.}\ \bibnamefont
  {Joura}}, \bibinfo {author} {\bibfnamefont {J.~K.}\ \bibnamefont
  {Freericks}},\ and\ \bibinfo {author} {\bibfnamefont {T.}~\bibnamefont
  {Pruschke}},\ }\bibfield  {title} {\bibinfo {title} {Steady-state
  nonequilibrium density of states of driven strongly correlated lattice models
  in infinite dimensions},\ }\href
  {https://doi.org/10.1103/PhysRevLett.101.196401} {\bibfield  {journal}
  {\bibinfo  {journal} {Phys. Rev. Lett.}\ }\textbf {\bibinfo {volume} {101}},\
  \bibinfo {pages} {196401} (\bibinfo {year} {2008})}\BibitemShut {NoStop}%
\bibitem [{\citenamefont {Tsuji}\ \emph {et~al.}(2008)\citenamefont {Tsuji},
  \citenamefont {Oka},\ and\ \citenamefont {Aoki}}]{Floquet2008}%
  \BibitemOpen
  \bibfield  {author} {\bibinfo {author} {\bibfnamefont {N.}~\bibnamefont
  {Tsuji}}, \bibinfo {author} {\bibfnamefont {T.}~\bibnamefont {Oka}},\ and\
  \bibinfo {author} {\bibfnamefont {H.}~\bibnamefont {Aoki}},\ }\bibfield
  {title} {\bibinfo {title} {Correlated electron systems periodically driven
  out of equilibrium: $\text{Floquet}+\text{DMFT}$ formalism},\ }\href
  {https://doi.org/10.1103/PhysRevB.78.235124} {\bibfield  {journal} {\bibinfo
  {journal} {Phys. Rev. B}\ }\textbf {\bibinfo {volume} {78}},\ \bibinfo
  {pages} {235124} (\bibinfo {year} {2008})}\BibitemShut {NoStop}%
\bibitem [{\citenamefont {Aron}\ \emph {et~al.}(2012)\citenamefont {Aron},
  \citenamefont {Kotliar},\ and\ \citenamefont {Weber}}]{camille2012}%
  \BibitemOpen
  \bibfield  {author} {\bibinfo {author} {\bibfnamefont {C.}~\bibnamefont
  {Aron}}, \bibinfo {author} {\bibfnamefont {G.}~\bibnamefont {Kotliar}},\ and\
  \bibinfo {author} {\bibfnamefont {C.}~\bibnamefont {Weber}},\ }\bibfield
  {title} {\bibinfo {title} {Dimensional crossover driven by an electric
  field},\ }\href {https://doi.org/10.1103/PhysRevLett.108.086401} {\bibfield
  {journal} {\bibinfo  {journal} {Phys. Rev. Lett.}\ }\textbf {\bibinfo
  {volume} {108}},\ \bibinfo {pages} {086401} (\bibinfo {year}
  {2012})}\BibitemShut {NoStop}%
\bibitem [{\citenamefont {Arrigoni}\ \emph {et~al.}(2013)\citenamefont
  {Arrigoni}, \citenamefont {Knap},\ and\ \citenamefont {von~der
  Linden}}]{Arrigoni2013}%
  \BibitemOpen
  \bibfield  {author} {\bibinfo {author} {\bibfnamefont {E.}~\bibnamefont
  {Arrigoni}}, \bibinfo {author} {\bibfnamefont {M.}~\bibnamefont {Knap}},\
  and\ \bibinfo {author} {\bibfnamefont {W.}~\bibnamefont {von~der Linden}},\
  }\bibfield  {title} {\bibinfo {title} {Nonequilibrium dynamical mean-field
  theory: An auxiliary quantum master equation approach},\ }\href
  {https://doi.org/10.1103/PhysRevLett.110.086403} {\bibfield  {journal}
  {\bibinfo  {journal} {Phys. Rev. Lett.}\ }\textbf {\bibinfo {volume} {110}},\
  \bibinfo {pages} {086403} (\bibinfo {year} {2013})}\BibitemShut {NoStop}%
\bibitem [{\citenamefont {Li}\ \emph {et~al.}(2015)\citenamefont {Li},
  \citenamefont {Aron}, \citenamefont {Kotliar},\ and\ \citenamefont
  {Han}}]{camille2015}%
  \BibitemOpen
  \bibfield  {author} {\bibinfo {author} {\bibfnamefont {J.}~\bibnamefont
  {Li}}, \bibinfo {author} {\bibfnamefont {C.}~\bibnamefont {Aron}}, \bibinfo
  {author} {\bibfnamefont {G.}~\bibnamefont {Kotliar}},\ and\ \bibinfo {author}
  {\bibfnamefont {J.~E.}\ \bibnamefont {Han}},\ }\bibfield  {title} {\bibinfo
  {title} {Electric-field-driven resistive switching in the dissipative
  {H}ubbard model},\ }\href {https://doi.org/10.1103/PhysRevLett.114.226403}
  {\bibfield  {journal} {\bibinfo  {journal} {Phys. Rev. Lett.}\ }\textbf
  {\bibinfo {volume} {114}},\ \bibinfo {pages} {226403} (\bibinfo {year}
  {2015})}\BibitemShut {NoStop}%
\bibitem [{\citenamefont {Eckstein}\ \emph {et~al.}(2010)\citenamefont
  {Eckstein}, \citenamefont {Oka},\ and\ \citenamefont {Werner}}]{Werner2010}%
  \BibitemOpen
  \bibfield  {author} {\bibinfo {author} {\bibfnamefont {M.}~\bibnamefont
  {Eckstein}}, \bibinfo {author} {\bibfnamefont {T.}~\bibnamefont {Oka}},\ and\
  \bibinfo {author} {\bibfnamefont {P.}~\bibnamefont {Werner}},\ }\bibfield
  {title} {\bibinfo {title} {Dielectric breakdown of {M}ott insulators in
  dynamical mean-field theory},\ }\href
  {https://doi.org/10.1103/PhysRevLett.105.146404} {\bibfield  {journal}
  {\bibinfo  {journal} {Phys. Rev. Lett.}\ }\textbf {\bibinfo {volume} {105}},\
  \bibinfo {pages} {146404} (\bibinfo {year} {2010})}\BibitemShut {NoStop}%
\bibitem [{\citenamefont {Aron}(2012)}]{camille2012b}%
  \BibitemOpen
  \bibfield  {author} {\bibinfo {author} {\bibfnamefont {C.}~\bibnamefont
  {Aron}},\ }\bibfield  {title} {\bibinfo {title} {Dielectric breakdown of a
  mott insulator},\ }\href {https://doi.org/10.1103/PhysRevB.86.085127}
  {\bibfield  {journal} {\bibinfo  {journal} {Phys. Rev. B}\ }\textbf {\bibinfo
  {volume} {86}},\ \bibinfo {pages} {085127} (\bibinfo {year}
  {2012})}\BibitemShut {NoStop}%
\bibitem [{\citenamefont {Mazza}\ \emph {et~al.}(2015)\citenamefont {Mazza},
  \citenamefont {Amaricci}, \citenamefont {Capone},\ and\ \citenamefont
  {Fabrizio}}]{Mazza2015}%
  \BibitemOpen
  \bibfield  {author} {\bibinfo {author} {\bibfnamefont {G.}~\bibnamefont
  {Mazza}}, \bibinfo {author} {\bibfnamefont {A.}~\bibnamefont {Amaricci}},
  \bibinfo {author} {\bibfnamefont {M.}~\bibnamefont {Capone}},\ and\ \bibinfo
  {author} {\bibfnamefont {M.}~\bibnamefont {Fabrizio}},\ }\bibfield  {title}
  {\bibinfo {title} {Electronic transport and dynamics in correlated
  heterostructures},\ }\href {https://doi.org/10.1103/PhysRevB.91.195124}
  {\bibfield  {journal} {\bibinfo  {journal} {Phys. Rev. B}\ }\textbf {\bibinfo
  {volume} {91}},\ \bibinfo {pages} {195124} (\bibinfo {year}
  {2015})}\BibitemShut {NoStop}%
\bibitem [{\citenamefont {Li}\ \emph {et~al.}(2017)\citenamefont {Li},
  \citenamefont {Aron}, \citenamefont {Kotliar},\ and\ \citenamefont
  {Han}}]{jongfilament2017}%
  \BibitemOpen
  \bibfield  {author} {\bibinfo {author} {\bibfnamefont {J.}~\bibnamefont
  {Li}}, \bibinfo {author} {\bibfnamefont {C.}~\bibnamefont {Aron}}, \bibinfo
  {author} {\bibfnamefont {G.}~\bibnamefont {Kotliar}},\ and\ \bibinfo {author}
  {\bibfnamefont {J.~E.}\ \bibnamefont {Han}},\ }\bibfield  {title} {\bibinfo
  {title} {Microscopic theory of resistive switching in ordered insulators:
  Electronic versus thermal mechanisms},\ }\href
  {https://doi.org/10.1021/acs.nanolett.7b00286} {\bibfield  {journal}
  {\bibinfo  {journal} {Nano Lett.}\ }\textbf {\bibinfo {volume} {17}},\
  \bibinfo {pages} {2994} (\bibinfo {year} {2017})},\ \bibinfo {note} {pMID:
  28394624}\BibitemShut {NoStop}%
\bibitem [{\citenamefont {Chiriac\`o}\ and\ \citenamefont
  {Millis}(2018)}]{millis2018}%
  \BibitemOpen
  \bibfield  {author} {\bibinfo {author} {\bibfnamefont {G.}~\bibnamefont
  {Chiriac\`o}}\ and\ \bibinfo {author} {\bibfnamefont {A.~J.}\ \bibnamefont
  {Millis}},\ }\bibfield  {title} {\bibinfo {title} {Voltage-induced
  metal-insulator transition in a one-dimensional charge density wave},\ }\href
  {https://doi.org/10.1103/PhysRevB.98.205152} {\bibfield  {journal} {\bibinfo
  {journal} {Phys. Rev. B}\ }\textbf {\bibinfo {volume} {98}},\ \bibinfo
  {pages} {205152} (\bibinfo {year} {2018})}\BibitemShut {NoStop}%
\bibitem [{\citenamefont {Han}\ \emph {et~al.}(2023)\citenamefont {Han},
  \citenamefont {Aron}, \citenamefont {Chen}, \citenamefont {Mansaray},
  \citenamefont {Han}, \citenamefont {Kim}, \citenamefont {Randle},\ and\
  \citenamefont {Bird}}]{CamilleJongAvalanche}%
  \BibitemOpen
  \bibfield  {author} {\bibinfo {author} {\bibfnamefont {J.~E.}\ \bibnamefont
  {Han}}, \bibinfo {author} {\bibfnamefont {C.}~\bibnamefont {Aron}}, \bibinfo
  {author} {\bibfnamefont {X.}~\bibnamefont {Chen}}, \bibinfo {author}
  {\bibfnamefont {I.}~\bibnamefont {Mansaray}}, \bibinfo {author}
  {\bibfnamefont {J.-H.}\ \bibnamefont {Han}}, \bibinfo {author} {\bibfnamefont
  {K.-S.}\ \bibnamefont {Kim}}, \bibinfo {author} {\bibfnamefont
  {M.}~\bibnamefont {Randle}},\ and\ \bibinfo {author} {\bibfnamefont {J.~P.}\
  \bibnamefont {Bird}},\ }\bibfield  {title} {\bibinfo {title} {Correlated
  insulator collapse due to quantum avalanche via in-gap ladder states},\
  }\href {https://doi.org/10.1038/s41467-023-38557-8} {\bibfield  {journal}
  {\bibinfo  {journal} {Nature Communications}\ }\textbf {\bibinfo {volume}
  {14}},\ \bibinfo {pages} {2936} (\bibinfo {year} {2023})}\BibitemShut
  {NoStop}%
\bibitem [{\citenamefont {Mazzocchi}\ \emph {et~al.}(2022)\citenamefont
  {Mazzocchi}, \citenamefont {Gazzaneo}, \citenamefont {Lotze},\ and\
  \citenamefont {Arrigoni}}]{arrigoni2022}%
  \BibitemOpen
  \bibfield  {author} {\bibinfo {author} {\bibfnamefont {T.~M.}\ \bibnamefont
  {Mazzocchi}}, \bibinfo {author} {\bibfnamefont {P.}~\bibnamefont {Gazzaneo}},
  \bibinfo {author} {\bibfnamefont {J.}~\bibnamefont {Lotze}},\ and\ \bibinfo
  {author} {\bibfnamefont {E.}~\bibnamefont {Arrigoni}},\ }\href
  {https://doi.org/10.48550/ARXIV.2207.01921} {\bibinfo {title} {Correlated
  {M}ott insulators in strong electric fields: {R}ole of phonons in heat
  dissipation}} (\bibinfo {year} {2022})\BibitemShut {NoStop}%
\bibitem [{\citenamefont {Dasari}\ \emph {et~al.}(2020)\citenamefont {Dasari},
  \citenamefont {Li}, \citenamefont {Werner},\ and\ \citenamefont
  {Eckstein}}]{Hunds2020}%
  \BibitemOpen
  \bibfield  {author} {\bibinfo {author} {\bibfnamefont {N.}~\bibnamefont
  {Dasari}}, \bibinfo {author} {\bibfnamefont {J.}~\bibnamefont {Li}}, \bibinfo
  {author} {\bibfnamefont {P.}~\bibnamefont {Werner}},\ and\ \bibinfo {author}
  {\bibfnamefont {M.}~\bibnamefont {Eckstein}},\ }\bibfield  {title} {\bibinfo
  {title} {Revealing {H}und's multiplets in {M}ott insulators under strong
  electric fields},\ }\href {https://doi.org/10.1103/PhysRevB.101.161107}
  {\bibfield  {journal} {\bibinfo  {journal} {Phys. Rev. B}\ }\textbf {\bibinfo
  {volume} {101}},\ \bibinfo {pages} {161107} (\bibinfo {year}
  {2020})}\BibitemShut {NoStop}%
\bibitem [{\citenamefont {Grandi}\ and\ \citenamefont
  {Eckstein}(2021)}]{eckstein2021}%
  \BibitemOpen
  \bibfield  {author} {\bibinfo {author} {\bibfnamefont {F.}~\bibnamefont
  {Grandi}}\ and\ \bibinfo {author} {\bibfnamefont {M.}~\bibnamefont
  {Eckstein}},\ }\bibfield  {title} {\bibinfo {title} {Ultrafast
  metal-to-insulator switching in a strongly correlated system},\ }\href@noop
  {} {\bibfield  {journal} {\bibinfo  {journal} {arXiv:2104.03644}\ } (\bibinfo
  {year} {2021})}\BibitemShut {NoStop}%
\bibitem [{\citenamefont {N\'ajera}\ \emph {et~al.}(2018)\citenamefont
  {N\'ajera}, \citenamefont {Civelli}, \citenamefont
  {Dobrosavljevi\ifmmode~\acute{c}\else \'{c}\fi{}},\ and\ \citenamefont
  {Rozenberg}}]{najera2}%
  \BibitemOpen
  \bibfield  {author} {\bibinfo {author} {\bibfnamefont {O.}~\bibnamefont
  {N\'ajera}}, \bibinfo {author} {\bibfnamefont {M.}~\bibnamefont {Civelli}},
  \bibinfo {author} {\bibfnamefont {V.}~\bibnamefont
  {Dobrosavljevi\ifmmode~\acute{c}\else \'{c}\fi{}}},\ and\ \bibinfo {author}
  {\bibfnamefont {M.~J.}\ \bibnamefont {Rozenberg}},\ }\bibfield  {title}
  {\bibinfo {title} {Multiple crossovers and coherent states in a mott-peierls
  insulator},\ }\href {https://doi.org/10.1103/PhysRevB.97.045108} {\bibfield
  {journal} {\bibinfo  {journal} {Phys. Rev. B}\ }\textbf {\bibinfo {volume}
  {97}},\ \bibinfo {pages} {045108} (\bibinfo {year} {2018})}\BibitemShut
  {NoStop}%
\bibitem [{\citenamefont {Lu}\ and\ \citenamefont {Robertson}(2019)}]{dist}%
  \BibitemOpen
  \bibfield  {author} {\bibinfo {author} {\bibfnamefont {H.}~\bibnamefont
  {Lu}}\ and\ \bibinfo {author} {\bibfnamefont {J.}~\bibnamefont {Robertson}},\
  }\bibfield  {title} {\bibinfo {title} {Density functional theory studies of
  the metal–insulator transition in vanadium dioxide alloys},\ }\href
  {https://doi.org/10.1002/pssb.201900210} {\bibfield  {journal} {\bibinfo
  {journal} {Phys. Status Solidi B}\ }\textbf {\bibinfo {volume} {256}}
  (\bibinfo {year} {2019})}\BibitemShut {NoStop}%
\bibitem [{\citenamefont {Biermann}\ \emph {et~al.}(2005)\citenamefont
  {Biermann}, \citenamefont {Poteryaev}, \citenamefont {Lichtenstein},\ and\
  \citenamefont {Georges}}]{biermann2005}%
  \BibitemOpen
  \bibfield  {author} {\bibinfo {author} {\bibfnamefont {S.}~\bibnamefont
  {Biermann}}, \bibinfo {author} {\bibfnamefont {A.}~\bibnamefont {Poteryaev}},
  \bibinfo {author} {\bibfnamefont {A.~I.}\ \bibnamefont {Lichtenstein}},\ and\
  \bibinfo {author} {\bibfnamefont {A.}~\bibnamefont {Georges}},\ }\bibfield
  {title} {\bibinfo {title} {Dynamical singlets and correlation-assisted
  peierls transition in ${\mathrm{v}\mathrm{o}}_{2}$},\ }\href
  {https://doi.org/10.1103/PhysRevLett.94.026404} {\bibfield  {journal}
  {\bibinfo  {journal} {Phys. Rev. Lett.}\ }\textbf {\bibinfo {volume} {94}},\
  \bibinfo {pages} {026404} (\bibinfo {year} {2005})}\BibitemShut {NoStop}%
\bibitem [{\citenamefont {Shao}\ \emph {et~al.}(2018)\citenamefont {Shao},
  \citenamefont {Cao}, \citenamefont {Luo},\ and\ \citenamefont
  {Jin}}]{shao2018}%
  \BibitemOpen
  \bibfield  {author} {\bibinfo {author} {\bibfnamefont {Z.}~\bibnamefont
  {Shao}}, \bibinfo {author} {\bibfnamefont {X.}~\bibnamefont {Cao}}, \bibinfo
  {author} {\bibfnamefont {H.}~\bibnamefont {Luo}},\ and\ \bibinfo {author}
  {\bibfnamefont {P.}~\bibnamefont {Jin}},\ }\bibfield  {title} {\bibinfo
  {title} {Recent progress in the phase-transition mechanism and modulation of
  vanadium dioxide materials},\ }\href@noop {} {\bibfield  {journal} {\bibinfo
  {journal} {NPG Asia Mater.}\ }\textbf {\bibinfo {volume} {10}},\ \bibinfo
  {pages} {581} (\bibinfo {year} {2018})}\BibitemShut {NoStop}%
\bibitem [{\citenamefont {Sommers}\ and\ \citenamefont
  {Doniach}(1978)}]{doniach}%
  \BibitemOpen
  \bibfield  {author} {\bibinfo {author} {\bibfnamefont {C.}~\bibnamefont
  {Sommers}}\ and\ \bibinfo {author} {\bibfnamefont {S.}~\bibnamefont
  {Doniach}},\ }\bibfield  {title} {\bibinfo {title} {First principles
  calculation of the intra-atomic correlation energy in vo2},\ }\href
  {https://doi.org/https://doi.org/10.1016/0038-1098(78)90343-5} {\bibfield
  {journal} {\bibinfo  {journal} {Solid State Commun.}\ }\textbf {\bibinfo
  {volume} {28}},\ \bibinfo {pages} {133} (\bibinfo {year} {1978})}\BibitemShut
  {NoStop}%
\bibitem [{\citenamefont {N\'ajera}\ \emph {et~al.}(2017)\citenamefont
  {N\'ajera}, \citenamefont {Civelli}, \citenamefont
  {Dobrosavljevi\ifmmode~\acute{c}\else \'{c}\fi{}},\ and\ \citenamefont
  {Rozenberg}}]{najera1}%
  \BibitemOpen
  \bibfield  {author} {\bibinfo {author} {\bibfnamefont {O.}~\bibnamefont
  {N\'ajera}}, \bibinfo {author} {\bibfnamefont {M.}~\bibnamefont {Civelli}},
  \bibinfo {author} {\bibfnamefont {V.}~\bibnamefont
  {Dobrosavljevi\ifmmode~\acute{c}\else \'{c}\fi{}}},\ and\ \bibinfo {author}
  {\bibfnamefont {M.~J.}\ \bibnamefont {Rozenberg}},\ }\bibfield  {title}
  {\bibinfo {title} {Resolving the {VO}$_{2}$ controversy: Mott mechanism
  dominates the insulator-to-metal transition},\ }\href
  {https://doi.org/10.1103/PhysRevB.95.035113} {\bibfield  {journal} {\bibinfo
  {journal} {Phys. Rev. B}\ }\textbf {\bibinfo {volume} {95}},\ \bibinfo
  {pages} {035113} (\bibinfo {year} {2017})}\BibitemShut {NoStop}%
\bibitem [{\citenamefont {Tomczak}\ \emph {et~al.}(2008)\citenamefont
  {Tomczak}, \citenamefont {Aryasetiawan},\ and\ \citenamefont
  {Biermann}}]{biermann2008}%
  \BibitemOpen
  \bibfield  {author} {\bibinfo {author} {\bibfnamefont {J.~M.}\ \bibnamefont
  {Tomczak}}, \bibinfo {author} {\bibfnamefont {F.}~\bibnamefont
  {Aryasetiawan}},\ and\ \bibinfo {author} {\bibfnamefont {S.}~\bibnamefont
  {Biermann}},\ }\bibfield  {title} {\bibinfo {title} {Effective bandstructure
  in the insulating phase versus strong dynamical correlations in metallic
  {VO}$_{2}$},\ }\href {https://doi.org/10.1103/PhysRevB.78.115103} {\bibfield
  {journal} {\bibinfo  {journal} {Phys. Rev. B}\ }\textbf {\bibinfo {volume}
  {78}},\ \bibinfo {pages} {115103} (\bibinfo {year} {2008})}\BibitemShut
  {NoStop}%
\bibitem [{\citenamefont {Eyert}(2011)}]{eyert2011}%
  \BibitemOpen
  \bibfield  {author} {\bibinfo {author} {\bibfnamefont {V.}~\bibnamefont
  {Eyert}},\ }\bibfield  {title} {\bibinfo {title} {{VO}$_2$: A novel view from
  band theory},\ }\href {https://doi.org/10.1103/PhysRevLett.107.016401}
  {\bibfield  {journal} {\bibinfo  {journal} {Phys. Rev. Lett.}\ }\textbf
  {\bibinfo {volume} {107}},\ \bibinfo {pages} {016401} (\bibinfo {year}
  {2011})}\BibitemShut {NoStop}%
\bibitem [{\citenamefont {Weber}\ \emph {et~al.}(2012)\citenamefont {Weber},
  \citenamefont {O'Regan}, \citenamefont {Hine}, \citenamefont {Payne},
  \citenamefont {Kotliar},\ and\ \citenamefont {Littlewood}}]{weber2012}%
  \BibitemOpen
  \bibfield  {author} {\bibinfo {author} {\bibfnamefont {C.}~\bibnamefont
  {Weber}}, \bibinfo {author} {\bibfnamefont {D.~D.}\ \bibnamefont {O'Regan}},
  \bibinfo {author} {\bibfnamefont {N.~D.~M.}\ \bibnamefont {Hine}}, \bibinfo
  {author} {\bibfnamefont {M.~C.}\ \bibnamefont {Payne}}, \bibinfo {author}
  {\bibfnamefont {G.}~\bibnamefont {Kotliar}},\ and\ \bibinfo {author}
  {\bibfnamefont {P.~B.}\ \bibnamefont {Littlewood}},\ }\bibfield  {title}
  {\bibinfo {title} {Vanadium dioxide: A {P}eierls-{M}ott insulator stable
  against disorder},\ }\href {https://doi.org/10.1103/PhysRevLett.108.256402}
  {\bibfield  {journal} {\bibinfo  {journal} {Phys. Rev. Lett.}\ }\textbf
  {\bibinfo {volume} {108}},\ \bibinfo {pages} {256402} (\bibinfo {year}
  {2012})}\BibitemShut {NoStop}%
\bibitem [{\citenamefont {Brito}\ \emph {et~al.}(2016)\citenamefont {Brito},
  \citenamefont {Aguiar}, \citenamefont {Haule},\ and\ \citenamefont
  {Kotliar}}]{brito2016}%
  \BibitemOpen
  \bibfield  {author} {\bibinfo {author} {\bibfnamefont {W.~H.}\ \bibnamefont
  {Brito}}, \bibinfo {author} {\bibfnamefont {M.~C.~O.}\ \bibnamefont
  {Aguiar}}, \bibinfo {author} {\bibfnamefont {K.}~\bibnamefont {Haule}},\ and\
  \bibinfo {author} {\bibfnamefont {G.}~\bibnamefont {Kotliar}},\ }\bibfield
  {title} {\bibinfo {title} {Metal-insulator transition in {VO}$_{2}$: A
  $\mathrm{DFT}+\mathrm{DMFT}$ perspective},\ }\href
  {https://doi.org/10.1103/PhysRevLett.117.056402} {\bibfield  {journal}
  {\bibinfo  {journal} {Phys. Rev. Lett.}\ }\textbf {\bibinfo {volume} {117}},\
  \bibinfo {pages} {056402} (\bibinfo {year} {2016})}\BibitemShut {NoStop}%
\bibitem [{\citenamefont {dos Santos}(1995)}]{bhm1}%
  \BibitemOpen
  \bibfield  {author} {\bibinfo {author} {\bibfnamefont {R.~R.}\ \bibnamefont
  {dos Santos}},\ }\bibfield  {title} {\bibinfo {title} {Magnetism and pairing
  in {H}ubbard bilayers},\ }\href {https://doi.org/10.1103/PhysRevB.51.15540}
  {\bibfield  {journal} {\bibinfo  {journal} {Phys. Rev. B}\ }\textbf {\bibinfo
  {volume} {51}},\ \bibinfo {pages} {15540} (\bibinfo {year}
  {1995})}\BibitemShut {NoStop}%
\bibitem [{\citenamefont {Kancharla}\ and\ \citenamefont
  {Okamoto}(2007)}]{bhm2}%
  \BibitemOpen
  \bibfield  {author} {\bibinfo {author} {\bibfnamefont {S.~S.}\ \bibnamefont
  {Kancharla}}\ and\ \bibinfo {author} {\bibfnamefont {S.}~\bibnamefont
  {Okamoto}},\ }\bibfield  {title} {\bibinfo {title} {Band insulator to {M}ott
  insulator transition in a bilayer {H}ubbard model},\ }\href@noop {}
  {\bibfield  {journal} {\bibinfo  {journal} {Phys. Rev. B}\ }\textbf {\bibinfo
  {volume} {75}},\ \bibinfo {pages} {193103} (\bibinfo {year}
  {2007})}\BibitemShut {NoStop}%
\bibitem [{\citenamefont {Rüger}\ \emph {et~al.}(2014)\citenamefont {Rüger},
  \citenamefont {Tocchio}, \citenamefont {Valent{\'{\i}}},\ and\ \citenamefont
  {Gros}}]{bhm3}%
  \BibitemOpen
  \bibfield  {author} {\bibinfo {author} {\bibfnamefont {R.}~\bibnamefont
  {Rüger}}, \bibinfo {author} {\bibfnamefont {L.~F.}\ \bibnamefont {Tocchio}},
  \bibinfo {author} {\bibfnamefont {R.}~\bibnamefont {Valent{\'{\i}}}},\ and\
  \bibinfo {author} {\bibfnamefont {C.}~\bibnamefont {Gros}},\ }\bibfield
  {title} {\bibinfo {title} {The phase diagram of the square lattice bilayer
  {H}ubbard model: a variational {M}onte {C}arlo study},\ }\href
  {https://doi.org/10.1088/1367-2630/16/3/033010} {\bibfield  {journal}
  {\bibinfo  {journal} {New J. Phys.}\ }\textbf {\bibinfo {volume} {16}},\
  \bibinfo {pages} {033010} (\bibinfo {year} {2014})}\BibitemShut {NoStop}%
\bibitem [{\citenamefont {Xu}\ \emph {et~al.}(2022)\citenamefont {Xu},
  \citenamefont {Kang}, \citenamefont {Watanabe}, \citenamefont {Taniguchi},
  \citenamefont {Mak},\ and\ \citenamefont {Shan}}]{bhm4}%
  \BibitemOpen
  \bibfield  {author} {\bibinfo {author} {\bibfnamefont {Y.}~\bibnamefont
  {Xu}}, \bibinfo {author} {\bibfnamefont {K.}~\bibnamefont {Kang}}, \bibinfo
  {author} {\bibfnamefont {K.}~\bibnamefont {Watanabe}}, \bibinfo {author}
  {\bibfnamefont {T.}~\bibnamefont {Taniguchi}}, \bibinfo {author}
  {\bibfnamefont {K.~F.}\ \bibnamefont {Mak}},\ and\ \bibinfo {author}
  {\bibfnamefont {J.}~\bibnamefont {Shan}},\ }\bibfield  {title} {\bibinfo
  {title} {A tunable bilayer {H}ubbard model in twisted {WS}e$_2$},\
  }\href@noop {} {\bibfield  {journal} {\bibinfo  {journal} {Nat.
  Nanotechnol.}\ ,\ \bibinfo {pages} {1}} (\bibinfo {year} {2022})}\BibitemShut
  {NoStop}%
\bibitem [{\citenamefont {B\"uttiker}(1985)}]{buttiker1985}%
  \BibitemOpen
  \bibfield  {author} {\bibinfo {author} {\bibfnamefont {M.}~\bibnamefont
  {B\"uttiker}},\ }\bibfield  {title} {\bibinfo {title} {Small normal-metal
  loop coupled to an electron reservoir},\ }\href
  {https://doi.org/10.1103/PhysRevB.32.1846} {\bibfield  {journal} {\bibinfo
  {journal} {Phys. Rev. B}\ }\textbf {\bibinfo {volume} {32}},\ \bibinfo
  {pages} {1846} (\bibinfo {year} {1985})}\BibitemShut {NoStop}%
\bibitem [{\citenamefont {Tsuji}\ \emph {et~al.}(2009)\citenamefont {Tsuji},
  \citenamefont {Oka},\ and\ \citenamefont {Aoki}}]{oka2009}%
  \BibitemOpen
  \bibfield  {author} {\bibinfo {author} {\bibfnamefont {N.}~\bibnamefont
  {Tsuji}}, \bibinfo {author} {\bibfnamefont {T.}~\bibnamefont {Oka}},\ and\
  \bibinfo {author} {\bibfnamefont {H.}~\bibnamefont {Aoki}},\ }\bibfield
  {title} {\bibinfo {title} {Nonequilibrium steady state of photoexcited
  correlated electrons in the presence of dissipation},\ }\href
  {https://doi.org/10.1103/PhysRevLett.103.047403} {\bibfield  {journal}
  {\bibinfo  {journal} {Phys. Rev. Lett.}\ }\textbf {\bibinfo {volume} {103}},\
  \bibinfo {pages} {047403} (\bibinfo {year} {2009})}\BibitemShut {NoStop}%
\bibitem [{\citenamefont {Amaricci}\ \emph {et~al.}(2012)\citenamefont
  {Amaricci}, \citenamefont {Weber}, \citenamefont {Capone},\ and\
  \citenamefont {Kotliar}}]{amaricci2012}%
  \BibitemOpen
  \bibfield  {author} {\bibinfo {author} {\bibfnamefont {A.}~\bibnamefont
  {Amaricci}}, \bibinfo {author} {\bibfnamefont {C.}~\bibnamefont {Weber}},
  \bibinfo {author} {\bibfnamefont {M.}~\bibnamefont {Capone}},\ and\ \bibinfo
  {author} {\bibfnamefont {G.}~\bibnamefont {Kotliar}},\ }\bibfield  {title}
  {\bibinfo {title} {Approach to a stationary state in a driven {H}ubbard model
  coupled to a thermostat},\ }\href
  {https://doi.org/10.1103/PhysRevB.86.085110} {\bibfield  {journal} {\bibinfo
  {journal} {Phys. Rev. B}\ }\textbf {\bibinfo {volume} {86}},\ \bibinfo
  {pages} {085110} (\bibinfo {year} {2012})}\BibitemShut {NoStop}%
\bibitem [{\citenamefont {Han}(2013)}]{jong2013a}%
  \BibitemOpen
  \bibfield  {author} {\bibinfo {author} {\bibfnamefont {J.~E.}\ \bibnamefont
  {Han}},\ }\bibfield  {title} {\bibinfo {title} {Solution of
  electric-field-driven tight-binding lattice coupled to fermion reservoirs},\
  }\href {https://doi.org/10.1103/PhysRevB.87.085119} {\bibfield  {journal}
  {\bibinfo  {journal} {Phys. Rev. B}\ }\textbf {\bibinfo {volume} {87}},\
  \bibinfo {pages} {085119} (\bibinfo {year} {2013})}\BibitemShut {NoStop}%
\bibitem [{\citenamefont {Han}\ and\ \citenamefont {Li}(2013)}]{jong2013}%
  \BibitemOpen
  \bibfield  {author} {\bibinfo {author} {\bibfnamefont {J.~E.}\ \bibnamefont
  {Han}}\ and\ \bibinfo {author} {\bibfnamefont {J.}~\bibnamefont {Li}},\
  }\bibfield  {title} {\bibinfo {title} {Energy dissipation in a
  dc-field-driven electron lattice coupled to fermion baths},\ }\href
  {https://doi.org/10.1103/PhysRevB.88.075113} {\bibfield  {journal} {\bibinfo
  {journal} {Phys. Rev. B}\ }\textbf {\bibinfo {volume} {88}},\ \bibinfo
  {pages} {075113} (\bibinfo {year} {2013})}\BibitemShut {NoStop}%
\bibitem [{\citenamefont {Murakami}\ and\ \citenamefont
  {Werner}(2018)}]{werner2018}%
  \BibitemOpen
  \bibfield  {author} {\bibinfo {author} {\bibfnamefont {Y.}~\bibnamefont
  {Murakami}}\ and\ \bibinfo {author} {\bibfnamefont {P.}~\bibnamefont
  {Werner}},\ }\bibfield  {title} {\bibinfo {title} {Nonequilibrium steady
  states of electric field driven {M}ott insulators},\ }\href
  {https://doi.org/10.1103/PhysRevB.98.075102} {\bibfield  {journal} {\bibinfo
  {journal} {Phys. Rev. B}\ }\textbf {\bibinfo {volume} {98}},\ \bibinfo
  {pages} {075102} (\bibinfo {year} {2018})}\BibitemShut {NoStop}%
\bibitem [{\citenamefont {Moeller}\ \emph {et~al.}(1999)\citenamefont
  {Moeller}, \citenamefont {Dobrosavljevi\ifmmode~\acute{c}\else \'{c}\fi{}},\
  and\ \citenamefont {Ruckenstein}}]{rkkyMott1999}%
  \BibitemOpen
  \bibfield  {author} {\bibinfo {author} {\bibfnamefont {G.}~\bibnamefont
  {Moeller}}, \bibinfo {author} {\bibfnamefont {V.}~\bibnamefont
  {Dobrosavljevi\ifmmode~\acute{c}\else \'{c}\fi{}}},\ and\ \bibinfo {author}
  {\bibfnamefont {A.~E.}\ \bibnamefont {Ruckenstein}},\ }\bibfield  {title}
  {\bibinfo {title} {Rkky interactions and the {M}ott transition},\ }\href
  {https://doi.org/10.1103/PhysRevB.59.6846} {\bibfield  {journal} {\bibinfo
  {journal} {Phys. Rev. B}\ }\textbf {\bibinfo {volume} {59}},\ \bibinfo
  {pages} {6846} (\bibinfo {year} {1999})}\BibitemShut {NoStop}%
\bibitem [{\citenamefont {Monien}\ \emph {et~al.}(1997)\citenamefont {Monien},
  \citenamefont {Elstner},\ and\ \citenamefont {Millis}}]{Millis1997}%
  \BibitemOpen
  \bibfield  {author} {\bibinfo {author} {\bibfnamefont {H.}~\bibnamefont
  {Monien}}, \bibinfo {author} {\bibfnamefont {N.}~\bibnamefont {Elstner}},\
  and\ \bibinfo {author} {\bibfnamefont {A.~J.}\ \bibnamefont {Millis}},\
  }\href {https://doi.org/10.48550/ARXIV.COND-MAT/9707051} {\bibinfo {title}
  {Possible explanation for the absence of bilayer splitting in ybco}}
  (\bibinfo {year} {1997})\BibitemShut {NoStop}%
\bibitem [{Note1()}]{Note1}%
  \BibitemOpen
  \bibinfo {note} {This difference in the nature of the insulating state is
  analogous to the difference between the insulating states of the SOHM
  computed by means of single-site DMFT versus cluster-DMFT. In the latter,
  short-range correlations are kept such as to allow for a lower entropy
  insulating ground state.}\BibitemShut {Stop}%
\bibitem [{\citenamefont {Park}\ \emph {et~al.}(2008)\citenamefont {Park},
  \citenamefont {Haule},\ and\ \citenamefont {Kotliar}}]{kotliar2008}%
  \BibitemOpen
  \bibfield  {author} {\bibinfo {author} {\bibfnamefont {H.}~\bibnamefont
  {Park}}, \bibinfo {author} {\bibfnamefont {K.}~\bibnamefont {Haule}},\ and\
  \bibinfo {author} {\bibfnamefont {G.}~\bibnamefont {Kotliar}},\ }\bibfield
  {title} {\bibinfo {title} {Cluster dynamical mean-field theory of the {M}ott
  transition},\ }\href {https://doi.org/10.1103/PhysRevLett.101.186403}
  {\bibfield  {journal} {\bibinfo  {journal} {Phys. Rev. Lett.}\ }\textbf
  {\bibinfo {volume} {101}},\ \bibinfo {pages} {186403} (\bibinfo {year}
  {2008})}\BibitemShut {NoStop}%
\bibitem [{\citenamefont {Biroli}\ \emph {et~al.}(2004)\citenamefont {Biroli},
  \citenamefont {Parcollet},\ and\ \citenamefont {Kotliar}}]{cdmft2004}%
  \BibitemOpen
  \bibfield  {author} {\bibinfo {author} {\bibfnamefont {G.}~\bibnamefont
  {Biroli}}, \bibinfo {author} {\bibfnamefont {O.}~\bibnamefont {Parcollet}},\
  and\ \bibinfo {author} {\bibfnamefont {G.}~\bibnamefont {Kotliar}},\
  }\bibfield  {title} {\bibinfo {title} {Cluster dynamical mean-field theories:
  Causality and classical limit},\ }\href
  {https://doi.org/10.1103/PhysRevB.69.205108} {\bibfield  {journal} {\bibinfo
  {journal} {Phys. Rev. B}\ }\textbf {\bibinfo {volume} {69}},\ \bibinfo
  {pages} {205108} (\bibinfo {year} {2004})}\BibitemShut {NoStop}%
\bibitem [{\citenamefont {Maier}\ \emph {et~al.}(2005)\citenamefont {Maier},
  \citenamefont {Jarrell}, \citenamefont {Pruschke},\ and\ \citenamefont
  {Hettler}}]{cdmft2005}%
  \BibitemOpen
  \bibfield  {author} {\bibinfo {author} {\bibfnamefont {T.}~\bibnamefont
  {Maier}}, \bibinfo {author} {\bibfnamefont {M.}~\bibnamefont {Jarrell}},
  \bibinfo {author} {\bibfnamefont {T.}~\bibnamefont {Pruschke}},\ and\
  \bibinfo {author} {\bibfnamefont {M.~H.}\ \bibnamefont {Hettler}},\
  }\bibfield  {title} {\bibinfo {title} {Quantum cluster theories},\ }\href
  {https://doi.org/10.1103/RevModPhys.77.1027} {\bibfield  {journal} {\bibinfo
  {journal} {Rev. Mod. Phys.}\ }\textbf {\bibinfo {volume} {77}},\ \bibinfo
  {pages} {1027} (\bibinfo {year} {2005})}\BibitemShut {NoStop}%
\bibitem [{\citenamefont {Tsuji}\ \emph {et~al.}(2014)\citenamefont {Tsuji},
  \citenamefont {Barmettler}, \citenamefont {Aoki},\ and\ \citenamefont
  {Werner}}]{NeqDCA2014}%
  \BibitemOpen
  \bibfield  {author} {\bibinfo {author} {\bibfnamefont {N.}~\bibnamefont
  {Tsuji}}, \bibinfo {author} {\bibfnamefont {P.}~\bibnamefont {Barmettler}},
  \bibinfo {author} {\bibfnamefont {H.}~\bibnamefont {Aoki}},\ and\ \bibinfo
  {author} {\bibfnamefont {P.}~\bibnamefont {Werner}},\ }\bibfield  {title}
  {\bibinfo {title} {Nonequilibrium dynamical cluster theory},\ }\href
  {https://doi.org/10.1103/PhysRevB.90.075117} {\bibfield  {journal} {\bibinfo
  {journal} {Phys. Rev. B}\ }\textbf {\bibinfo {volume} {90}},\ \bibinfo
  {pages} {075117} (\bibinfo {year} {2014})}\BibitemShut {NoStop}%
\bibitem [{\citenamefont {Georges}\ and\ \citenamefont
  {Kotliar}(1992)}]{GeorgesKotliar1992}%
  \BibitemOpen
  \bibfield  {author} {\bibinfo {author} {\bibfnamefont {A.}~\bibnamefont
  {Georges}}\ and\ \bibinfo {author} {\bibfnamefont {G.}~\bibnamefont
  {Kotliar}},\ }\bibfield  {title} {\bibinfo {title} {{H}ubbard model in
  infinite dimensions},\ }\href {https://doi.org/10.1103/PhysRevB.45.6479}
  {\bibfield  {journal} {\bibinfo  {journal} {Phys. Rev. B}\ }\textbf {\bibinfo
  {volume} {45}},\ \bibinfo {pages} {6479} (\bibinfo {year}
  {1992})}\BibitemShut {NoStop}%
\bibitem [{\citenamefont {Zhang}\ \emph {et~al.}(1993)\citenamefont {Zhang},
  \citenamefont {Rozenberg},\ and\ \citenamefont {Kotliar}}]{kotliar1993}%
  \BibitemOpen
  \bibfield  {author} {\bibinfo {author} {\bibfnamefont {X.~Y.}\ \bibnamefont
  {Zhang}}, \bibinfo {author} {\bibfnamefont {M.~J.}\ \bibnamefont
  {Rozenberg}},\ and\ \bibinfo {author} {\bibfnamefont {G.}~\bibnamefont
  {Kotliar}},\ }\bibfield  {title} {\bibinfo {title} {{M}ott transition in the
  $d$=\ensuremath{\infty} {H}ubbard model at zero temperature},\ }\href
  {https://doi.org/10.1103/PhysRevLett.70.1666} {\bibfield  {journal} {\bibinfo
   {journal} {Phys. Rev. Lett.}\ }\textbf {\bibinfo {volume} {70}},\ \bibinfo
  {pages} {1666} (\bibinfo {year} {1993})}\BibitemShut {NoStop}%
\bibitem [{\citenamefont {Davies}\ and\ \citenamefont
  {Wilkins}(1988)}]{wilkins1988}%
  \BibitemOpen
  \bibfield  {author} {\bibinfo {author} {\bibfnamefont {J.~H.}\ \bibnamefont
  {Davies}}\ and\ \bibinfo {author} {\bibfnamefont {J.~W.}\ \bibnamefont
  {Wilkins}},\ }\bibfield  {title} {\bibinfo {title} {Narrow electronic bands
  in high electric fields: Static properties},\ }\href
  {https://doi.org/10.1103/PhysRevB.38.1667} {\bibfield  {journal} {\bibinfo
  {journal} {Phys. Rev. B}\ }\textbf {\bibinfo {volume} {38}},\ \bibinfo
  {pages} {1667} (\bibinfo {year} {1988})}\BibitemShut {NoStop}%
\bibitem [{\citenamefont {Aron}\ \emph {et~al.}(2013)\citenamefont {Aron},
  \citenamefont {Weber},\ and\ \citenamefont {Kotliar}}]{camille2013}%
  \BibitemOpen
  \bibfield  {author} {\bibinfo {author} {\bibfnamefont {C.}~\bibnamefont
  {Aron}}, \bibinfo {author} {\bibfnamefont {C.}~\bibnamefont {Weber}},\ and\
  \bibinfo {author} {\bibfnamefont {G.}~\bibnamefont {Kotliar}},\ }\bibfield
  {title} {\bibinfo {title} {Impurity model for non-equilibrium steady
  states},\ }\href {https://doi.org/10.1103/PhysRevB.87.125113} {\bibfield
  {journal} {\bibinfo  {journal} {Phys. Rev. B}\ }\textbf {\bibinfo {volume}
  {87}},\ \bibinfo {pages} {125113} (\bibinfo {year} {2013})}\BibitemShut
  {NoStop}%
\bibitem [{\citenamefont {Bakalov}\ \emph {et~al.}(2016)\citenamefont
  {Bakalov}, \citenamefont {Nasr~Esfahani}, \citenamefont {Covaci},
  \citenamefont {Peeters}, \citenamefont {Tempere},\ and\ \citenamefont
  {Locquet}}]{bakalov2016}%
  \BibitemOpen
  \bibfield  {author} {\bibinfo {author} {\bibfnamefont {P.}~\bibnamefont
  {Bakalov}}, \bibinfo {author} {\bibfnamefont {D.}~\bibnamefont
  {Nasr~Esfahani}}, \bibinfo {author} {\bibfnamefont {L.}~\bibnamefont
  {Covaci}}, \bibinfo {author} {\bibfnamefont {F.~M.}\ \bibnamefont {Peeters}},
  \bibinfo {author} {\bibfnamefont {J.}~\bibnamefont {Tempere}},\ and\ \bibinfo
  {author} {\bibfnamefont {J.-P.}\ \bibnamefont {Locquet}},\ }\bibfield
  {title} {\bibinfo {title} {Electric-field-driven mott metal-insulator
  transition in correlated thin films: An inhomogeneous dynamical mean-field
  theory approach},\ }\href {https://doi.org/10.1103/PhysRevB.93.165112}
  {\bibfield  {journal} {\bibinfo  {journal} {Phys. Rev. B}\ }\textbf {\bibinfo
  {volume} {93}},\ \bibinfo {pages} {165112} (\bibinfo {year}
  {2016})}\BibitemShut {NoStop}%
\bibitem [{\citenamefont {Okamoto}(2007)}]{Okamoto2007}%
  \BibitemOpen
  \bibfield  {author} {\bibinfo {author} {\bibfnamefont {S.}~\bibnamefont
  {Okamoto}},\ }\bibfield  {title} {\bibinfo {title} {Nonequilibrium transport
  and optical properties of model metal--mott-insulator--metal
  heterostructures},\ }\href {https://doi.org/10.1103/PhysRevB.76.035105}
  {\bibfield  {journal} {\bibinfo  {journal} {Phys. Rev. B}\ }\textbf {\bibinfo
  {volume} {76}},\ \bibinfo {pages} {035105} (\bibinfo {year}
  {2007})}\BibitemShut {NoStop}%
\bibitem [{\citenamefont {Potthoff}\ and\ \citenamefont
  {Nolting}(1999{\natexlab{a}})}]{Potthoff1999a}%
  \BibitemOpen
  \bibfield  {author} {\bibinfo {author} {\bibfnamefont {M.}~\bibnamefont
  {Potthoff}}\ and\ \bibinfo {author} {\bibfnamefont {W.}~\bibnamefont
  {Nolting}},\ }\bibfield  {title} {\bibinfo {title} {Metallic surface of a
  mott insulator--mott insulating surface of a metal},\ }\href
  {https://doi.org/10.1103/PhysRevB.60.7834} {\bibfield  {journal} {\bibinfo
  {journal} {Phys. Rev. B}\ }\textbf {\bibinfo {volume} {60}},\ \bibinfo
  {pages} {7834} (\bibinfo {year} {1999}{\natexlab{a}})}\BibitemShut {NoStop}%
\bibitem [{\citenamefont {Potthoff}\ and\ \citenamefont
  {Nolting}(1999{\natexlab{b}})}]{Potthoff1999b}%
  \BibitemOpen
  \bibfield  {author} {\bibinfo {author} {\bibfnamefont {M.}~\bibnamefont
  {Potthoff}}\ and\ \bibinfo {author} {\bibfnamefont {W.}~\bibnamefont
  {Nolting}},\ }\bibfield  {title} {\bibinfo {title} {Surface metal-insulator
  transition in the hubbard model},\ }\href
  {https://doi.org/10.1103/PhysRevB.59.2549} {\bibfield  {journal} {\bibinfo
  {journal} {Phys. Rev. B}\ }\textbf {\bibinfo {volume} {59}},\ \bibinfo
  {pages} {2549} (\bibinfo {year} {1999}{\natexlab{b}})}\BibitemShut {NoStop}%
\bibitem [{\citenamefont {{Kotliar, G.}}(1999)}]{MeanFieldKotliar1}%
  \BibitemOpen
  \bibfield  {author} {\bibinfo {author} {\bibnamefont {{Kotliar, G.}}},\
  }\bibfield  {title} {\bibinfo {title} {Landau theory of the {M}ott transition
  in the fully frustrated {H}ubbard model in infinite dimensions},\ }\href
  {https://doi.org/10.1007/s100510050914} {\bibfield  {journal} {\bibinfo
  {journal} {Eur. Phys. J. B}\ }\textbf {\bibinfo {volume} {11}},\ \bibinfo
  {pages} {27} (\bibinfo {year} {1999})}\BibitemShut {NoStop}%
\bibitem [{\citenamefont {Kotliar}\ \emph {et~al.}(2000)\citenamefont
  {Kotliar}, \citenamefont {Lange},\ and\ \citenamefont
  {Rozenberg}}]{MeanFieldKotliar2}%
  \BibitemOpen
  \bibfield  {author} {\bibinfo {author} {\bibfnamefont {G.}~\bibnamefont
  {Kotliar}}, \bibinfo {author} {\bibfnamefont {E.}~\bibnamefont {Lange}},\
  and\ \bibinfo {author} {\bibfnamefont {M.~J.}\ \bibnamefont {Rozenberg}},\
  }\bibfield  {title} {\bibinfo {title} {Landau theory of the finite
  temperature {M}ott transition},\ }\href
  {https://doi.org/10.1103/PhysRevLett.84.5180} {\bibfield  {journal} {\bibinfo
   {journal} {Phys. Rev. Lett.}\ }\textbf {\bibinfo {volume} {84}},\ \bibinfo
  {pages} {5180} (\bibinfo {year} {2000})}\BibitemShut {NoStop}%
\bibitem [{\citenamefont {Chitra}\ and\ \citenamefont
  {Kotliar}(2001)}]{MeanFieldKotliar3}%
  \BibitemOpen
  \bibfield  {author} {\bibinfo {author} {\bibfnamefont {R.}~\bibnamefont
  {Chitra}}\ and\ \bibinfo {author} {\bibfnamefont {G.}~\bibnamefont
  {Kotliar}},\ }\bibfield  {title} {\bibinfo {title} {Effective-action approach
  to strongly correlated fermion systems},\ }\href
  {https://doi.org/10.1103/PhysRevB.63.115110} {\bibfield  {journal} {\bibinfo
  {journal} {Phys. Rev. B}\ }\textbf {\bibinfo {volume} {63}},\ \bibinfo
  {pages} {115110} (\bibinfo {year} {2001})}\BibitemShut {NoStop}%
\bibitem [{\citenamefont {Kotliar}(2021)}]{MeanFieldKotliar4}%
  \BibitemOpen
  \bibfield  {author} {\bibinfo {author} {\bibfnamefont {G.}~\bibnamefont
  {Kotliar}},\ }\bibinfo {title} {14 -- {T}he {M}ott {T}ransition},\ in\ \href
  {https://doi.org/doi:10.1515/9780691219530-016} {\emph {\bibinfo {booktitle}
  {More is Different, Fifty Years of Condensed Matter Physics}}},\ \bibinfo
  {editor} {edited by\ \bibinfo {editor} {\bibfnamefont {N.-P.}\ \bibnamefont
  {Ong}}\ and\ \bibinfo {editor} {\bibfnamefont {R.}~\bibnamefont {Bhatt}}}\
  (\bibinfo  {publisher} {Princeton University Press},\ \bibinfo {address}
  {Princeton},\ \bibinfo {year} {2021})\ pp.\ \bibinfo {pages}
  {211--236}\BibitemShut {NoStop}%
\bibitem [{\citenamefont {D\'iaz}\ and\ \citenamefont {Aron}(2022)}]{futureus}%
  \BibitemOpen
  \bibfield  {author} {\bibinfo {author} {\bibfnamefont {M.~I.}\ \bibnamefont
  {D\'iaz}}\ and\ \bibinfo {author} {\bibfnamefont {C.}~\bibnamefont {Aron}},\
  }\bibfield  {title} {\bibinfo {title} {Effective field theory of electric
  field-driven resistive switching}} (\bibinfo {year} {2022}),\ \bibinfo {note}
  {in preparation}\BibitemShut {NoStop}%
\bibitem [{\citenamefont {Stoliar}\ \emph {et~al.}(2013)\citenamefont
  {Stoliar}, \citenamefont {Cario}, \citenamefont {Janod}, \citenamefont
  {Corraze}, \citenamefont {Guillot-Deudon}, \citenamefont {Salmon-Bourmand},
  \citenamefont {Guiot}, \citenamefont {Tranchant},\ and\ \citenamefont
  {Rozenberg}}]{MarceloResistorNetworks1}%
  \BibitemOpen
  \bibfield  {author} {\bibinfo {author} {\bibfnamefont {P.}~\bibnamefont
  {Stoliar}}, \bibinfo {author} {\bibfnamefont {L.}~\bibnamefont {Cario}},
  \bibinfo {author} {\bibfnamefont {E.}~\bibnamefont {Janod}}, \bibinfo
  {author} {\bibfnamefont {B.}~\bibnamefont {Corraze}}, \bibinfo {author}
  {\bibfnamefont {C.}~\bibnamefont {Guillot-Deudon}}, \bibinfo {author}
  {\bibfnamefont {S.}~\bibnamefont {Salmon-Bourmand}}, \bibinfo {author}
  {\bibfnamefont {V.}~\bibnamefont {Guiot}}, \bibinfo {author} {\bibfnamefont
  {J.}~\bibnamefont {Tranchant}},\ and\ \bibinfo {author} {\bibfnamefont
  {M.}~\bibnamefont {Rozenberg}},\ }\bibfield  {title} {\bibinfo {title}
  {Universal electric-field-driven resistive transition in narrow-gap {M}ott
  insulators},\ }\href {https://doi.org/https://doi.org/10.1002/adma.201301113}
  {\bibfield  {journal} {\bibinfo  {journal} {Adv. Mater.}\ }\textbf {\bibinfo
  {volume} {25}},\ \bibinfo {pages} {3222} (\bibinfo {year}
  {2013})}\BibitemShut {NoStop}%
\bibitem [{\citenamefont {Adda}\ \emph {et~al.}(2022)\citenamefont {Adda},
  \citenamefont {Lee}, \citenamefont {Kalcheim}, \citenamefont {Salev},
  \citenamefont {Rocco}, \citenamefont {Vargas}, \citenamefont {Ghazikhanian},
  \citenamefont {Li}, \citenamefont {Albright}, \citenamefont {Rozenberg},\
  and\ \citenamefont {Schuller}}]{MarceloResistorNetworks2}%
  \BibitemOpen
  \bibfield  {author} {\bibinfo {author} {\bibfnamefont {C.}~\bibnamefont
  {Adda}}, \bibinfo {author} {\bibfnamefont {M.-H.}\ \bibnamefont {Lee}},
  \bibinfo {author} {\bibfnamefont {Y.}~\bibnamefont {Kalcheim}}, \bibinfo
  {author} {\bibfnamefont {P.}~\bibnamefont {Salev}}, \bibinfo {author}
  {\bibfnamefont {R.}~\bibnamefont {Rocco}}, \bibinfo {author} {\bibfnamefont
  {N.~M.}\ \bibnamefont {Vargas}}, \bibinfo {author} {\bibfnamefont
  {N.}~\bibnamefont {Ghazikhanian}}, \bibinfo {author} {\bibfnamefont {C.-P.}\
  \bibnamefont {Li}}, \bibinfo {author} {\bibfnamefont {G.}~\bibnamefont
  {Albright}}, \bibinfo {author} {\bibfnamefont {M.}~\bibnamefont
  {Rozenberg}},\ and\ \bibinfo {author} {\bibfnamefont {I.~K.}\ \bibnamefont
  {Schuller}},\ }\bibfield  {title} {\bibinfo {title} {Direct observation of
  the electrically triggered insulator-metal transition in
  ${\mathrm{v}}_{3}{\mathrm{o}}_{5}$ far below the transition temperature},\
  }\href {https://doi.org/10.1103/PhysRevX.12.011025} {\bibfield  {journal}
  {\bibinfo  {journal} {Phys. Rev. X}\ }\textbf {\bibinfo {volume} {12}},\
  \bibinfo {pages} {011025} (\bibinfo {year} {2022})}\BibitemShut {NoStop}%
\end{thebibliography}%

\end{document}